\documentclass[pdflatex,sn-mathphys-ay,onecolumn,a4paper]{sn-jnl}

\usepackage{graphicx}%
\usepackage{multirow}%
\usepackage{amsmath,amssymb,amsfonts}%
\usepackage{amsthm}%
\usepackage{mathrsfs}%
\usepackage[title]{appendix}%
\usepackage{xcolor}%
\usepackage{textcomp}%
\usepackage{manyfoot}%
\usepackage{booktabs}%
\usepackage{algorithm}%
\usepackage{algorithmicx}%
\usepackage{algpseudocode}%
\usepackage{listings}%
\usepackage{changepage}
\usepackage{caption}
\usepackage{subcaption}
\usepackage{natbib}

\begin{document}

\title[Seasonal bias-correction of daily precipitation over France using a stitch model designed for robust extremes representation]{Seasonal bias-correction of daily precipitation over France using a stitch model designed for robust extremes representation}


\newcommand{\phil}[1]{{\color{cyan}#1}}
\newcommand{\ele}[1]{{\color{red}#1}}
\newcommand{\thomas}[1]{{\color{green}#1}}
\author*[1,2]{\fnm{Ear} \sur{Philippe}}\email{philippe.ear@hydroclimat.com}
\equalcont{These authors contributed equally to this work.}

\author[1]{\fnm{Di Bernardino} \sur{Elena}}
\equalcont{These authors contributed equally to this work.}

\author[1]{\fnm{Laloë} \sur{Thomas}}
\equalcont{These authors contributed equally to this work.}

\author[2]{\fnm{Troin} \sur{Magali}}
\equalcont{These authors contributed equally to this work.}

\author[2]{\fnm{Lambert} \sur{Adrien}}
\equalcont{These authors contributed equally to this work.}

\affil[1]{\orgdiv{Laboratoire Jean-Alexandre Dieudonné}, \orgname{Université Côte d'Azur}, \orgaddress{\city{Nice}, \postcode{06108}, \country{France}}}

\affil[2]{\orgname{Hydroclimat}, \orgaddress{\city{Aubagne}, \postcode{13400}, \country{France}}}


\abstract{Highly resoluted and accurate daily precipitation data are required for impact models to perform adequately and to correctly measure high-risk events' impact. In order to produce such data, bias-correction is often needed. Most of those {statistical} methods correct the probability distributions of daily precipitation by modeling them using either empirical or parametric distributions. A recent semi-parametric model based on a penalized Berk-Jones (BJ) statistical test  which allows for an automatic and personalized splicing of parametric and non-parametric  has been developed. This method, called Stitch-BJ model,  was found to be able to model daily precipitation correctly and showed interesting potential in a bias-correction setting. In the present study, we will consolidate these results by taking into account the seasonal properties of daily precipitation in an out-of-sample context, and by considering dry days probabilities in our methodology.  We evaluate the performance of the Stitch-BJ method  in this seasonal bias-correction setting against more classical models such as the Gamma, Exponentiated Weibull (ExpW), Extended Generalized Pareto (EGP) or empirical distributions.
The Stitch-BJ  distribution was able to consistently perform as good or better than all the other models over the validation set, including the empirical distribution, which is often used due to its robustness.}


\maketitle

\section{Introduction}
\label{sec:1}
Highly resoluted and robust daily precipitation data are key for hydrological models, impact models or to produce climate indicators \citep{casanueva_daily_2016,shayeghi_assessing_2024}. Insufficiently resoluted models are not able to produce realistic extreme events due to the smoothing effect of models and them not being convection-permitting \citep{fosser_convection-permitting_2024} . Such extreme events can often lead to dramatic catastrophes with important economic and human impacts if not correctly assessed \citep{alfieri_european_2015,sangati_influence_2009}. However, highly resoluted and robust data are quite scarce and often suffer from either short time series or small available areas. Most common datasets are issued from Global Circulation Models (GCM) which provide global coverage over a long period but at a coarse resolution (typically from 50\textit{km} to 200\textit{km}), which is not resoluted enough for impact studies or to produce realistic extremes \citep{henckes_benefit_2018,prein_precipitation_2016}. Regional Climate Models (RCM) are able to bridge this resolution gap to some extent, with some RCM producing data with a resolution as high as 2x2\textit{km}. However, data from GCM and RCM suffer from bias from multiple sources \citep{wang_global_2014} which need to be corrected \citep{san_daily_2023,xu_evaluation_2015}, and increasing a model resolution may not increase its skill in modeling extreme events compared to its lower resolution counterpart \citep{bador_impact_2020}.

Multiple bias correction methods exist in the literature \citep{deque_frequency_2007,michelangeli_probabilistic_2009} but the most used is often referred as the Quantile Mapping (QM) bias correction method, which is paired to empirical distributions and referred to as Empirical Quantile Mapping (EQM). However, other studies have used parametric distributions instead of the empirical one \citep{li_bias_2021,naveau_modeling_2016,mamalakis_parametric_2017} with promising results. The use of the empirical distribution is often justified by being distribution-free and its robustness. Variation to the EQM method while being fully empirical have also been explored \citep{velasquez_new_2020,byun_improved_2024}. Moreover, classical parametric distributions often fail to correctly model daily precipitation over large and highly resoluted study areas. This is especially true for study areas with a variety of climate zones, such as metropolitain France, containing mountainous regions, plains, as well as regions with a mediterranean climate \citep{joly_types_2010,strohmenger_koppengeiger_2024}.
Empirical models cannot however extrapolate values outside the sample's minimum and maximum, which prevent from producing new extremes that are likely to happen, especially in a climate change context \citep{boe_statistical_2007, deque_frequency_2007,byun_improved_2024}.

Let us remind the quantile mapping method :
\begin{equation}
    \hat{x}_{obs,fut} = F^{-1}_{obs,ref}(F_{mod,ref}(x_{mod,fut})).
    \label{eq:quantile_mapping}
\end{equation}

where $F_{data,period}$ refers to the cumulative distribution function (CDF) of the considered dataset \textit{data} for the period \textit{period}. {Here the datasets considered are the CERRA-Land reanalysis for \textit{obs} and the ERA5-Land reanalaysis for \textit{mod}, while the \textit{ref} period is from the 01/01/1985 to 31/12/2009 and the \textit{fut} period is from 01/01/2010 to 31/12/2020. More details on the datasets can be found in Section \ref{sec:2_1}.}
In the past few years, new parametric and semi-parametric models have been studied to bridge the gap between classical parametric distributions with their abilities to extrapolate extreme values, and the robustness and adaptability of empirical distributions. Such models may be fully parametric, with the presence of a stitch (merging of two distributional models at a given cutting point) or not \citep{naveau_modeling_2016,mamalakis_parametric_2017,derdour_bias_2022}, or semi-parametric \citep{langousis_assessing_2016,holthuijzen_robust_2022,trentini_novel_2023,ear_semi-parametric_2025}.
In order to correctly evaluate the performance of a model, an out-of-sample validation is required, with a separation of training and validation data. Additionally, data is also often separated into months or seasons to remove intra-annual seasonality \citep{katiraie-boroujerdy_bias_2020} and increase confidence on the data's stationarity. Finally, most studies only focus on correcting the wet days intensity, but only a few are also considering dry days probabilities \citep{vrac_bias_2016}. {The study from \cite{ear_semi-parametric_2025} focused on introducing a semi-parametric model referred as to the Stitch-BJ model, but was not applied in a more applicative bias-correction context. In the present paper, we will study the bias-correction performance of the Stitch-BJ model, by comparing it to multiple parametric and non-parametric distributions and applying multiple modifications, while taking into account the following operational constraints :}
\begin{enumerate}
    \item The CERRA-Land and ERA5-Land datasets were separated into a training and validation period (01/01/1985 - 31/12/2009 and 01/01/2010 - 31/12/2020) as discussed in Section \ref{sec:2_1}. This separation makes it possible to include the empirical distribution in the {bias-correction} performance comparison. As already remarked, in this study, CERRA-Land is used in Equation \eqref{eq:quantile_mapping} as \textit{obs} data and ERA5-Land as \textit{mod} data;
    \item A separation using meteorological seasons DJF (i.e. December-January-February), MAM, JJA and SON was used in order to take daily precipitation's seasonality into account and { increase the time series' stationarity} (see Section \ref{sec:2_1} for details);
    \item Correction of dry days probability is included in the bias correction using the Singularity Stochastic Removal from \cite{vrac_bias_2016} as described in Section \ref{sec:2_2}.
\end{enumerate}

We will then study the bias-correction performance using the quantile mapping in \eqref{eq:quantile_mapping} paired with {the Gamma, Exponentiated Weibull, Extended Generalized Pareto, empirical and Stitch-BJ models, all described in Section \ref{sec:2_3}.}

\textit{Structure of the paper.} In Section \ref{sec:2_1}, we present the datasets for daily precipitation over France used in this study and explain how and why they were divided for the bias-correction evaluation. Sections  \ref{sec:2_2} and \ref{sec:2_3} describe the methods used for respectively the dry days' proportion correction and the distributions for wet days modeling.   In Sections \ref{sec:3_1},  \ref{sec:3_2} and in Appendix \ref{app:rmse}, the seasonal bias correction results are presented where we used the CERRA-Land dataset to correct the ERA5-Land reanalysis with respect to several metrics. Section \ref{sec:3_4} compares the differences in performance depending on the season. An analysis of a selected location is presented in Section \ref{sec:3_5}. Finally, the conclusion and discussion on the performances and limitations are included in Section \ref{sec:4}.
\section{Materials and Methods}
\label{sec:2}

\subsection{Daily precipitation datasets}
\label{sec:2_1}

In this study, we consider the whole of metropolitan France, covering over $550000km^2$. Two reanalyses were used to evaluate the bias-correction performance of our models:
\begin{itemize}
    \item The ERA5-Land (ERA5L) reanalysis \citep{munoz-sabater_era5-land_2021} with a resolution of $0.1^{\circ}\times 0.1^{\circ}$, spanning from 01/01/1950 to 31/12/2021;
    \item The Copernicus Regional Reanalysis for Europe (CERRA-Land, CERRAL) reanalysis \citep{verrelle_cerra-land_2021} with a resolution of $5.5 \times 5.5 km$ and spanning from 01/01/1984 to 31/07/2021.
\end{itemize}
Both reanalysis were produced by the European Center for Medium-Range Weather Forecast (ECMWF) and are freely available through their \hyperlink{link:data}{Climate Data Store} (see the \textit{Data Availability Statement} section at the end of the present paper).

The ERA5L reanalysis is used as the model to be corrected (\textit{mod} data as in \eqref{eq:quantile_mapping}),  while CERRAL is used as a gridded observational dataset (\textit{obs} data as in \eqref{eq:quantile_mapping}). The aim is to use bias-correction methods to match as closely ERA5L data to CERRAL data. The datasets used here are not free of bias \citep{pelosi_performance_2023,guo_does_2024}, they have been chosen in this study for illustration purposes but can be swapped for any other datasets that may have a different resolution or bias. Datasets were chosen due to their similar time-period availability as well as their similar spatial coverage for bias-correction performance evaluation.

The CERRAL data was interpolated into a regular longitude/latitude grid (approximately $0.05^{\circ}\times 0.05^{\circ}$) using  the linear scattered interpolant (\texttt{scatteredInterpolant}) from \texttt{MATLAB}. The ERA5L data was interpolated to the same grid using the \texttt{CloughTocher2D} interpolator from \texttt{Python}'s library \texttt{SciPy}.

\begin{figure}[!h]
\centering
 \begin{subfigure}[b]{0.45\textwidth}   
    \centering
    \includegraphics[width=\textwidth]{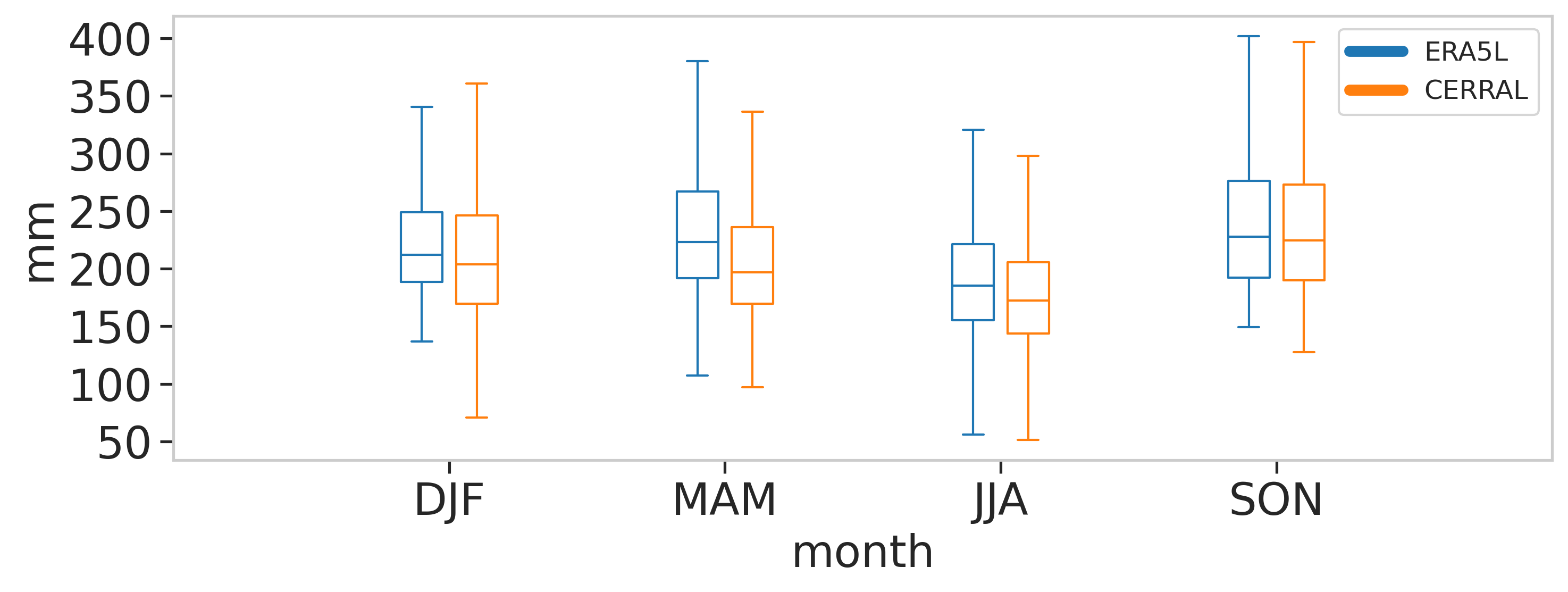}
    \caption{}
\end{subfigure}
\hfill
 \begin{subfigure}[b]{0.45\textwidth}   
    \centering
    \includegraphics[width=\textwidth]{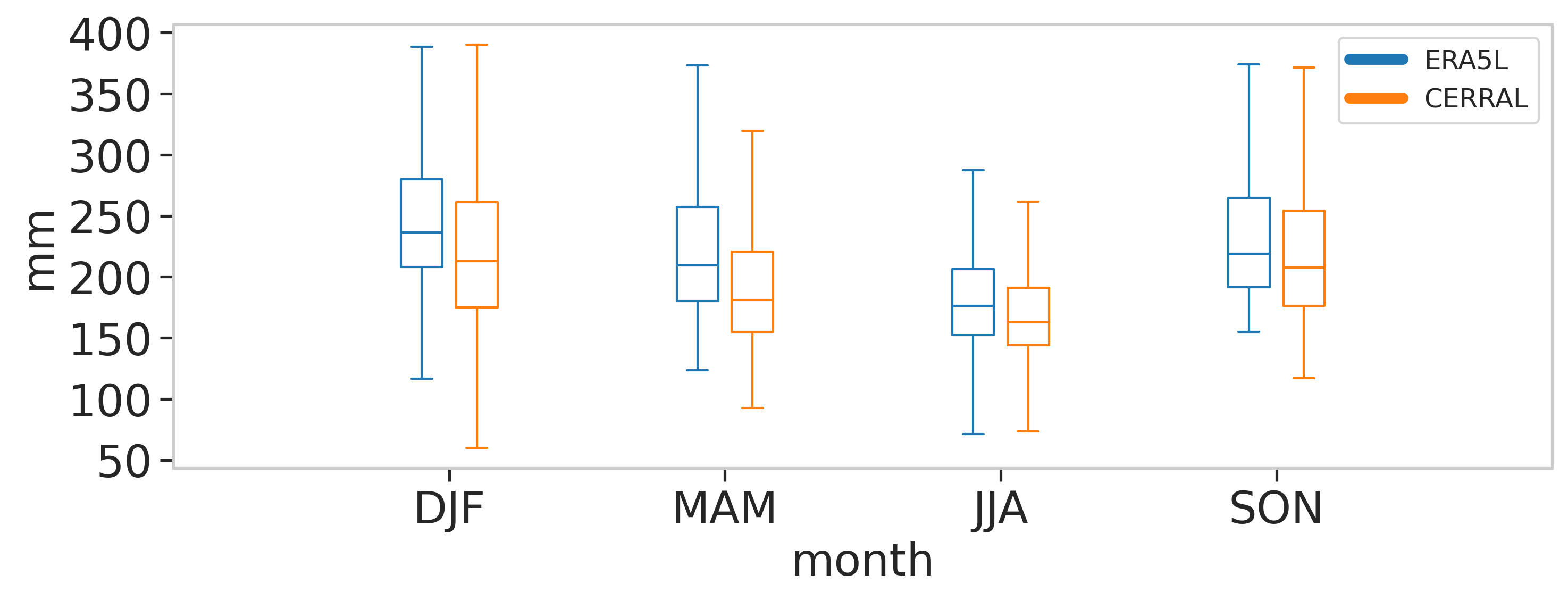}
    \caption{}
\end{subfigure}
\caption{Yearly mean of seasonal cumulated precipitation for ERA5L and CERRAL over the training (a) and validation (b) periods.}
\label{fig:monthly_cum_pr}
\end{figure}

We selected a common period for both reanalyses, starting from 01/01/1985 to 31/12/2020. Both reanalyses were separated using the meteorological seasons : DJF, MAM, JJA, SON as in \cite{katiraie-boroujerdy_bias_2020} and \cite{gutierrez_intercomparison_2019}. Discussion on the impact of subsampling on quantile mapping performance can be found in \cite{reiter_does_2018}.
The seasonality must be taken into account because of the intra-annual non-stationarity of daily precipitation. In our study area (Metropolitain France), summers see in general the least amount of precipitation, while autumn sees the most precipitation \citep{chaouche_analyses_2010}, as seen in Figure \ref{fig:monthly_cum_pr}. 

We performed two-sample t-tests between the means of each season for the period 01-12-1985 to 30/11/2009 to try to detect significative differences in the inter-seasonal mean. Our results in Table \ref{tab:percentage_seasons} show an important rejection rate of the test at a significance level $\alpha=5\%$ for all pairings, with a minimum of 35\% of locations rejected (JJA-MAM on ERA5L data), and a maximum of 80\% (SON-MAM on ERA5L data) of rejected locations. Such differences in means contradict the constant mean hypothesis for weak-stationarity in time series. 

\begin{table}[h]
\centering

\resizebox{.5\columnwidth}{!}{
\begin{subtable}{\textwidth}
\begin{tabular}{|c|c|c|c|c|}
\hline
\textbf{}    & \textbf{DJF}  & \textbf{MAM}  & \textbf{JJA}  & \textbf{SON}  \\ \hline
\textbf{DJF} & x             & \textbf{0.54} & \textbf{0.52} & \textbf{0.50} \\ \hline
\textbf{MAM} & \textit{0.50} & x             & \textbf{0.35} & \textbf{0.80} \\ \hline
\textbf{JJA} & \textit{0.60} & \textit{0.49} & x             & \textbf{0.55} \\ \hline
\textbf{SON} & \textit{0.51} & \textit{0.79} & \textit{0.49} & x             \\ \hline
\end{tabular}
\end{subtable}
}
\caption{Proportion of rejected locations of the Student's t-test for seasonal means differences. Note that values in \textbf{bold} correspond to ERA5L results, while values in \textit{italic} correspond to CERRAL results.}
\label{tab:percentage_seasons}
\end{table}

However, when testing the intra-seasonal means with a monthly division, we manage to greatly reduce the rejection rate for our Student's t-tests as shown in Table \ref{tab:percentage_months}. In the table, each season is decomposed into its 3 months component e.g; DJF is decomposed into December, January and February, and are then referred to as respectively 1, 2 and 3.  The two samples Student's t-test is then applied on every pair of months for the given season. In the example, to get the proportion of rejection for December against February, one would need to look at the DJF table, on either the first column and last line for CERRAL data, or the last column and first line for ERA5L data. The maximum rejection proportion is at 33\% (for March against April in the MAM subtable on ERA5L data) which is lower than the lowest rejection rate from Table \ref{tab:percentage_seasons}. While a constant mean is not sufficient for stationarity, separation into seasons seems to be an acceptable precautionary step to take to improve the homogeneity of our data.

\begin{table}[h]
\centering
\begin{tabular}{@{}cccclcccc@{}}
\textbf{DJF}         & \textbf{1}           & \textbf{2}           & \textbf{3}           &  & \textbf{MAM}         & \textbf{1}           & \textbf{2}           & \textbf{3}           \\ \cmidrule(r){1-4} \cmidrule(l){6-9} 
\textbf{1}           & x                    & \textbf{0.05}        & \textbf{0.32}        &  & \textbf{1}           & x                    & \textbf{0.33}        & \textbf{0.26}        \\
\textbf{2}           & \textit{0.10}        & x                    & \textbf{0.17}        &  & \textbf{2}           & \textit{0.19}        & x                    & \textbf{0.12}        \\
\textbf{3}           & \textit{0.25}        & \textit{0.06}        & x                    &  & \textbf{3}           & \textit{0.24}        & \textit{0.18}        & x                    \\
\multicolumn{1}{l}{} & \multicolumn{1}{l}{} & \multicolumn{1}{l}{} & \multicolumn{1}{l}{} &  & \multicolumn{1}{l}{} & \multicolumn{1}{l}{} & \multicolumn{1}{l}{} & \multicolumn{1}{l}{} \\
\textbf{JJA}         & \textbf{1}           & \textbf{2}           & \textbf{3}           &  & \textbf{SON}         & \textbf{1}           & \textbf{2}           & \textbf{3}           \\ \cmidrule(r){1-4} \cmidrule(l){6-9} 
\textbf{1}           & x                    & \textbf{0.11}        & \textbf{0.11}        &  & \textbf{1}           & x                    & \textbf{0.14}        & \textbf{0.30}        \\
\textbf{2}           & \textit{0.17}        & x                    & \textbf{0.03}        &  & \textbf{2}           & \textit{0.13}        & x                    & \textbf{0.15}        \\
\textbf{3}           & \textit{0.10}        & \textit{0.08}        & x                    &  & \textbf{3}           & \textit{0.25}        & \textit{0.18}        & x                   
\end{tabular}
\captionsetup{width=\textwidth}

\caption{Proportion of rejected locations of the Student's t-test for monthly means differences. Note that values in \textbf{bold} correspond to ERA5L results, while values in \textit{italic} correspond to CERRAL results.}
\label{tab:percentage_months}
\end{table}

Moreover, a separation into a training and validation set was done with the period from 01/01/1985 to 31/12/2009 (25 years) used for training and the period 01/01/2010 to 31/12/2020 (11 years) is used for validation. 
Figure \ref{fig:monthly_cum_pr} not only shows the intra-annual variability of daily precipitation, justifying the seasonal separation of our datasets but also the wet bias (in terms on total cumulated intensity here) the ERA5L dataset suffers compared to the CERRAL data. This can be noticed in panel (b) where for all seasons, the CERRAL dataset produced a median annual cumulated precipitation lower than ERA5L. More details on the wet or dry bias different datasets may suffer and how to correct them are discussed in Section \ref{sec:2_2}.

{The JJA season produced much less total precipitation (15\% to 30\% less) than the other seasons for both training and validation period. This further shows the need to seasonally divide our data as stationary time-series should have a constant mean throughout.}

This separation was chosen to keep a sufficient amount of wet days in the dry season for the training period since the fitting ability of parametric distributions is greatly dependent on the number of available samples \citep{braunstein_how_1992}. Using 25 years for the training period, some locations can receive as few as 104 (resp. 126) rainy days during the JJA season for respectively the ERA5L and CERRAL datasets.
\subsection{On the correction of number of dry days and rain's probability}

Most of the daily precipitation data coming from climate models (GCM, RCM or reanalysis) suffer from the drizzle effect \citep{chen_convective--total_2021,gutowski_temporalspatial_2003,argueso_precipitation_2013} with too many low precipitations occurrences. The drizzle effect, {which happens with numerical models producing a high number of rainy days of very small intensity,} can have an important impact on the total precipitation amount over an extended period. The spatial and temporal distribution of precipitation for a given location is also affected, with daily precipitation-induced indicators such as the Cumulative Dry Days (CDD) being greatly impacted by this drizzle effect \citep{maraun_bias_2013}.

While both considered datasets share similarities in their model and forcing data, the differences in number of wet days and overall daily rain probability can be important. These differences can be due to both intrinsic differences from the model, but most likely from the difference in spatial resolution.

\begin{figure}[!h]
\centering
 \begin{subfigure}[b]{0.45\textwidth}   
    \centering
    \includegraphics[width=\textwidth]{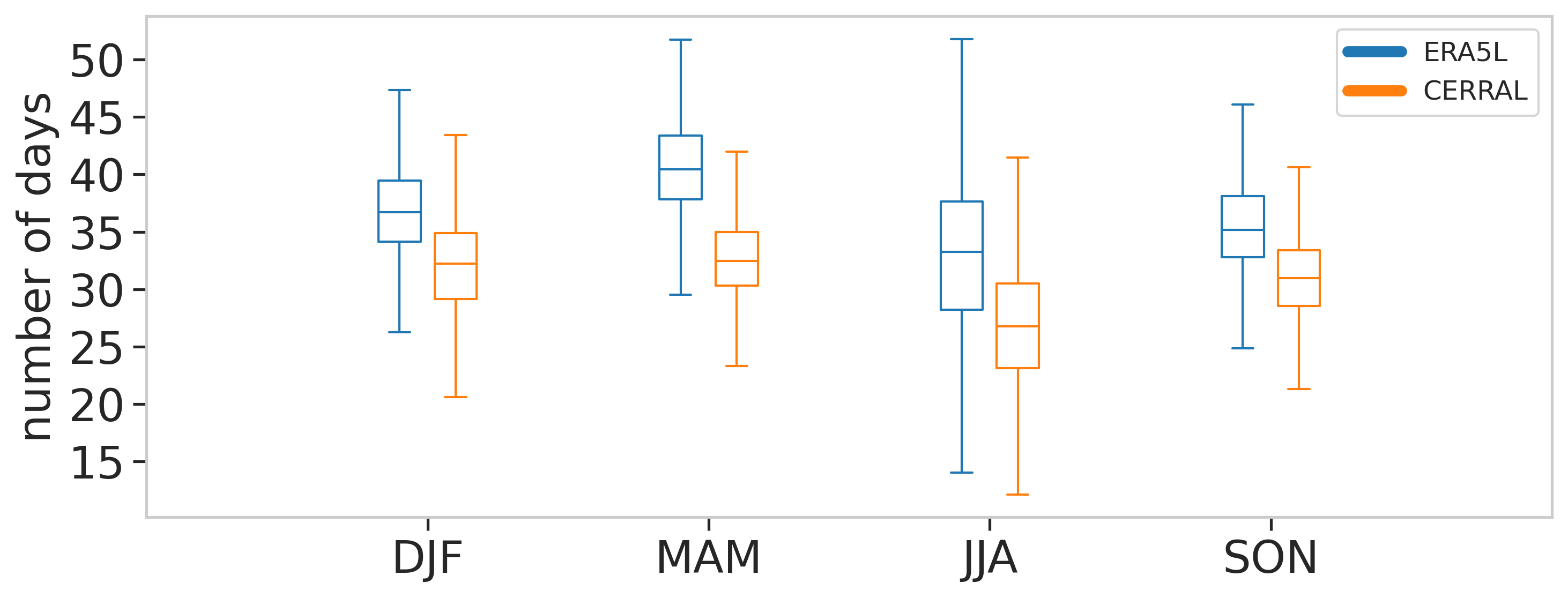}
    \caption{}
\end{subfigure}
\hfill
 \begin{subfigure}[b]{0.45\textwidth}   
    \centering
    \includegraphics[width=\textwidth]{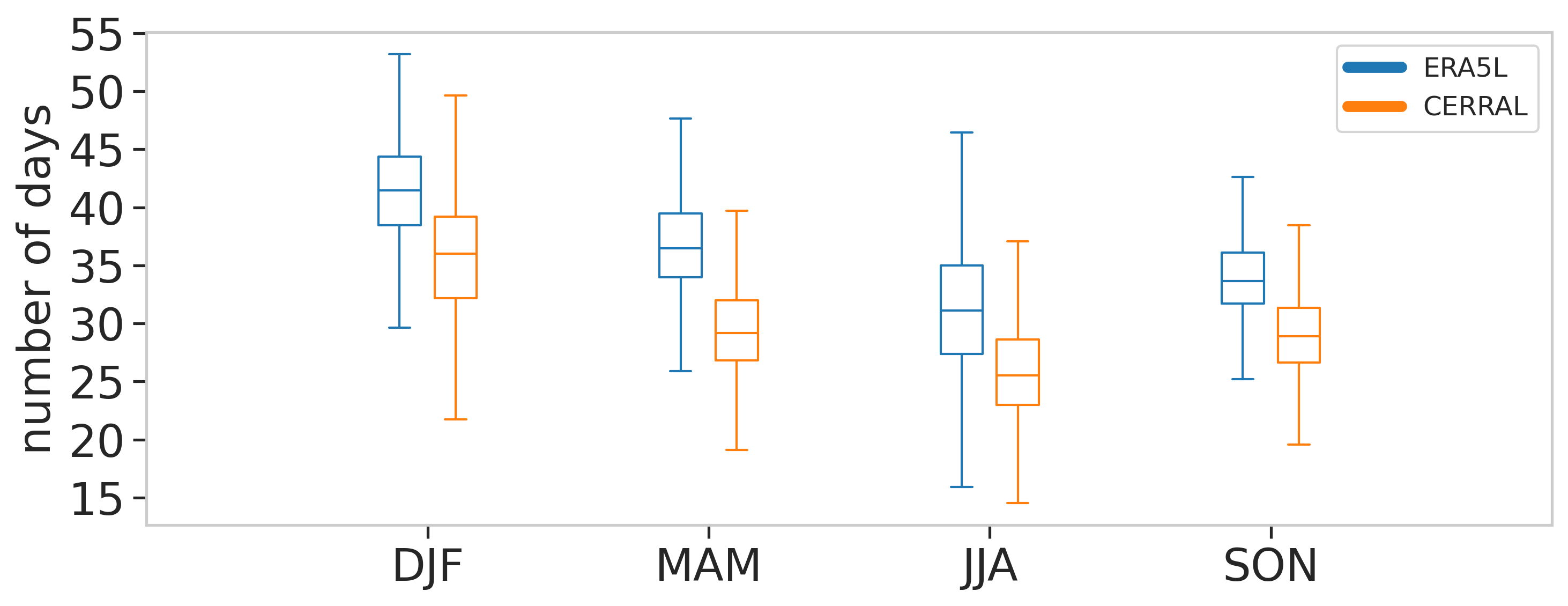}
    \caption{}
\end{subfigure}
\caption{Yearly mean of number of wet days (>1mm) for ERA5L and CERRAL per season for the training (a) and validation (b) periods.}
\label{fig:BP_mult_nb_rain}
\end{figure}

In Figure \ref{fig:BP_mult_nb_rain}, we have the number of wet days (days with more than 1\textit{mm} of rain) for each season, for both datasets in the training (a) and validation (b) period. There is a clear wet bias from the ERA5L model, with not only a higher median of the number of wet days but also a higher extreme number of wet days. 
The median differences of number of wet days between ERA5L and CERRAL is roughly 5 to 8 days per year, but can reach up to 35 days on some locations. This results in almost 1000 more rainy days during the training period for such locations.

\begin{figure}[!h]
\centering
 \begin{subfigure}[b]{0.45\textwidth}   
    \centering
    \includegraphics[width=\textwidth]{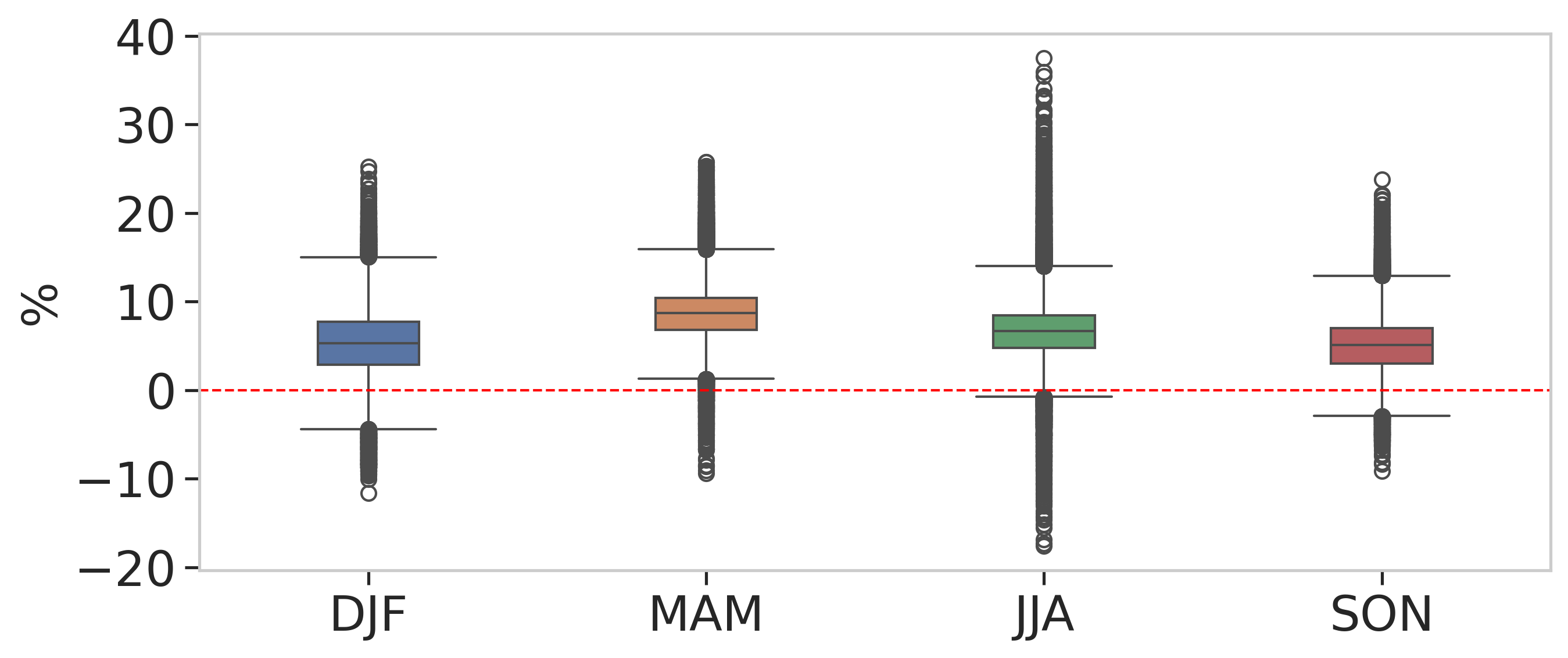}
    \caption{}
\end{subfigure}
\hfill
 \begin{subfigure}[b]{0.45\textwidth}   
    \centering
    \includegraphics[width=\textwidth]{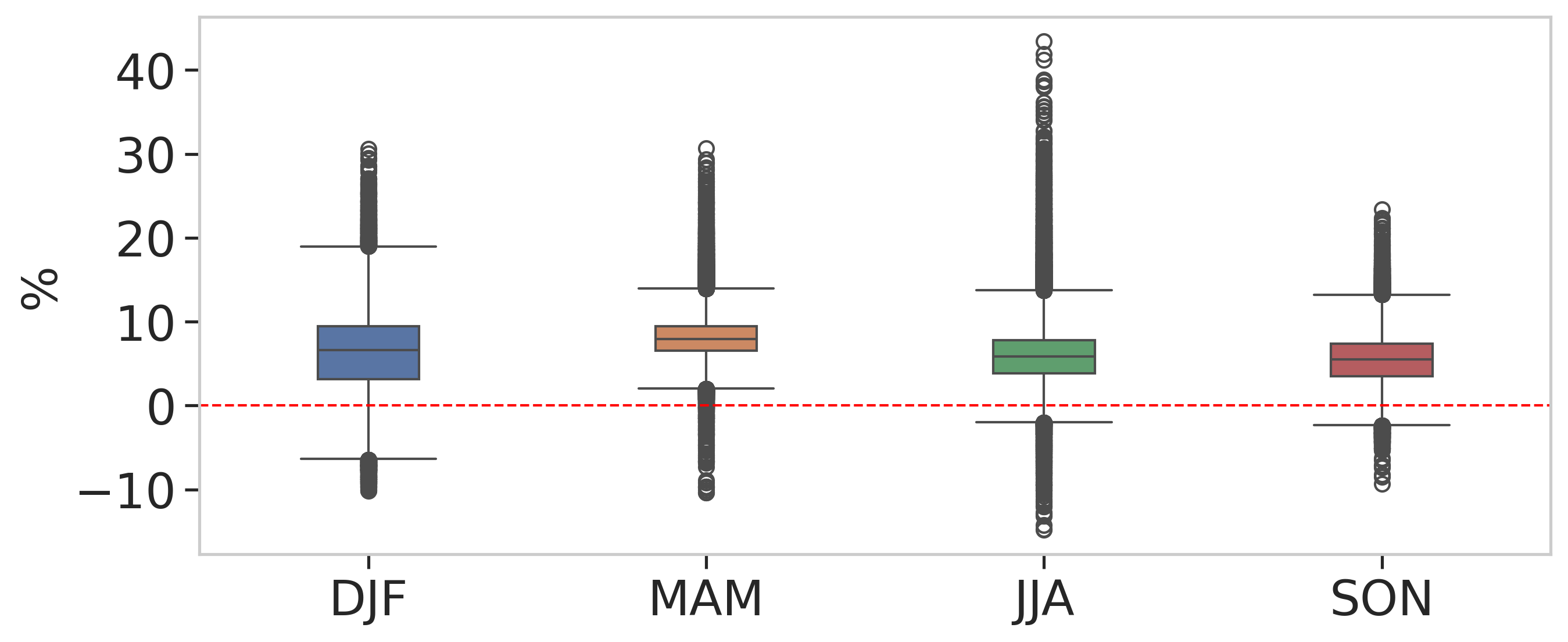}
    \caption{}
\end{subfigure}
\caption{Boxplot of differences in probability of rain (days with >1\textit{mm} of rain) of ERA5L against CERRAL by season for the training (a) and validation (b) periods.}
\label{fig:BP_mult_diff_prob_rain}
\end{figure}

Let us consider the difference in the probability of rain in Figure \ref{fig:BP_mult_diff_prob_rain}, where a positive value means a higher probability of rain in ERA5L compared to CERRAL for a given location.

In Figure \ref{fig:BP_mult_diff_prob_rain}, we can see a 5\% higher probability of rain across all seasons for ERA5L, with some locations with 20 to 40\% more rainy days in ERA5L than CERRAL. On the other hand, CERRAL locations have at most around 15\% more rain than ERA5L, and this only happens rarely. {Results are very similar for the training and validation period which seems to indicate that the evolution in rain probabilities evolved similarly in both datasets between the two periods.}

If left as it is, not only the resulting bias-corrected time series will keep the wet bias, but the amount of rain might also be greatly superior to the target time series. The wet or dry bias must then be corrected to reduce the number of bias sources and produce bias-corrected data as trustworthy as possible.
{
Multiple methods are available to correct differences in dry days probability with the threshold adaptation method \citep{schmidli_downscaling_2006,lavaysse_statistical_2012}, positive correction method \citep{mao_stochastic_2015}, or a more direct approach \citep{piani_statistical_2010,vigaud_probabilistic_2013}. A short intercomparison was done by \cite{vrac_bias_2016} where a new method in the direct approach family, able to correct both wet and dry bias was also introduced as the Singularity Stochastic Removal method (SSR). The positive threshold method is the most widely used with various thresholds considered by the scientific community \citep{semenov_comparison_1998,ambrosino_rainfall-derived_2014, vaittinada_ayar_intercomparison_2016, bouvier_generating_2003,ear_semi-parametric_2025}. However, it is not able to take into account dry bias or the specificity of each location.
}
Dry bias is much more difficult to correct than wet bias since the correction is two-fold: the intensity needs to be chosen as well as the temporal location of the correction.
Randomly adding wet days to the time series may break temporal correlation structure, precipitation being notably autocorrelated \citep{khan_spatio-temporal_2007,abbott_long_2016}.

The SSR method \citep{vrac_bias_2016} corrects both wet and dry bias, location-wise. However, for the dry bias correction, randomly chosen dry days are turned into wet days (thus the stochastic part) without any regard to the autocorrelation of daily precipitation data. 

While the SSR method does not take autocorrelation into account, it allows for a simple method to correct dry days proportion, with no distinction between adding or removing wet days which could be the case for other methods.

The SSR method applies the following steps:

\begin{enumerate}
    \item Select a threshold $th$ such that any value above $th$ is considered a wet day and any value below is considered a dry day. This can either be a common threshold or the minimum positive value of all datasets;
    \item Set all days below $th$ (null days) to a random uniformly taken between $0$ and $th$;
    \item Perform the bias correction technique;
    \item Set the bias-corrected data lower than $th$ to 0.
\end{enumerate}

In the case where a distribution has been fitted to the wet days' data, the method will be applied as such, considering the quantile mapping for the bias correction method (see Equation \eqref{eq:quantile_mapping}).

Let $F_{CERRAL,ref},F_{ERA5L,ref}$  be the fitted models to respectively the reference observation data and reference model data.
Let $x_{ERA5L,fut}$ be the to-be-corrected future model data.
Let $\alpha_{CERRAL,ref}$ and $\alpha_{ERA5L,ref}$ be the probability of dry days respectively for the reference observed and reference modeled.

Quantile mapping methods are usually only applied to correct wet days distributions \citep{themesl_empirical-statistical_2012,ajaaj_comparison_2016,lafon_bias_2013}. To include the dry day proportion correction, let us define:

\begin{equation}
\widetilde{F}_{ERA5L,ref}(x) =
\begin{cases}
    F_{ERA5L,ref}(x)\left(1-\alpha_{ERA5L,ref}\right) + \alpha_{ERA5L,ref} & x\geq th, \\
    \frac{\alpha_{ERA5L,ref}}{th} x & x<th,  
\end{cases}
\label{eq:ssr_param1}
\end{equation}

and 

\begin{equation}
    \widetilde{F}^{-1}_{CERRAL,ref}(x) =
    \begin{cases}
        F^{-1}_{CERRAL,ref}\left(\frac{p-\alpha_{CERRAL,ref}}{1-\alpha_{CERRAL,ref}}\right) & p \geq \alpha_{CERRAL,ref},\\
        \frac{th}{\alpha_{CERRAL,ref}} p & p < \alpha_{CERRAL,ref}.
    \end{cases}
    \label{eq:ssr_param2}
\end{equation}

In Equations \eqref{eq:ssr_param1} and \eqref{eq:ssr_param2}, any probability lower than $\alpha_{CERRAL,ref}$ will result in a value lower than $th$.
The SSR method will both correct the dry days proportion with regards to the reference period, but also keep the evolution of dry days proportion between the reference and future period.

\label{sec:2_2}

\subsection{Parametric,  semi-parametric and non-parametric models}
\label{sec:2_3}

Multiple distributional models were used in the bias correction procedure to compare their performances. We mostly used the same distributions as in \citep{ear_semi-parametric_2025} but we also added the empirical distribution for  comparison. 
Distributions were fitted using the Maximum Likelihood Estimator (MLE)  for all parametric models. Fitting was done location-wise, seasonally and using only wet days (>1mm). When available, a 1\textit{mm} location parameter was used, or a 1\textit{mm} shift was applied to the fitting procedure for distributions without one.  A left-censor has been used for the EGPD as in \citep{naveau_modeling_2016}, and the threshold has been determined through trials not shown here.

In the present study, the distributions used are the Gamma, Exponentiated Weibull, Extended Generalized Pareto, Stitch-BJ distribution and empirical model.
\smallskip

\textbf{Gamma distribution}
The Gamma distribution is one of the most used parametric distribution for daily rainfall \citep{martinez-villalobos_why_2019, husak_use_2007}.
The distribution is defined by its cumulative distribution function (CDF) as follows:

\begin{equation*}
    F(x) = \frac{1}{\Gamma(x)}\gamma\left(k,\frac{x}{\theta}\right),
\end{equation*}

with $x>0, k>0$ and $\theta>0$. $\Gamma$ is the gamma function and $\gamma$ is the lower incomplete gamma function.
\smallskip

\textbf{Exponentiated Weibull distribution.}  
The Exponentiated Weibull (ExpW) generalizes the Weibull distribution and while it has historically been used for failure rates and survival data modelisation, \citep{khan_exponentiated_2018,mudholkar_generalization_1996}, the ExpW has also been used for precipitation modeling \citep{sharma_exponentiated_2022}.
This distribution is defined by its CDF:
\begin{equation*}
    F(x) = \left[1-e^{-(x/\lambda)^k}\right]^\alpha, 
\end{equation*}
with $x>0, k>0, \alpha>0$ and $\lambda>0$.
With $\alpha=1$, we retrieve the original Weibull distribution.
\smallskip

\textbf{Extended Generalized Pareto distribution.}  
Here we recall a family of distributions which has theoretical properties in line with  Extreme Value Theory for both the lower and upper heavy-tail behavior (see  \cite{naveau_modeling_2016}). The Extended Generalized Pareto distribution (EGP)  has recently been used in multiple studies for rainfall modeling due to its flexibility and theoretical guarantees \citep{tencaliec_flexible_2019,rivoire_comparison_2021,haruna_modeling_2023}. We will only use the type 1 EGPD with the same 3\textit{mm} left-censoring as used in \cite{ear_semi-parametric_2025}. The CDF of the type 1 EGPD is:

\begin{equation*}
    F(x) = \begin{cases}
        \left(1-(1+\xi x/\sigma) ^{-\frac{1}{\xi}}\right)^\kappa &,\xi>0, \\
        \left(1-e^{-x/\sigma}\right)^\kappa &,\xi=0,
    \end{cases}
\end{equation*}
with $x>0, \sigma>0, \xi \geq 0$ and $\kappa >0$.
Notice that for $\xi=0$, the model matches the previous ExpW model with $k=1$. 
\smallskip

\textbf{Empirical model.} 
The empirical distribution is often used in climate studies {for its robustness}  \citep{enayati_bias_2021}. The CDF can be written as follows:

\begin{equation*}
    F_n(x) = \frac{1}{n}\sum_{i=1}^n\mathbf{1}_{\{x_{(i)}\leq x\}}
\end{equation*}

with $x_{(1)},...,x_{(n)}$ the order statistics of the considered sample, and $n$ the sample size.
\smallskip

\textbf{Stitch-BJ model.} The Stitch-BJ model is a semi-parametric distributional model based on a penalized version of the Berk-Jones statistical test (PBJ) recently introduced in \cite{ear_semi-parametric_2025}.

The model is a flexible distribution which allows for an automatic stitch between the EGP, ExpW and empirical distributions using the PBJ. For more information, we refer to the original article where details can be found, as well as an analytical form of the resulting final CDF.

\begin{figure}[!h]
\centering
 \begin{subfigure}[b]{0.45\textwidth}   
    \centering
    \includegraphics[width=\textwidth]{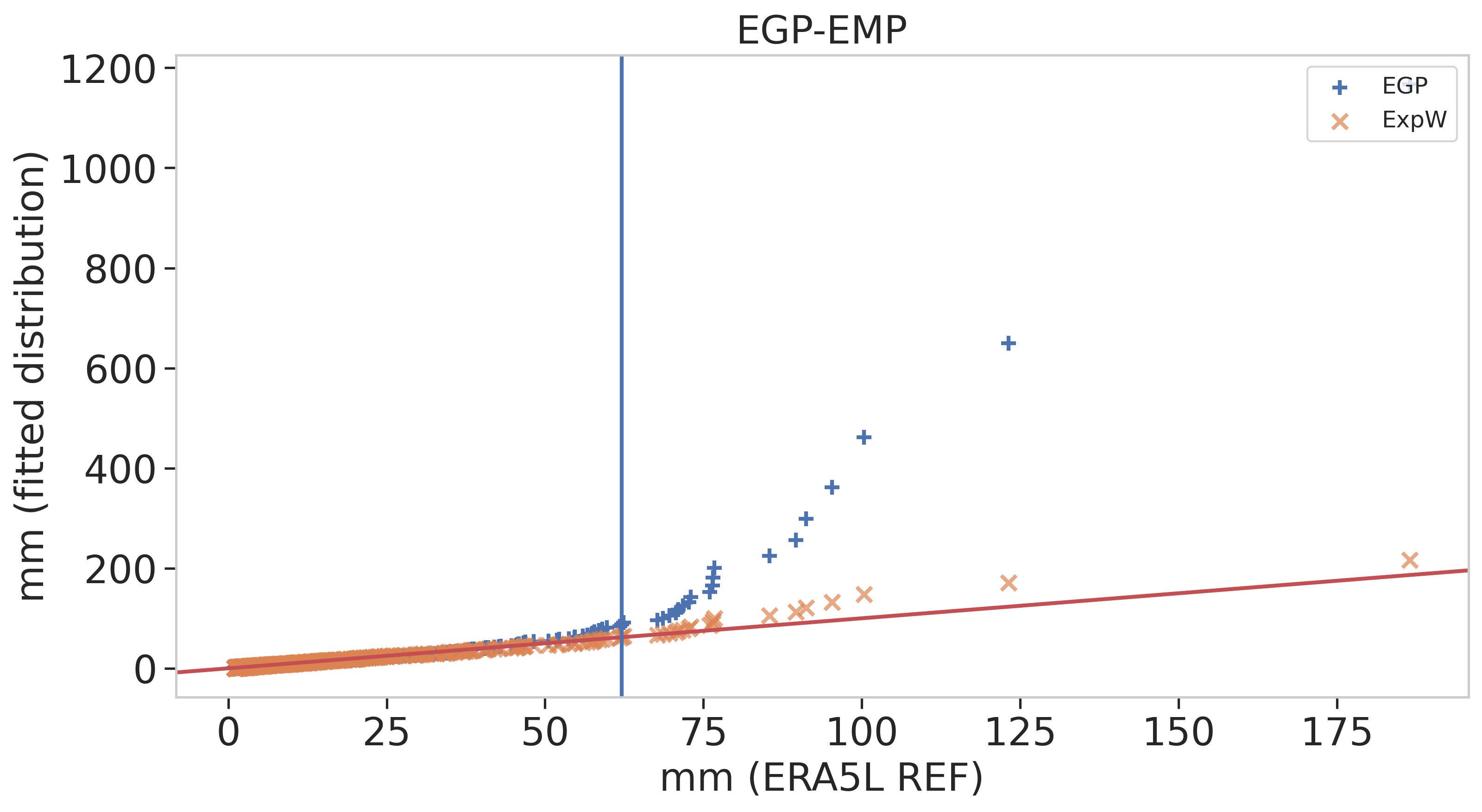}
    \caption{}
\end{subfigure}
\hfill
 \begin{subfigure}[b]{0.45\textwidth}   
    \centering
    \includegraphics[width=\textwidth]{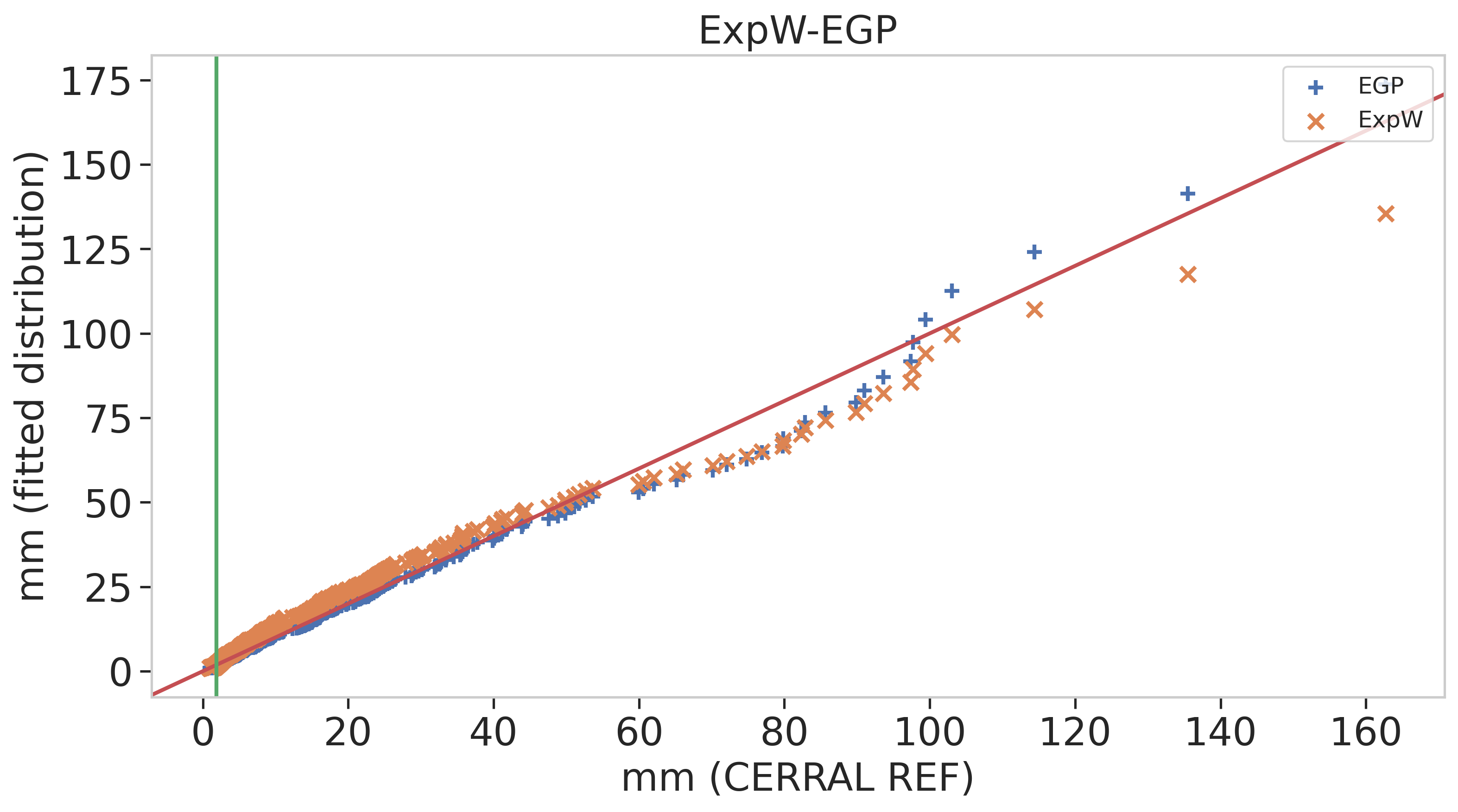}
    \caption{}
\end{subfigure}
\caption{Example of replacement indexes for the fitting of the Stitch-BJ model for a given location on ERA5L (a) and CERRAL (b) in the reference period (01/ 01/1985 to 31/12/2009).}
\label{fig:mecha_stitch}
\end{figure}

To easily understand the working mechanism of the Stitch-BJ, a simple example is given in Figure \ref{fig:mecha_stitch} where the quantile-quantile plot (QQplot) of the fitted EGP and ExpW distributions against the actual quantiles of respectively ERA5L (panel a) and CERRAL (panel b) are shown for a selected location. In both panels, the vertical line indicates an upper and lower rejection index (respectively in blue and green), named $i_u$ and $i_l$. These indexes are derived from the Penalized Berk-Jones statistical test and indicate the point from which the considered parametric distribution is deemed unfit. 

On panel (a) of Figure \ref{fig:mecha_stitch},  the final distribution can be expressed as :
\begin{equation}
    F(x) = \begin{cases}
        F_{EGP}(x) & \text{if $x<F^{-1}_{EGP}(\frac{i_u}{n})$} \\
        F_n(x) & \text{else,}
    \end{cases}
    \label{eq:stitch1}
\end{equation}

with $n$ the number of wet days for the considered time series. The upper tail of the EGP distribution has been replaced with the empirical distribution. {Note that while the ExpW seems to be well fitted in panel (a) of Figure \ref{fig:mecha_stitch}, errors on the upper tail can be up to 48\textit{mm} compared to the reference data, which is why it was also rejected on the upper tail.}

On panel (b), the lower tail of the EGP has been replaced with the ExpW distribution which resulted in a better fitting distribution.  Using the PBJ test, the final stitch distribution can then be expressed as :
\begin{equation}
    F(x) = \begin{cases}
        F_{ExpW}(x) & \text{if $x<F^{-1}_{EGP}(\frac{i_l}{n})$} \\
        F_{EGP}(x) & \text{else.}
    \end{cases}
    \label{eq:stitch2}
\end{equation}

Note that Equations \eqref{eq:stitch1} and \eqref{eq:stitch2} are simple illustrations related to specific situations in panel (a) and (b) of Figure \ref{fig:mecha_stitch}. Complete general analytical forms with monotonicity guarantees are available in  \cite{ear_semi-parametric_2025}.

\begin{figure}[!h]
\centering
 \begin{subfigure}[b]{0.45\textwidth}   
    \centering
    \includegraphics[width=\textwidth]{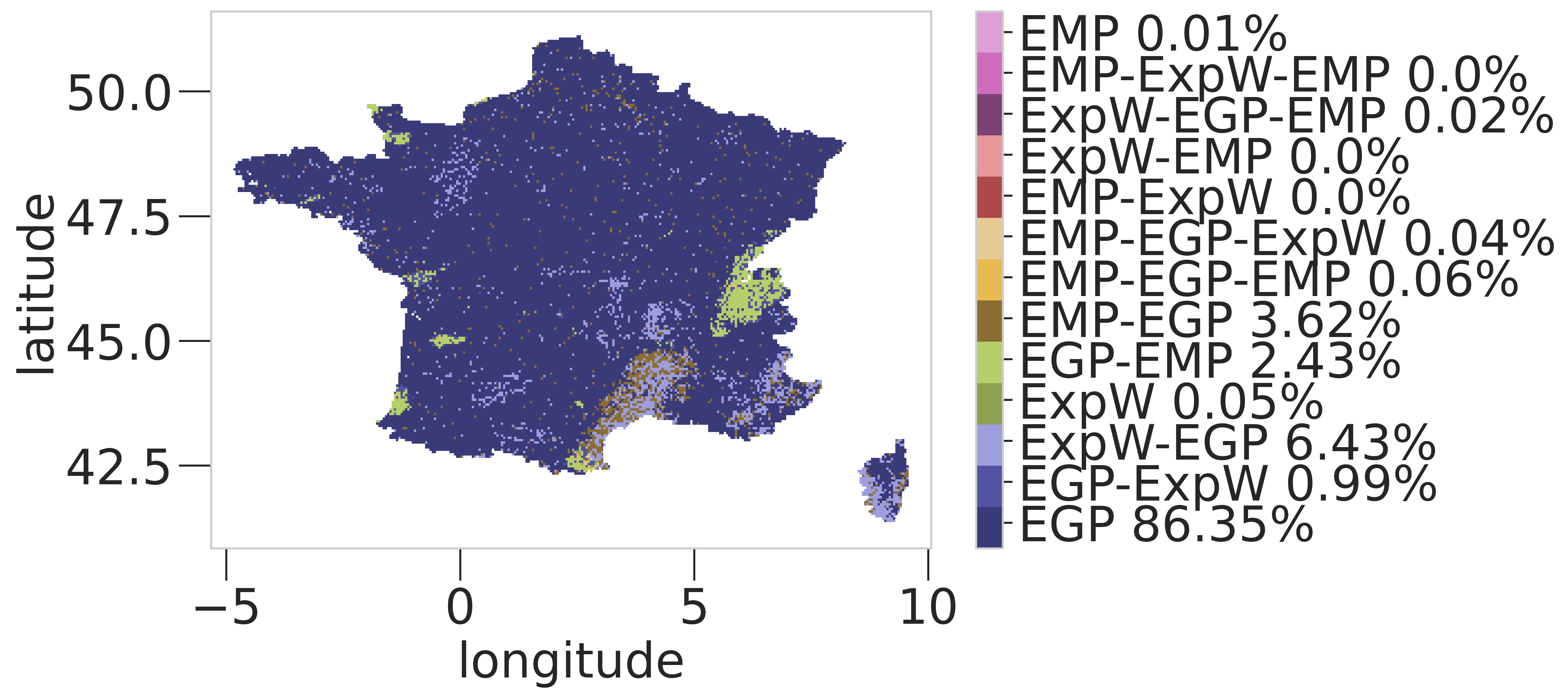}
    \caption{}
\end{subfigure}
\hfill
 \begin{subfigure}[b]{0.45\textwidth}   
    \centering
    \includegraphics[width=\textwidth]{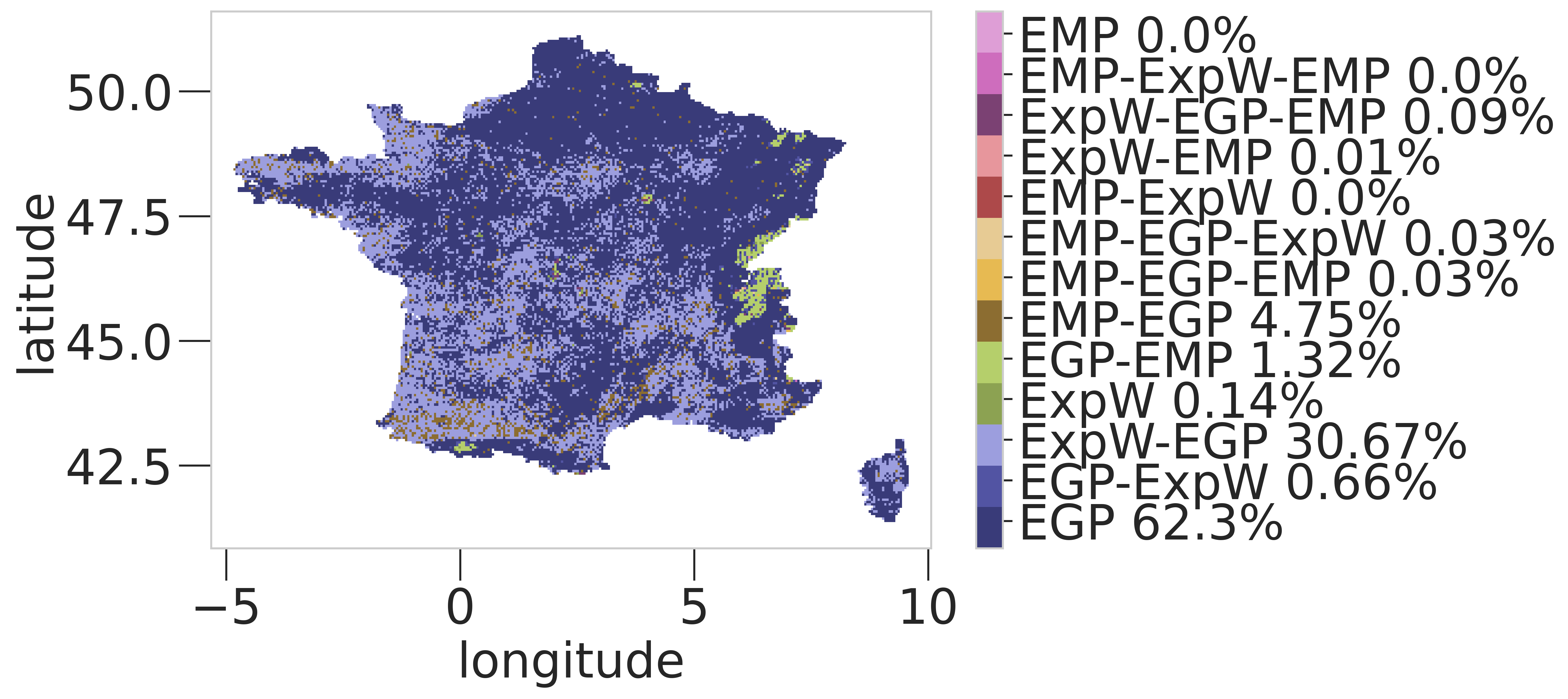}
    \caption{}
\end{subfigure}
\vskip\baselineskip

 \begin{subfigure}[b]{0.45\textwidth}   
    \centering
    \includegraphics[width=\textwidth]{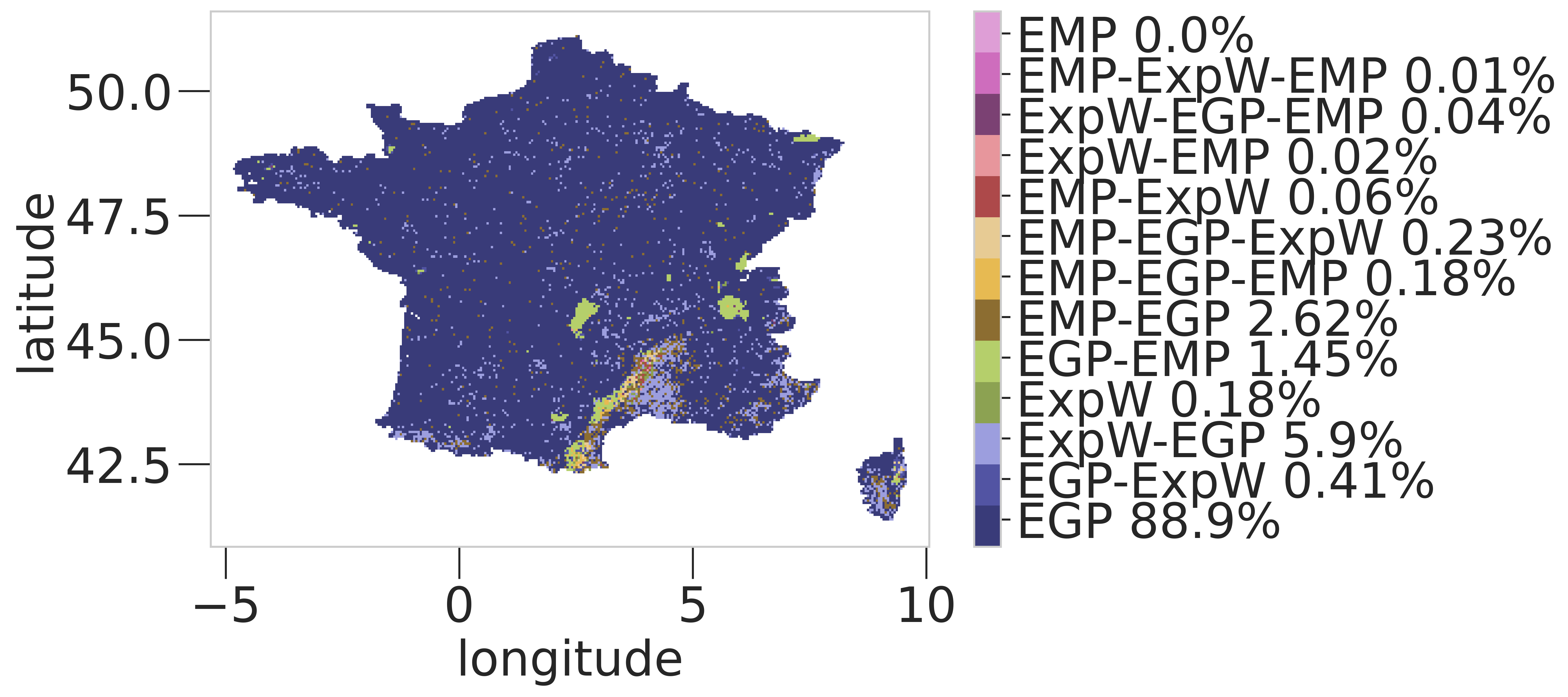}
    \caption{}
\end{subfigure}
\hfill
 \begin{subfigure}[b]{0.45\textwidth}   
    \centering
    \includegraphics[width=\textwidth]{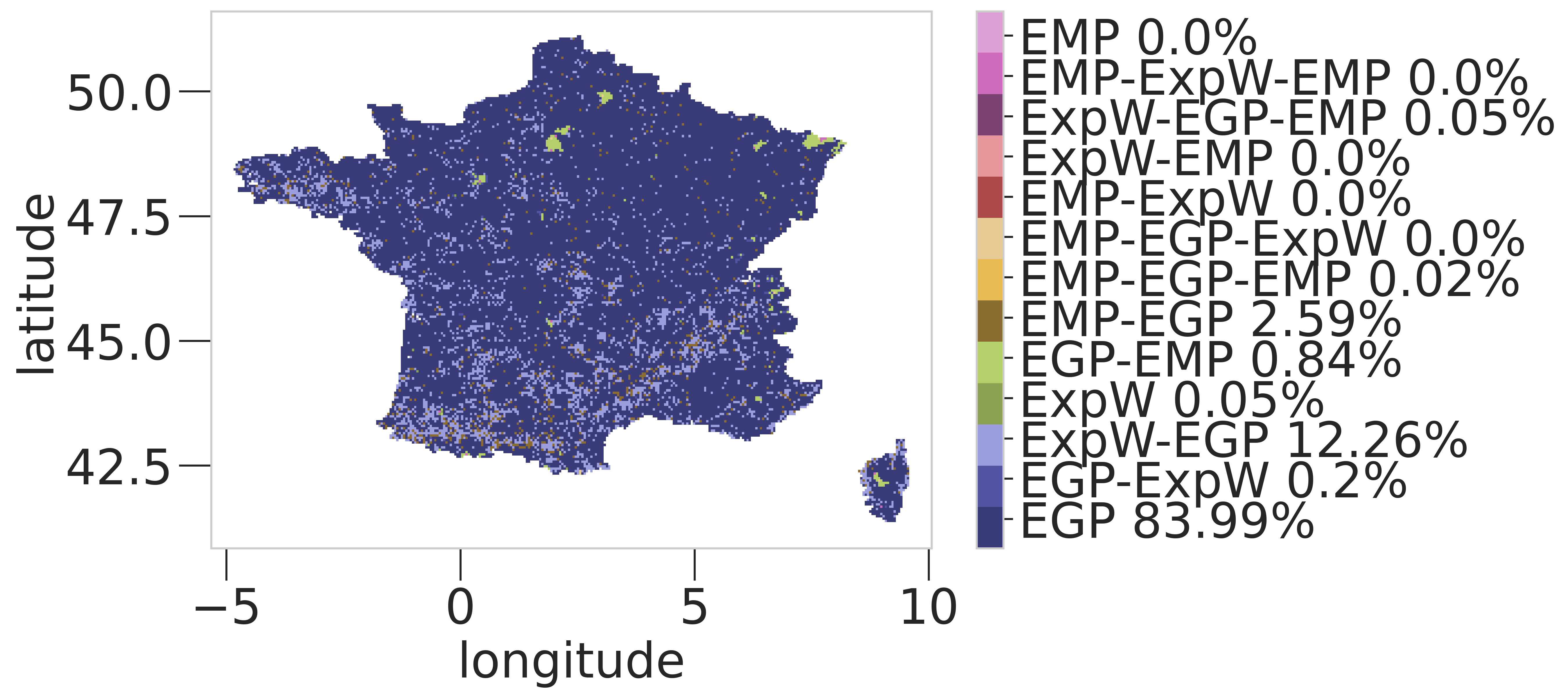}
    \caption{}
\end{subfigure}
\caption{Replacement map for the Stitch-BJ fitted on the reference period of ERA5L (a, b) and CERRAL (c, d) for DJF and JJA.} 
\label{fig:replacement_ERA5L_CERRAL}
\end{figure}

\begin{figure}[!h]
\centering
 \begin{subfigure}[b]{0.45\textwidth}   
    \centering
    \includegraphics[width=\textwidth]{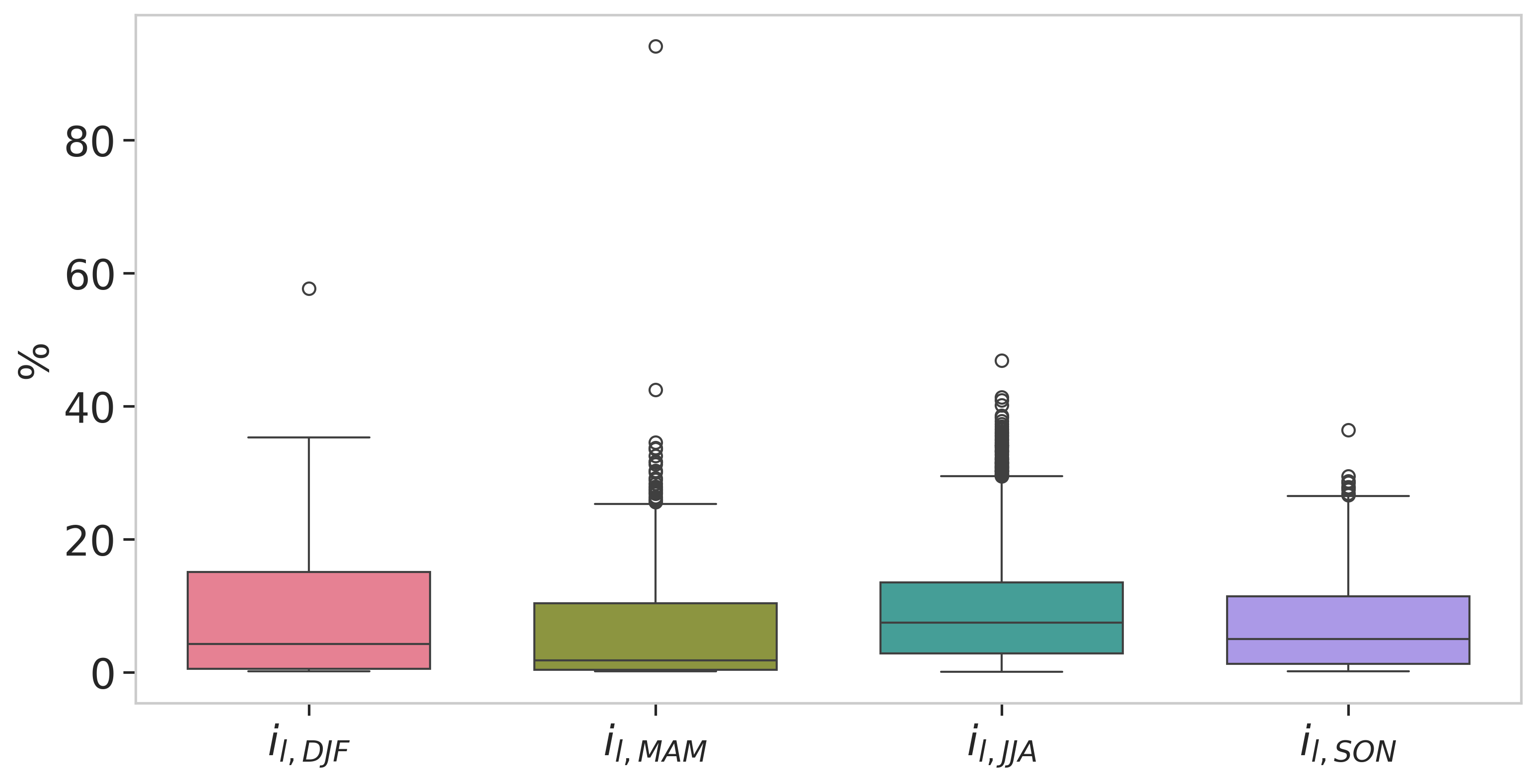}
    \caption{}
\end{subfigure}
\hfill
 \begin{subfigure}[b]{0.45\textwidth}   
    \centering
    \includegraphics[width=\textwidth]{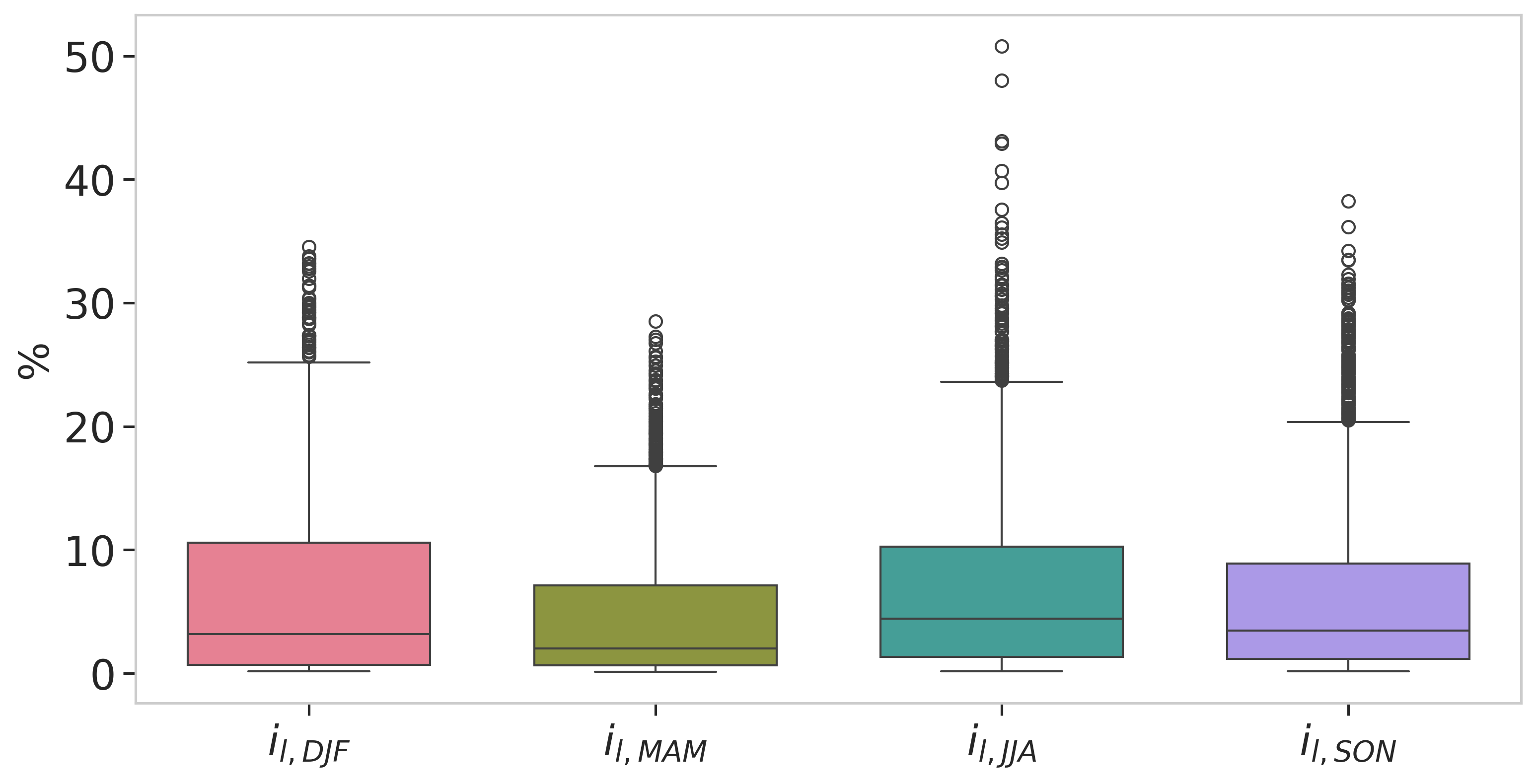}
    \caption{}
\end{subfigure}
\vskip\baselineskip

 \begin{subfigure}[b]{0.45\textwidth}   
    \centering
    \includegraphics[width=\textwidth]{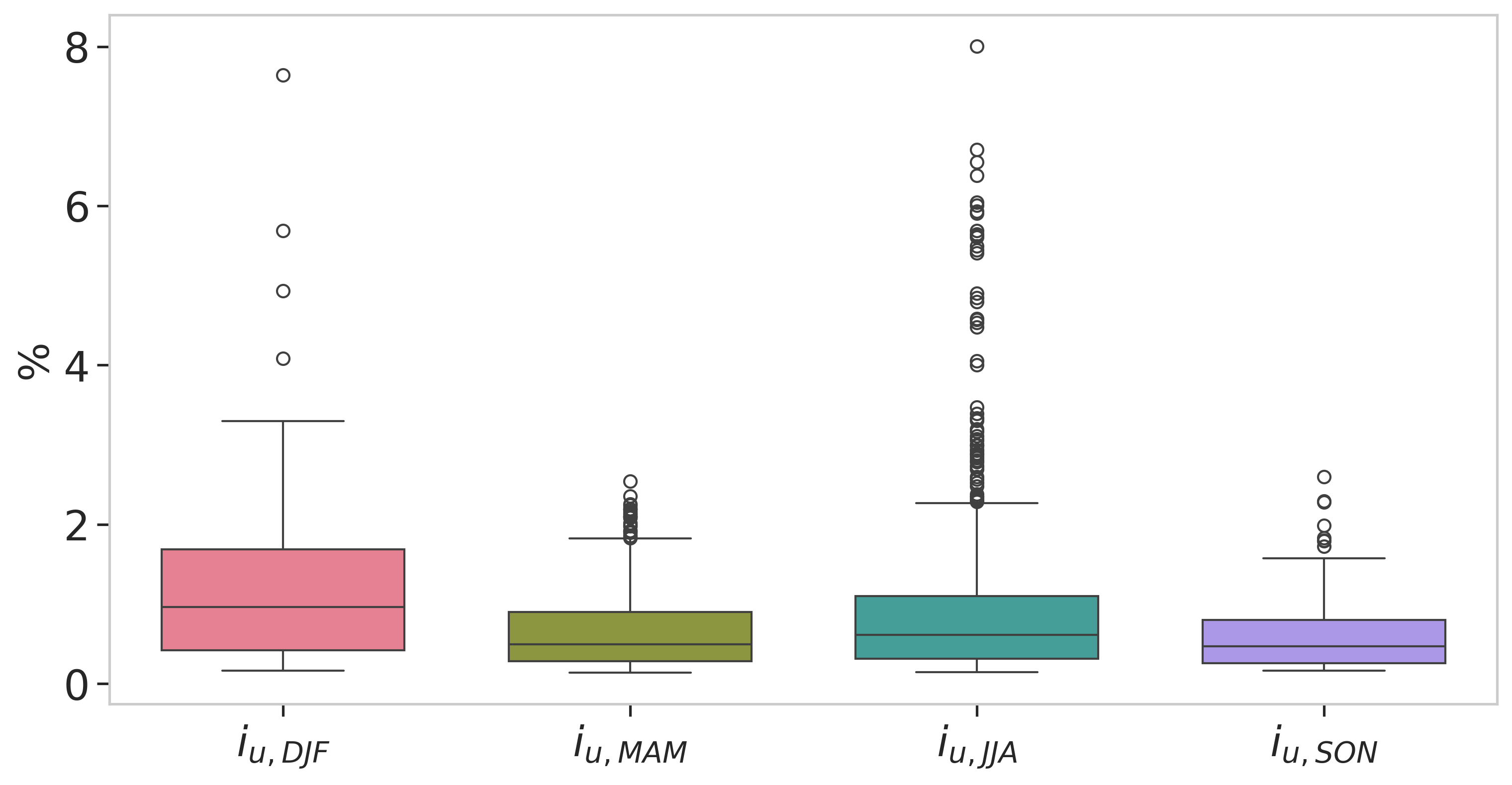}
    \caption{}
\end{subfigure}
\hfill
 \begin{subfigure}[b]{0.45\textwidth}   
    \centering
    \includegraphics[width=\textwidth]{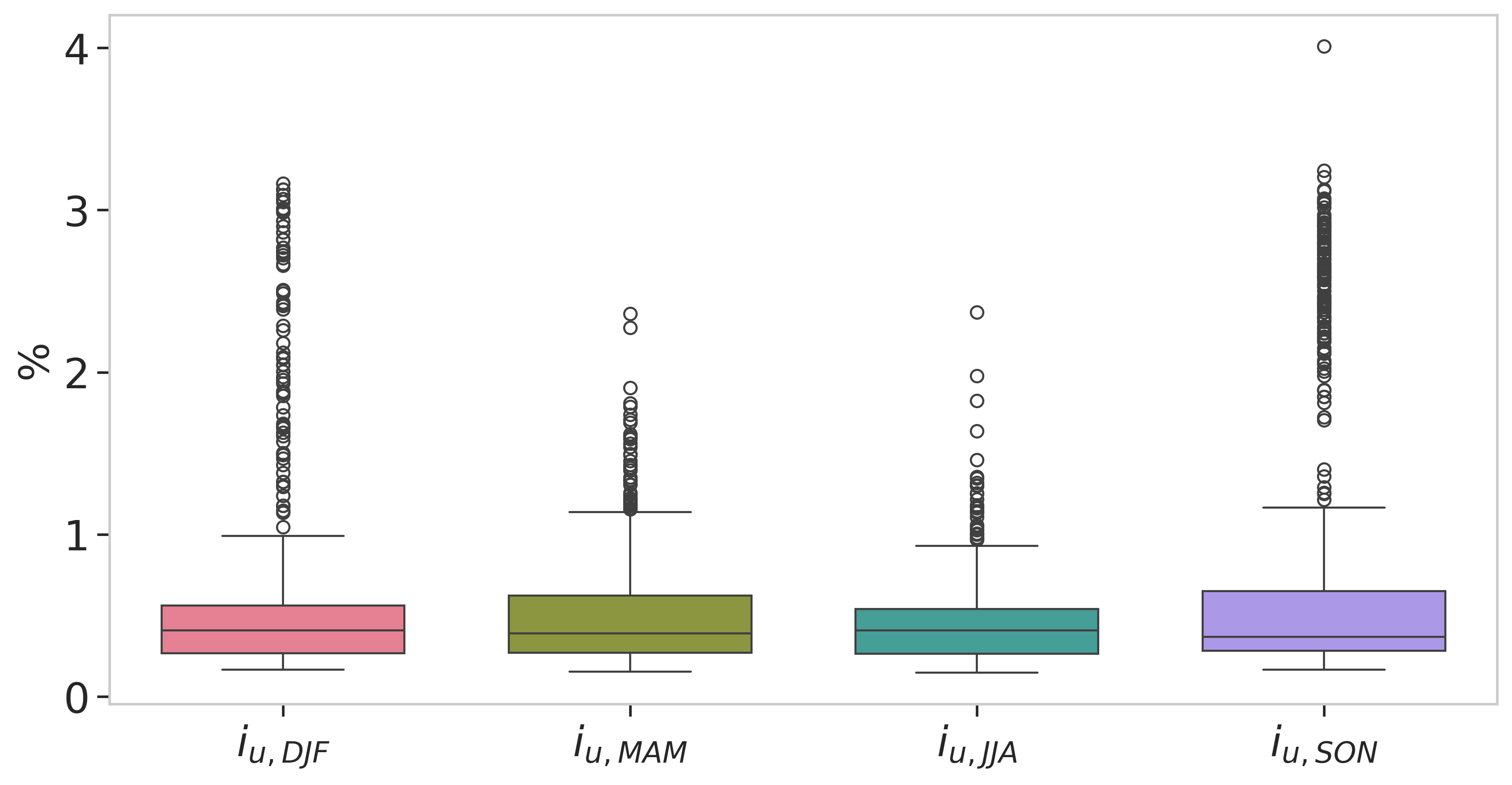}
    \caption{}
\end{subfigure}
\caption{Boxplots of percentages of the lower  (first row) and upper tail (second row) being replaced by the stitching algorithm for ERA5L (a, c) and CERRAL (b, d).} 
\label{fig:prop_lower_upper_replace}
\end{figure}

Using this methodology, we fitted the Stitch-BJ model to the daily precipitation data of ERA5L and CERRAL. From these fitted models, both a spatial analysis of the stitching type and a global analysis of the stitching proportion {can be obtained} in Figures \ref{fig:replacement_ERA5L_CERRAL} and \ref{fig:prop_lower_upper_replace}. 

For both DJF and JJA season on ERA5L, the vast majority (resp 86.35\% and 62.3\%) of locations used a pure EGP distribution as shown in Figure \ref{fig:replacement_ERA5L_CERRAL} panels (a, b). Considering all fully parametric models, stitch between parametric models included, the proportion goes up to around 94\% for both seasons. This shows that most locations where able to get a satisfying fit out of purely parametric distributions. The remaining locations used mainly an empirical lower tail replacement (3.6\% and 4.75\% of the total locations respectively), while almost no locations used a purely empirical distribution. Spatially, most of the locations using the empirical distribution for either the lower or upper tail replacement are located in historically complex climate regions such as the Cévènnes, the Alps or the Corsica \citep{vautard_extreme_2015,emmanuel_method_2017,cortes-hernandez_evaluation_2024,estermann_projections_2025}. For the JJA season, these locations are a bit more spread out but still have a large concentration around the Alps and the Cévènnes, with some replacement along the Pyrénnées.

Replacement maps for CERRAL in Figure \ref{fig:replacement_ERA5L_CERRAL} panel (c, d) are similar to those for ERA5L in panels (a, b) with most of the locations for both DJF and JJA seasons using pure EGP distributions (88.9\% and 84\% respectively). The use of an empirical lower or upper tail is even rarer with respectively 95.39\% and 96.5\% of locations using a purely parametric distribution. Use of the empirical distribution for the upper tail is mainly located around the Cévènnes for the DJF season, while for JJA locations seems to be spread out on all the study areas. No locations used a fully empirical distribution for both DJF and JJA season while fitted on the CERRAL dataset.

For locations where Stitch-BJ was used, one might want to know how much of the original distributions' the stitching replaced. Overall, the median replacement affected 5\% of the distribution for the lower tail, and less than 1\% for the upper tail. More details can be found in Figure \ref{fig:prop_lower_upper_replace} with boxplots of the percentage of replacement for the lower and upper tail per season, for respectively ERA5L (panels a and c) and CERRAL (panels b and d).
The lower replacement index $i_l$ concerned a larger portion of the distribution compared to $i_u$ but its impact is much smaller due to the small intensity of rain affected. Overall, $i_l$ for CERRAL were smaller, both for the median and outliers, than for ERA5L, throughout all seasons. For the upper tail index $i_u$, similar observations can be made with CERRAL upper replacement portion being roughly equal to or lower than for ERA5L.

The Stitch-BJ model is heavily flexible, allowing to automatically replace a portion of the EGP distribution with other models, either parametric or empirical, using   p-values results from the PBJ goodness-of-fit statistical tests. This allows for a better fit to precipitation distribution which can be extremely variable depending on the season and the location chosen.  

In the next section, bias correction performance of all methods will be assessed using the CERRAL validation period.

\section{Bias-correction results on the period 2010-2020}
\label{sec:3}

In this section, we compare the performance of the newly introduced Stitch-BJ against classical parametric distributions (EGP, ExpW, Gamma) and the empirical distribution (referred to as \textit{emp}).

As explained in Section \ref{sec:2_1}, both datasets ERA5L and CERRAL were first mapped on the same regular longitude/latitude grid (approximately $0.05^{\circ}\times0.05^{\circ}$) and the following periods have been used for training and validation:
\begin{itemize}
    \item Training period: 01/01/1985 to 31/12/2009;
    \item Validation period: 01/01/2010 to 31/12/2020;
\end{itemize}
along with a separation using the meteorological seasons.
Only the results of the correction of the validation period are shown in this section {as they are similar to the ones from the previous study}. In this section, the following metrics were computed on a pixel-by-pixel basis and we noted $q^n_i$ the quantile $\frac{i}{n}$ of the empirical distribution of the target data i.e. $F_n^{-1}(\frac{i}{n})$, and $q^F_i$ the same quantile for a given distribution $F$, i.e. $F^{-1}(\frac{i}{n})$.

\textbf{Mean Absolute Error}

\begin{equation*}
    MAE_F = \frac{1}{n} \sum_{i=1}^n |q^F_i-q^n_i  |.
\end{equation*}

\textbf{Mean Absolute Error over the 95th percentile}

\begin{equation*}
    MAE95sup_F =\frac{1}{n-\lceil{95\%\rceil}\times n} \sum_{i=\lceil{95\%\rceil}\times n}^n |q^F_i-q^n_i  |,
\end{equation*}

where $\lceil \cdot \rceil$ is the ceiling function.

\textbf{Root Mean Squared Error}

\begin{equation*}
    RMSE_F = \sqrt{\frac{1}{n} \sum_{i=1}^n (q^F_i-q^n_i)^2}.
\end{equation*}

The number of quantiles $n$ was chosen to be $50$ to ensure all locations had enough wet days without giving an artificially high weight to extremes.

Differences of metrics were also computed to increase visibility on the spatial improvements of one distribution over another. For two distributional models $F$ and $G$ respectively, we compute the differences metrics as follows:
\begin{align*}
    MAEdiff_{F,G} &= MAE_F - MAE_G, \\
    MAE95supdiff_{F,G} &= MAE95sup_F - MAE95sup_G,\\
    RMSEdiff_{F,G} &= RMSE_F - RMSE_G.
\end{align*}

Note that in the following sections, Gamma's error maps will not be shown. We will display only Gamma's results in boxplots due to  the very high errors. We do include for some figures the interpolated but non-corrected future model data, referred to as ERA5L in the legends. Moreover, the colour ranges are fixed between maps of the same metrics to represent the $99.9th$ quantile error of the Stitch-BJ method.  Finally, for differences of metrics, the scale is not fixed and may vary from distribution to distribution.

\subsection{Mean Absolute Error}
\label{sec:3_1}
\begin{figure}[!h]
\centering
 \begin{subfigure}[b]{0.24\textwidth}   
    \centering
    \includegraphics[width=\textwidth]{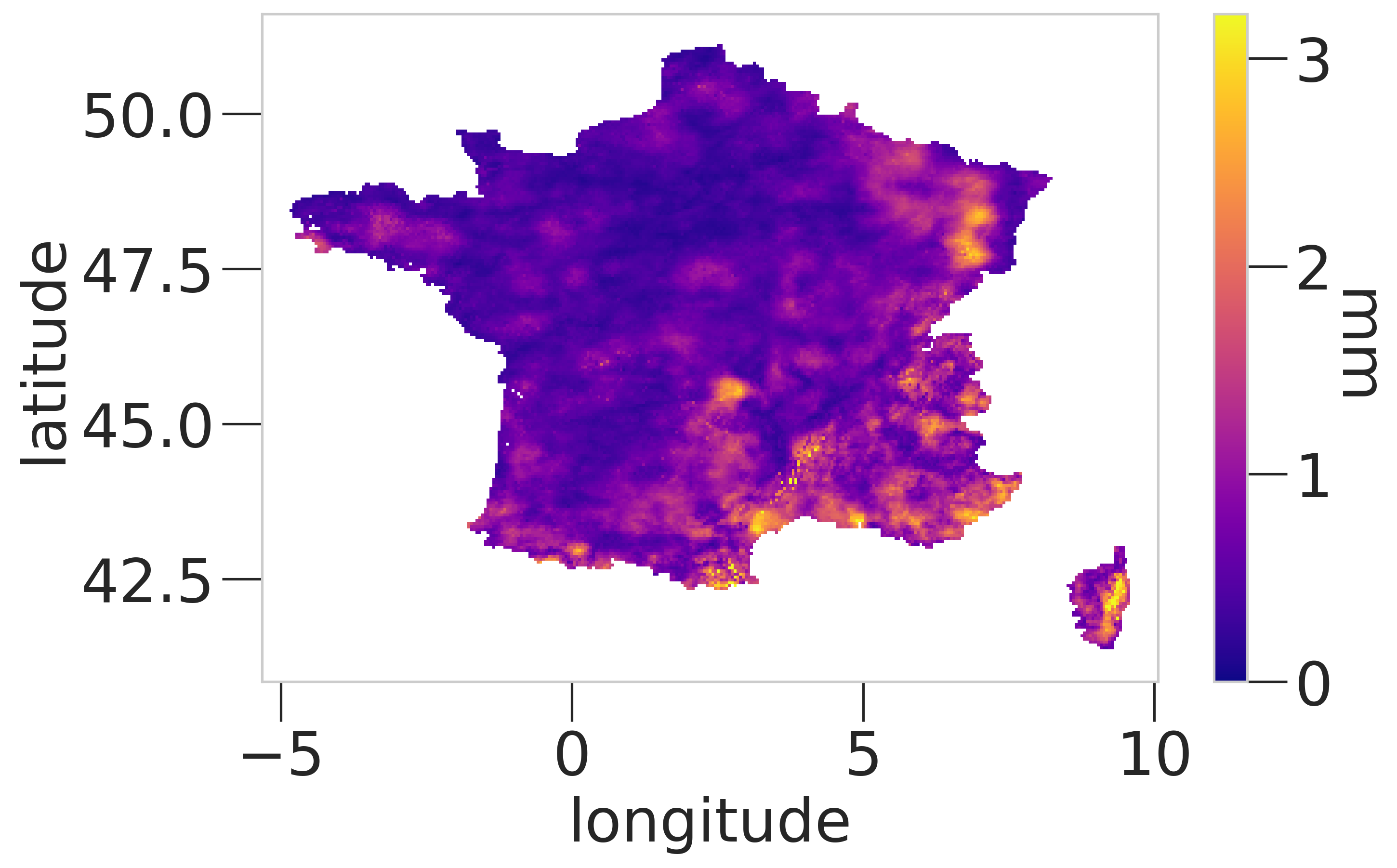}
    \caption{}
\end{subfigure}
\hfill
 \begin{subfigure}[b]{0.24\textwidth}   
    \centering
    \includegraphics[width=\textwidth]{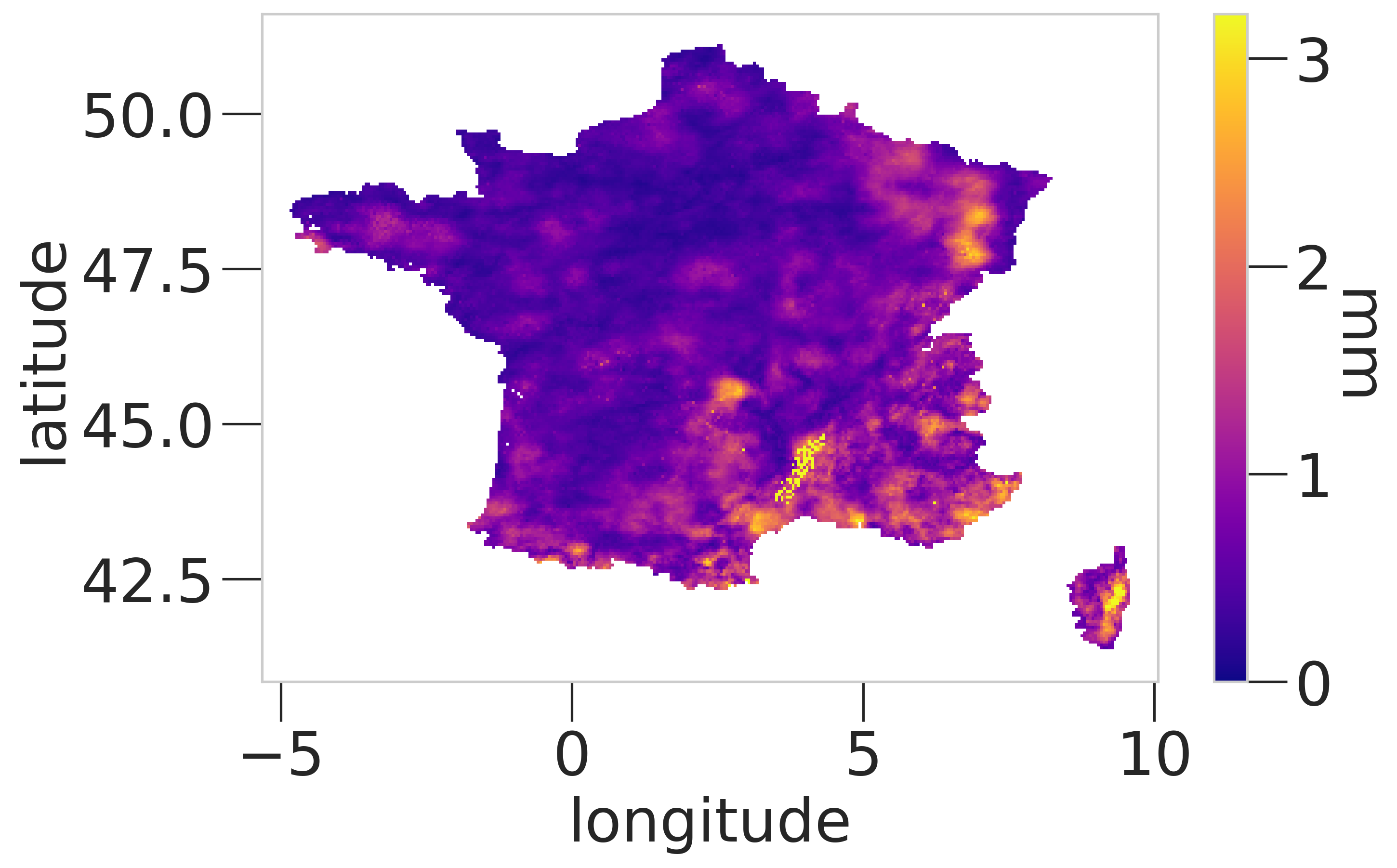}
    \caption{}
\end{subfigure}
\hfill
\begin{subfigure}[b]{0.24\textwidth}   
    \centering
    \includegraphics[width=\textwidth]{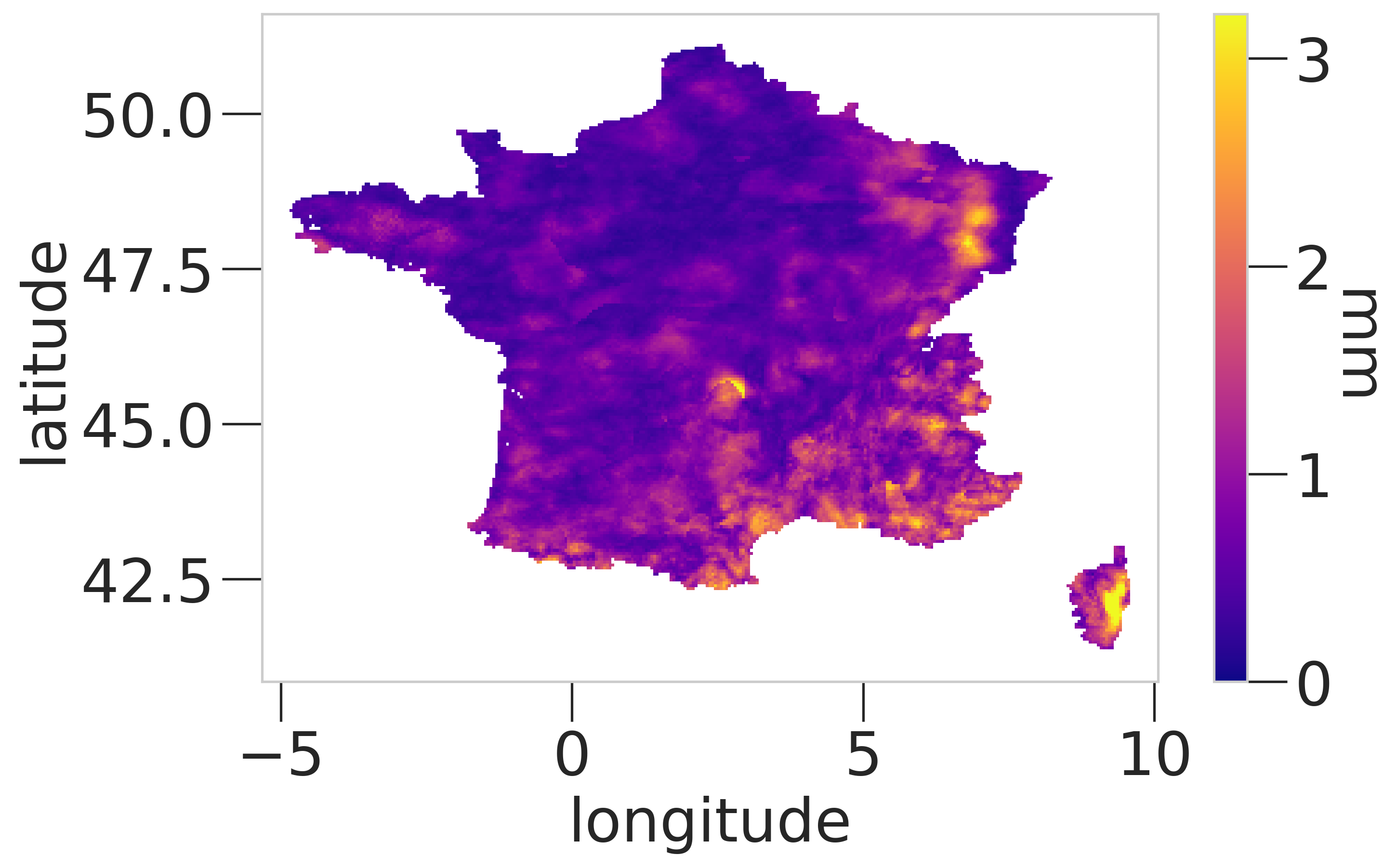}
    \caption{}
\end{subfigure}
\hfill
 \begin{subfigure}[b]{0.24\textwidth}   
    \centering
    \includegraphics[width=\textwidth]{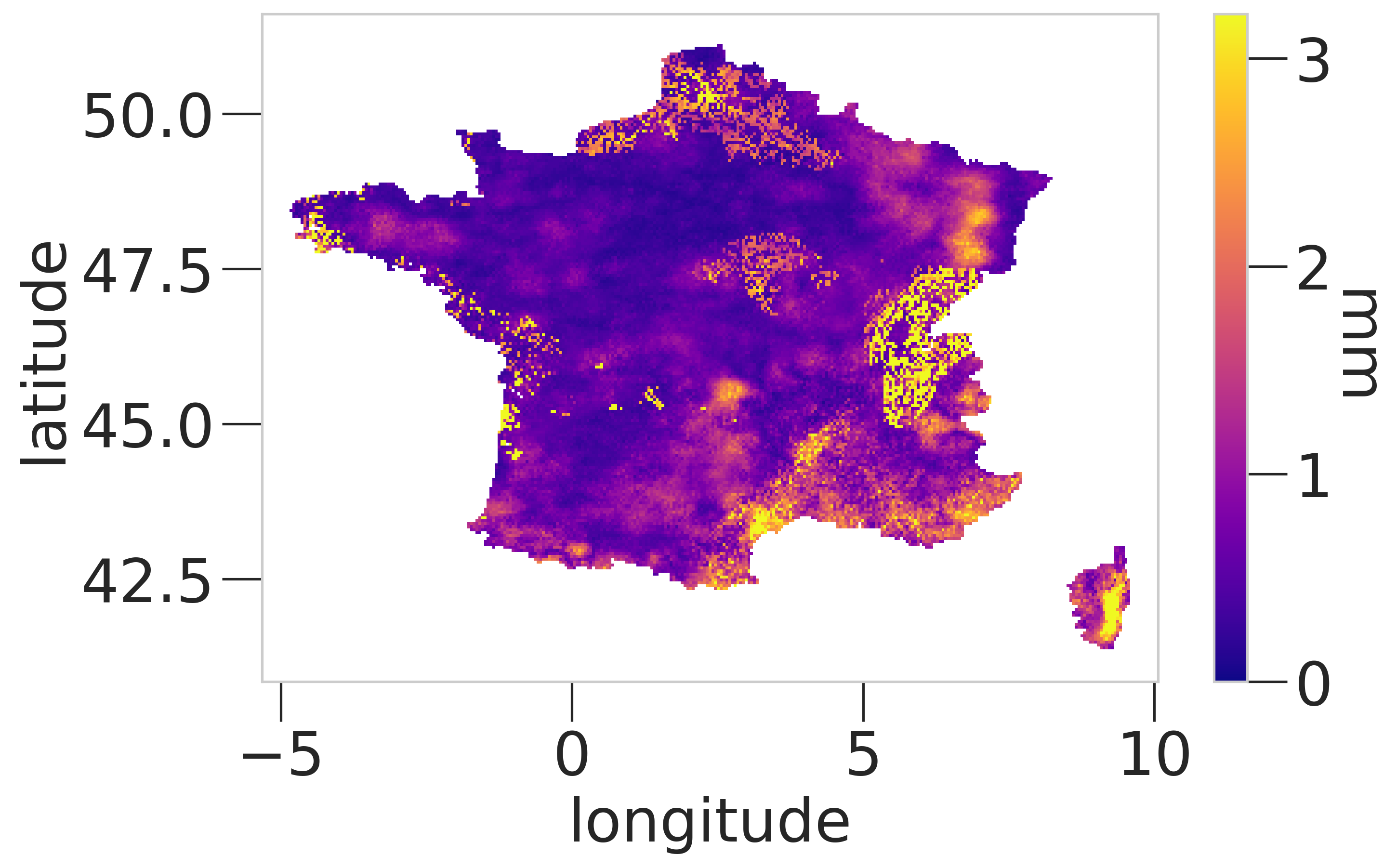}
    \caption{}
\end{subfigure}
\vskip\baselineskip
 \begin{subfigure}[b]{0.24\textwidth}   
    \centering
    \includegraphics[width=\textwidth]{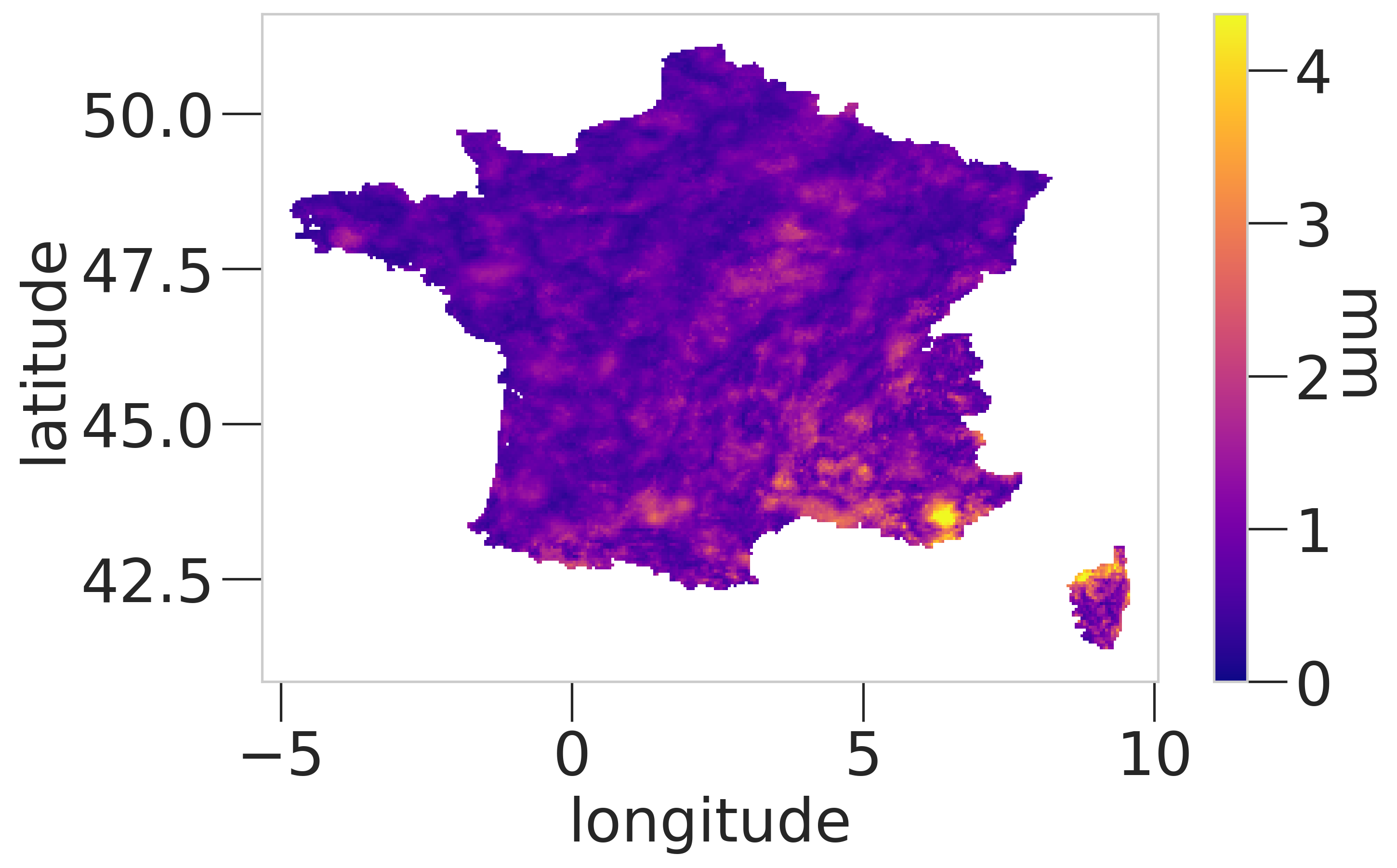}
    \caption{}
\end{subfigure}
\hfill
 \begin{subfigure}[b]{0.24\textwidth}   
    \centering
    \includegraphics[width=\textwidth]{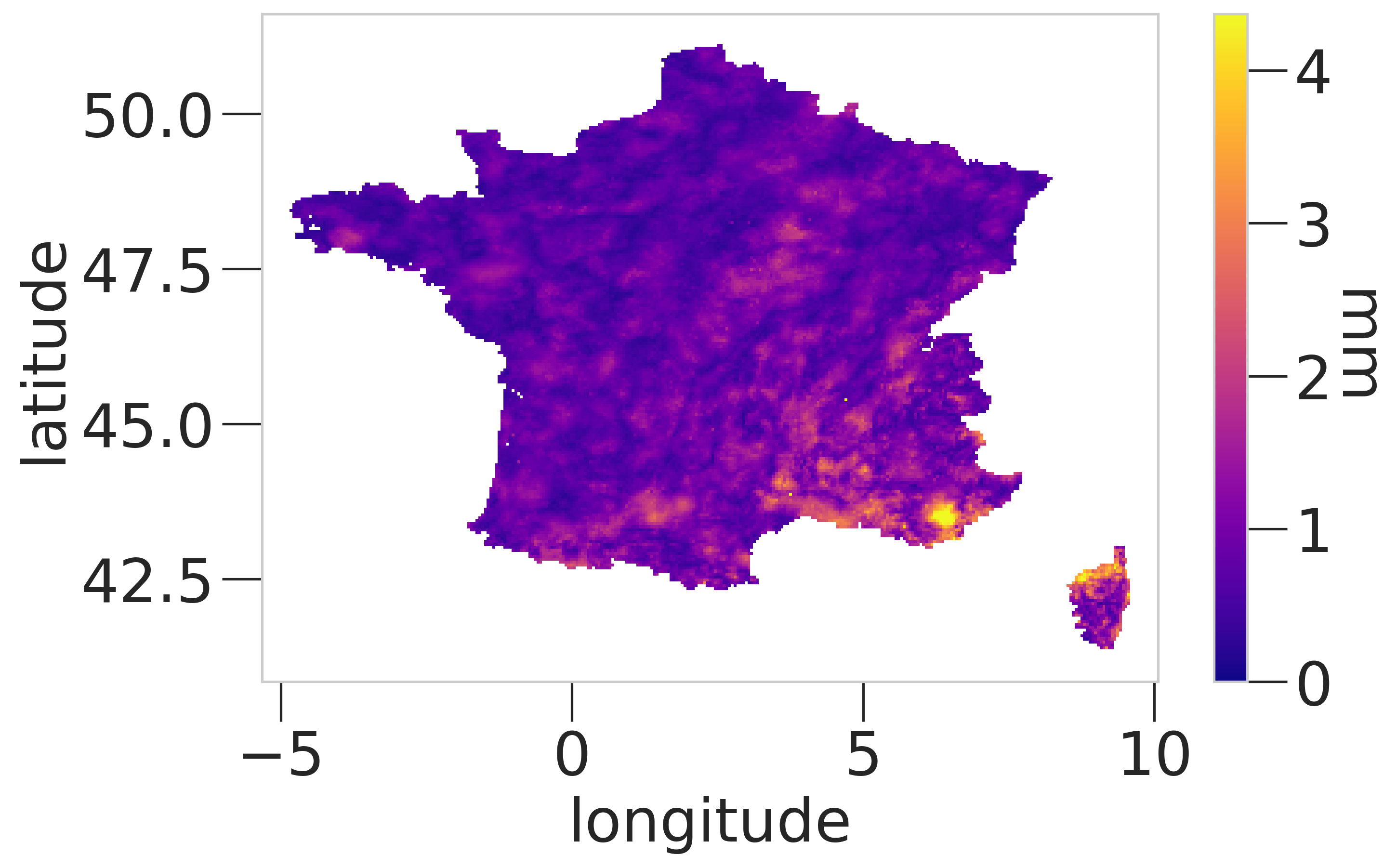}
    \caption{}
\end{subfigure}
\hfill
\begin{subfigure}[b]{0.24\textwidth}   
    \centering
    \includegraphics[width=\textwidth]{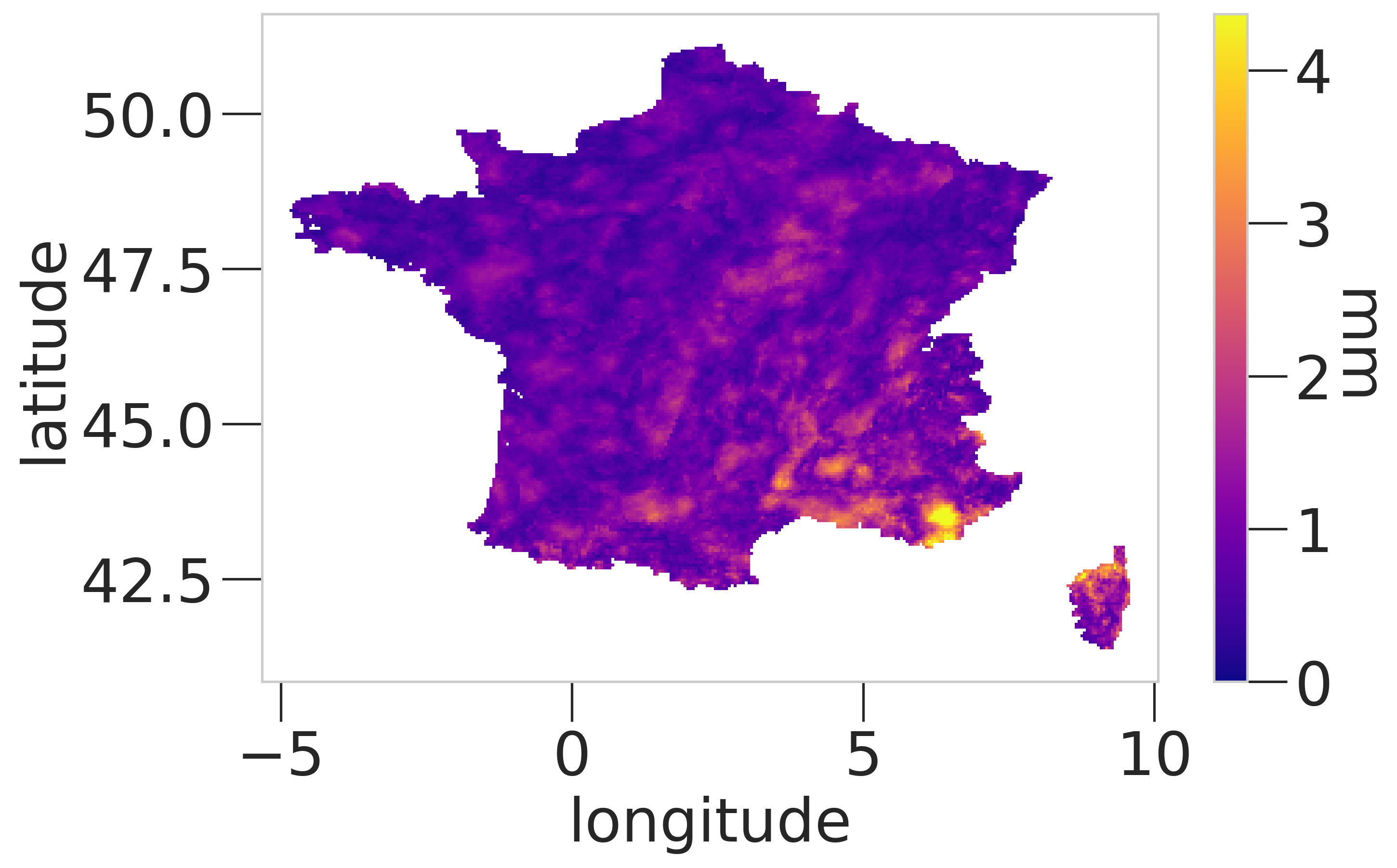}
    \caption{}
\end{subfigure}
\hfill
 \begin{subfigure}[b]{0.24\textwidth}   
    \centering
    \includegraphics[width=\textwidth]{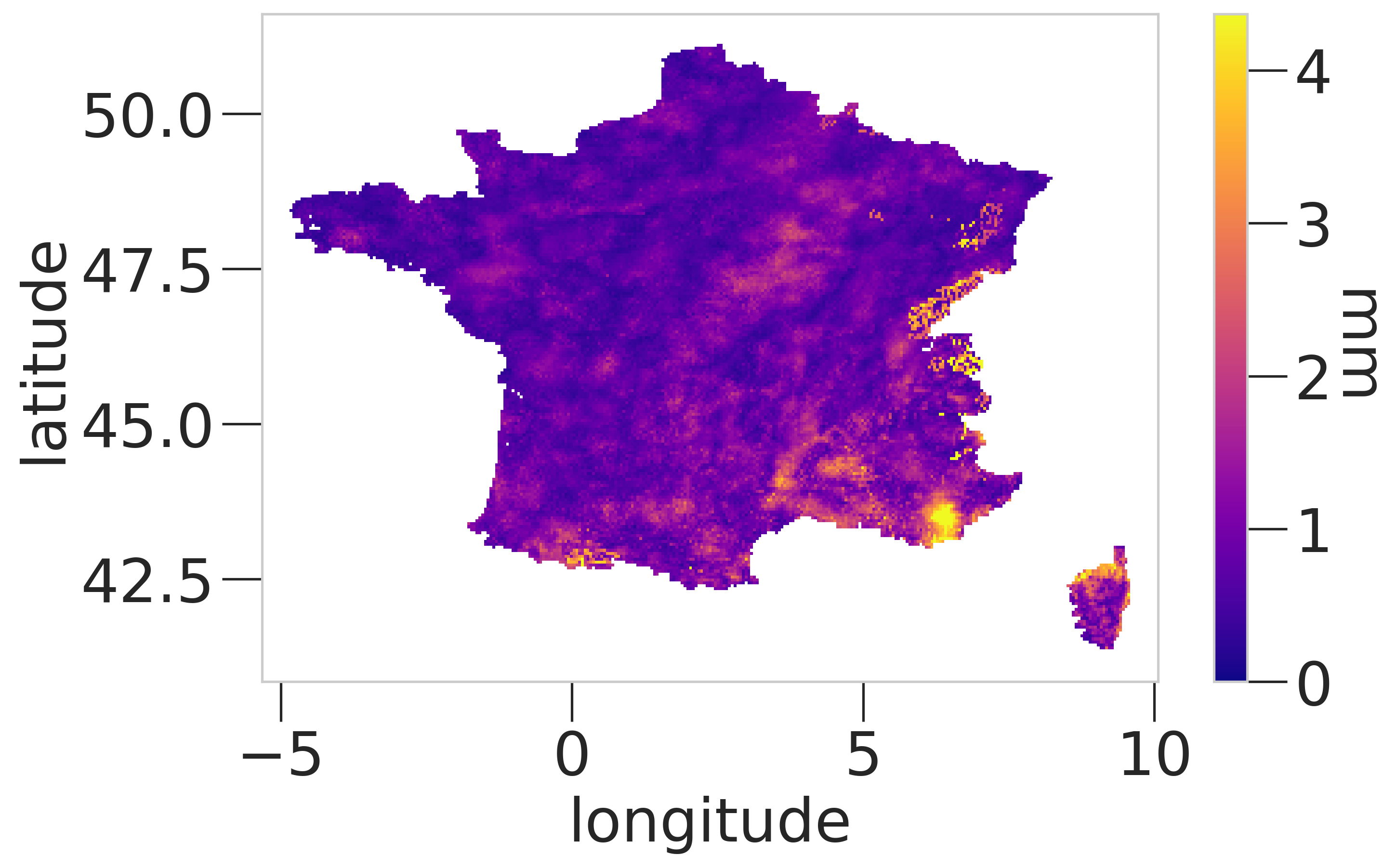}
    \caption{}
\end{subfigure}
\caption{Maps of MAE for DJF (first row) and JJA (second row) for the following models: Stitch-BJ panels (panels a,e), EGP (panels b,f), \textit{emp} (panels c,g) and ExpW (panels d,h).} 
\label{fig:MAE_DJF_JJA}
\end{figure}

Differences of MAE for the bias correction performance shown in Figure \ref{fig:MAE_DJF_JJA} are quite difficult to assess. Both the spatial pattern and intensity of error are very similar for all chosen distributions. However, some extreme MAE for the EGP and ExpW models can be seen for the DJF season (panels b and c), respectively near the Cévènnes and the Alps region. Those mountainous areas are known to be difficult to model due to their strong orographic features, paired with intense rain events during the SON and DJF seasons.
Performances for the JJA season (panels e, f ,g and h) are mostly identical which may be caused by the lower amount of rain and fewer extreme events.

\begin{figure}[!h]
\centering
 \begin{subfigure}[b]{0.32\textwidth}   
    \centering
    \includegraphics[width=\textwidth]{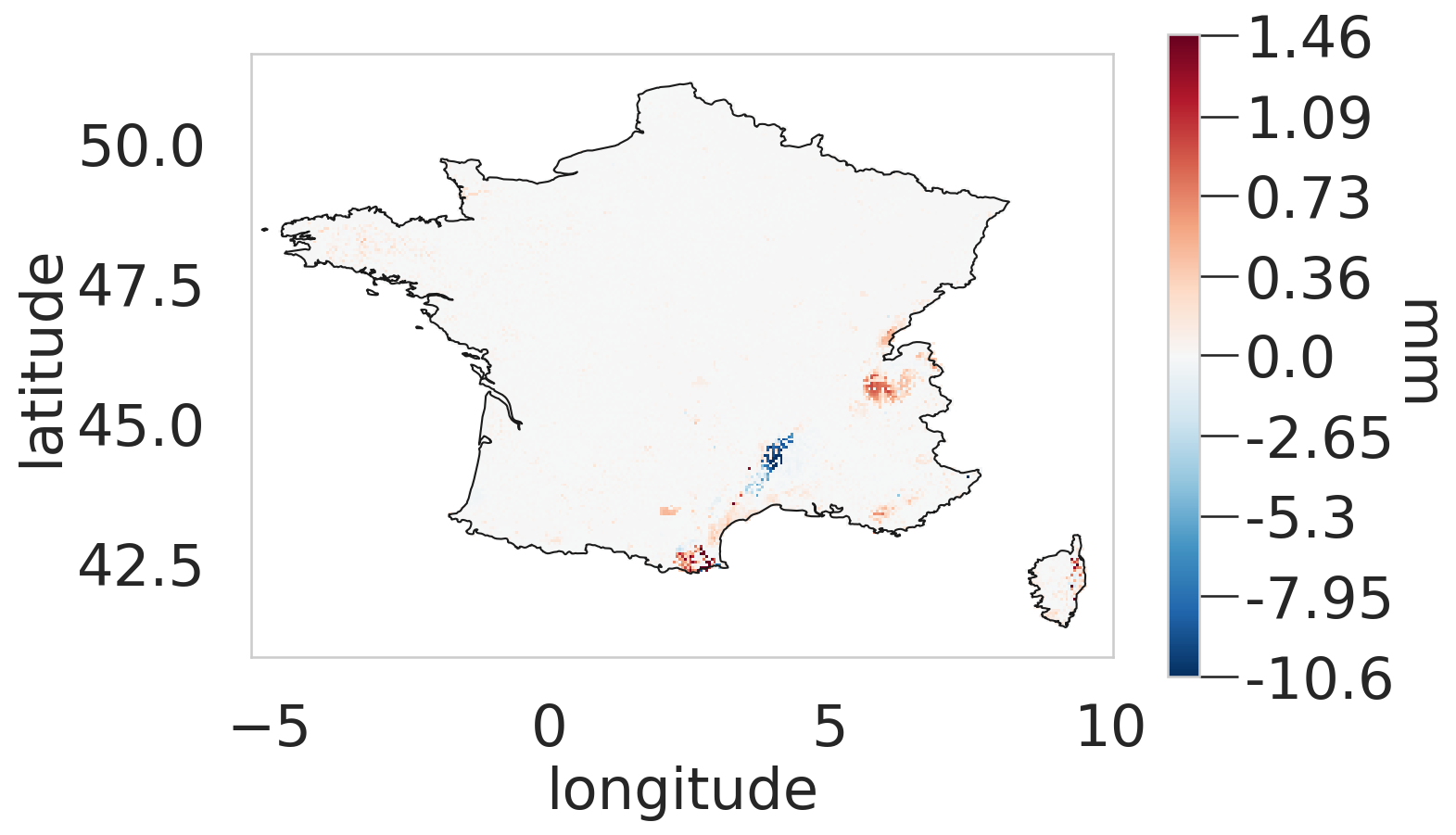}
    \caption{}
\end{subfigure}
\hfill
 \begin{subfigure}[b]{0.32\textwidth}   
    \centering
    \includegraphics[width=\textwidth]{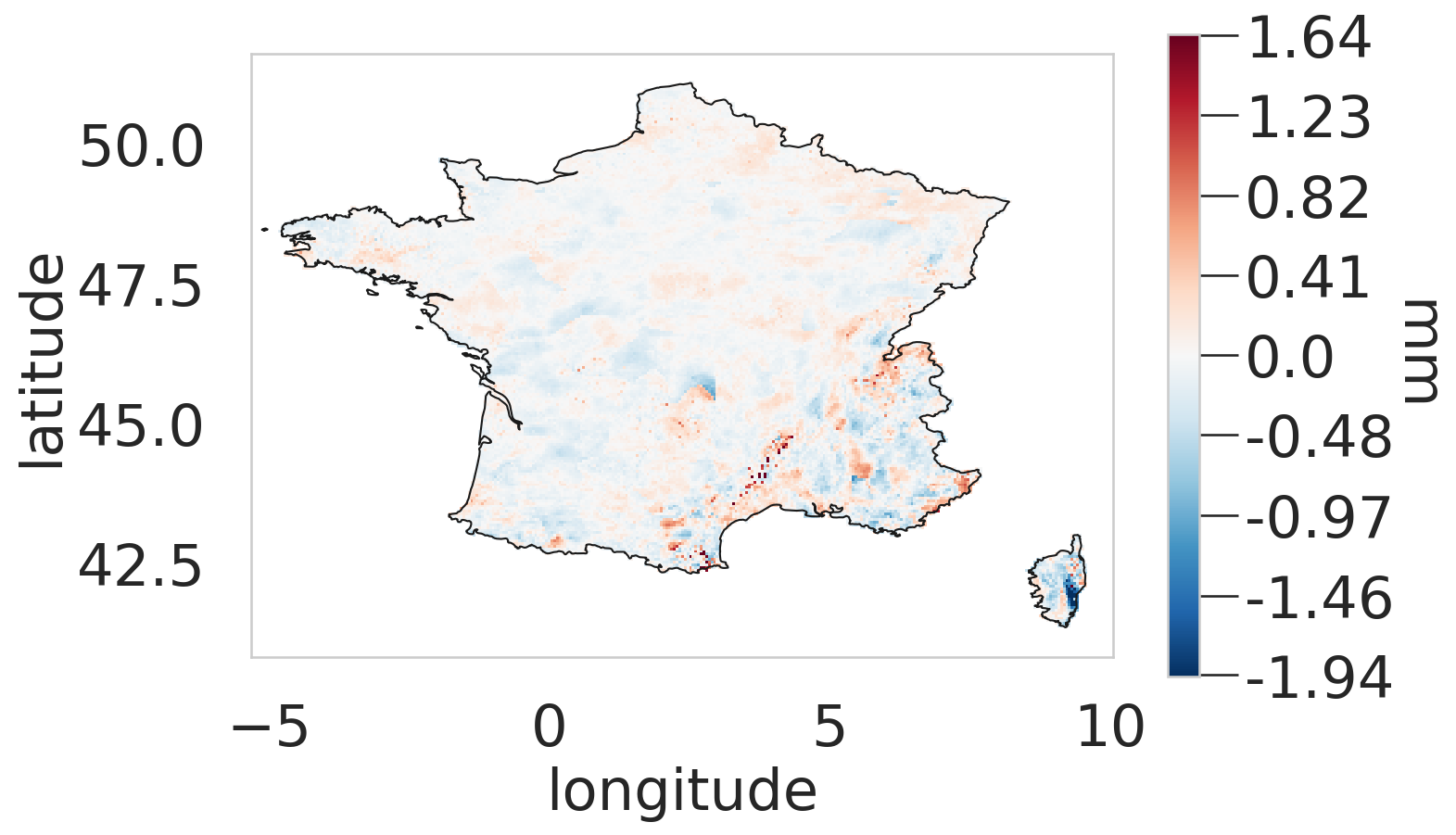}
    \caption{}
\end{subfigure}
\hfill
\begin{subfigure}[b]{0.32\textwidth}   
    \centering
    \includegraphics[width=\textwidth]{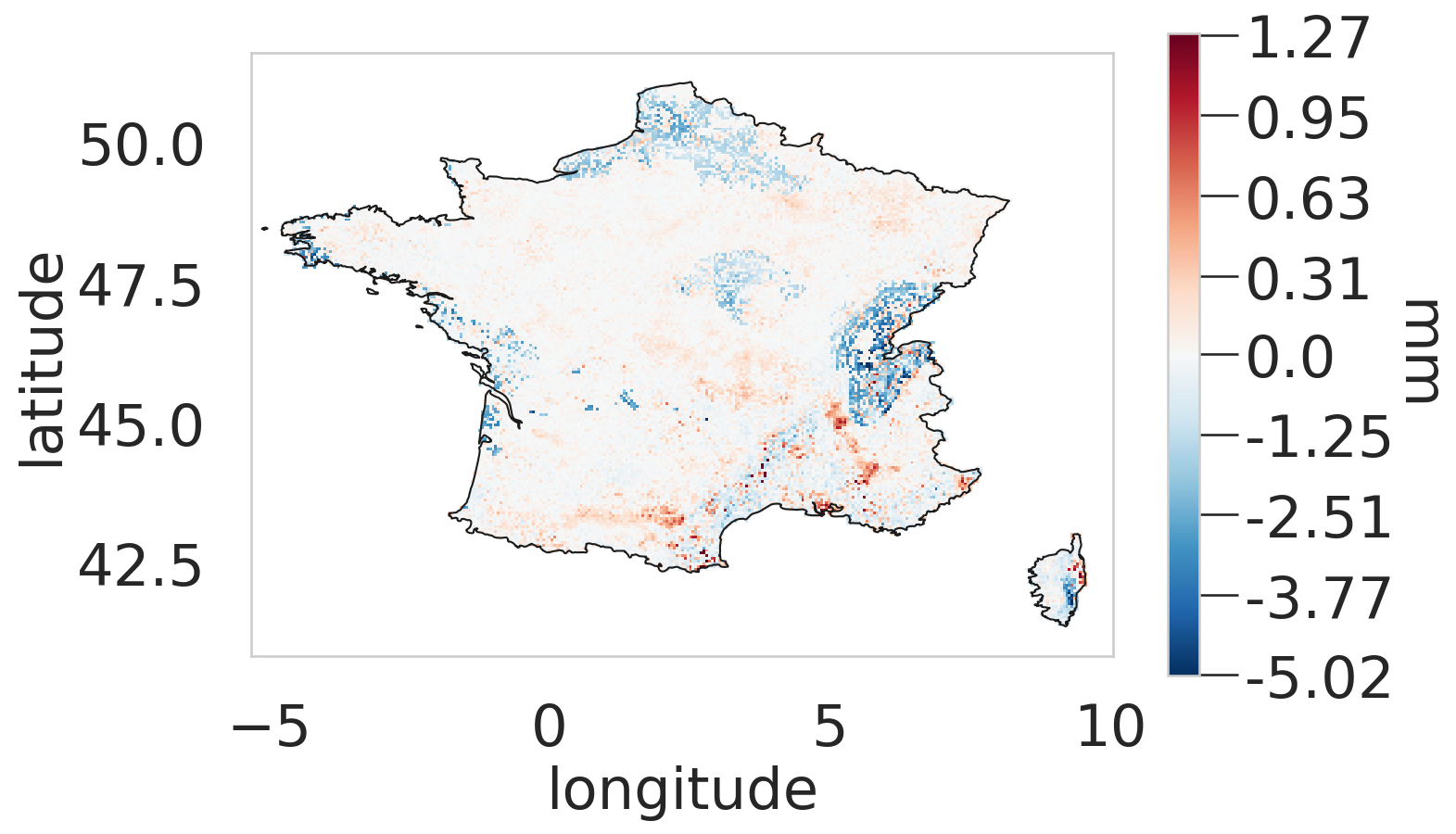}
    \caption{}
\end{subfigure}

\vskip\baselineskip
 \begin{subfigure}[b]{0.32\textwidth}   
    \centering
    \includegraphics[width=\textwidth]{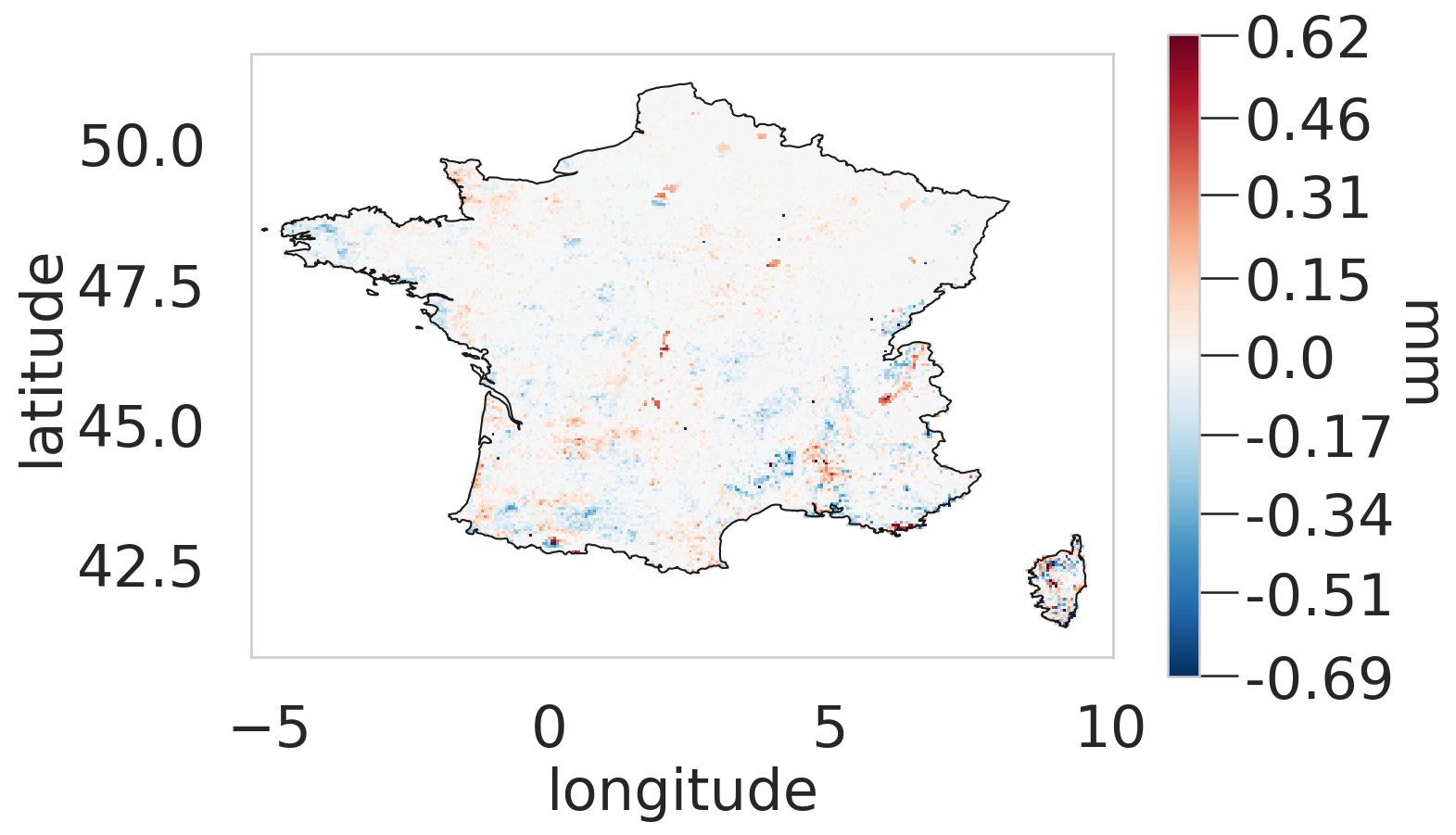}
    \caption{}
\end{subfigure}
\hfill
 \begin{subfigure}[b]{0.32\textwidth}   
    \centering
    \includegraphics[width=\textwidth]{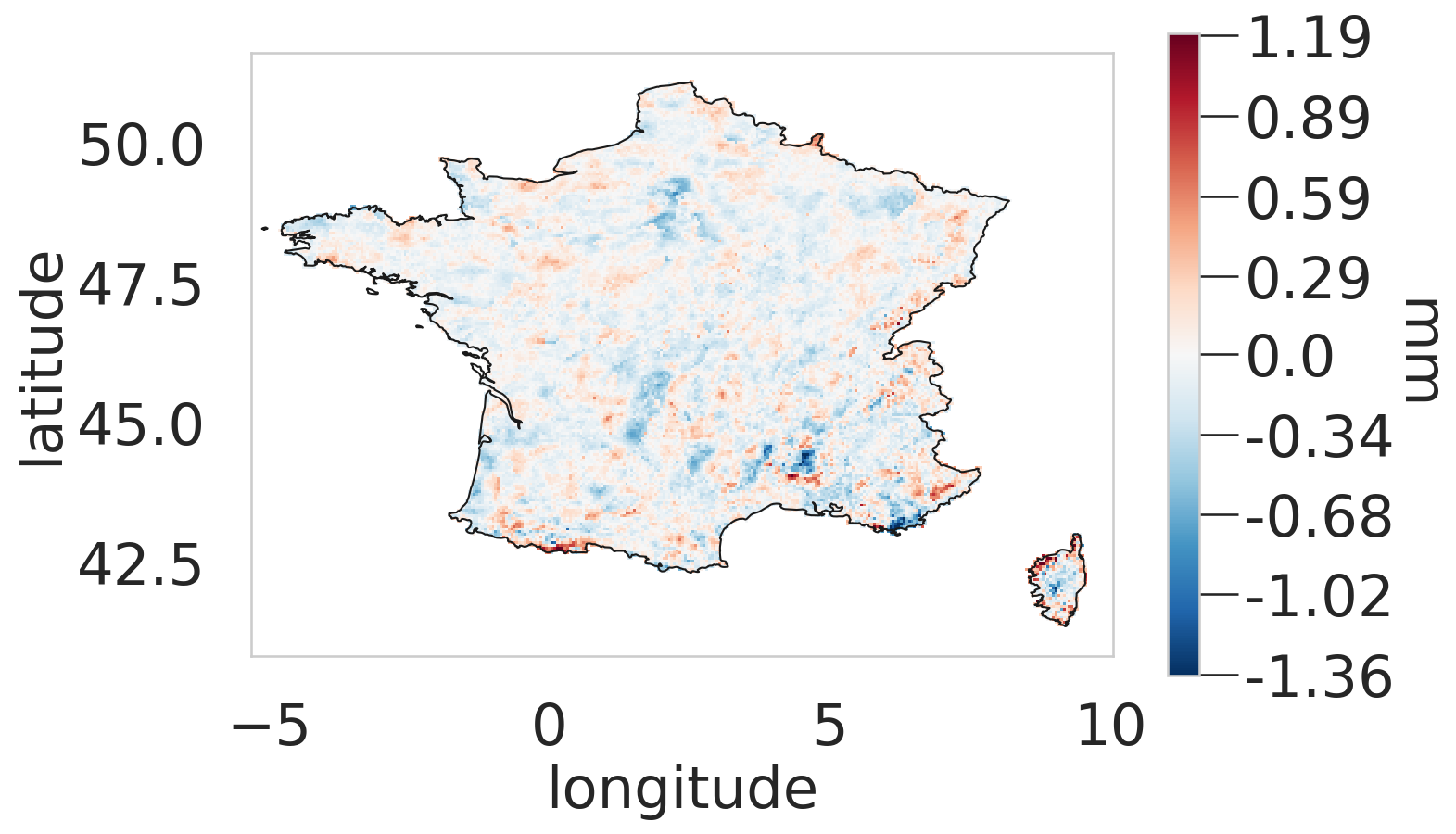}
    \caption{}
\end{subfigure}
\hfill
\begin{subfigure}[b]{0.32\textwidth}   
    \centering
    \includegraphics[width=\textwidth]{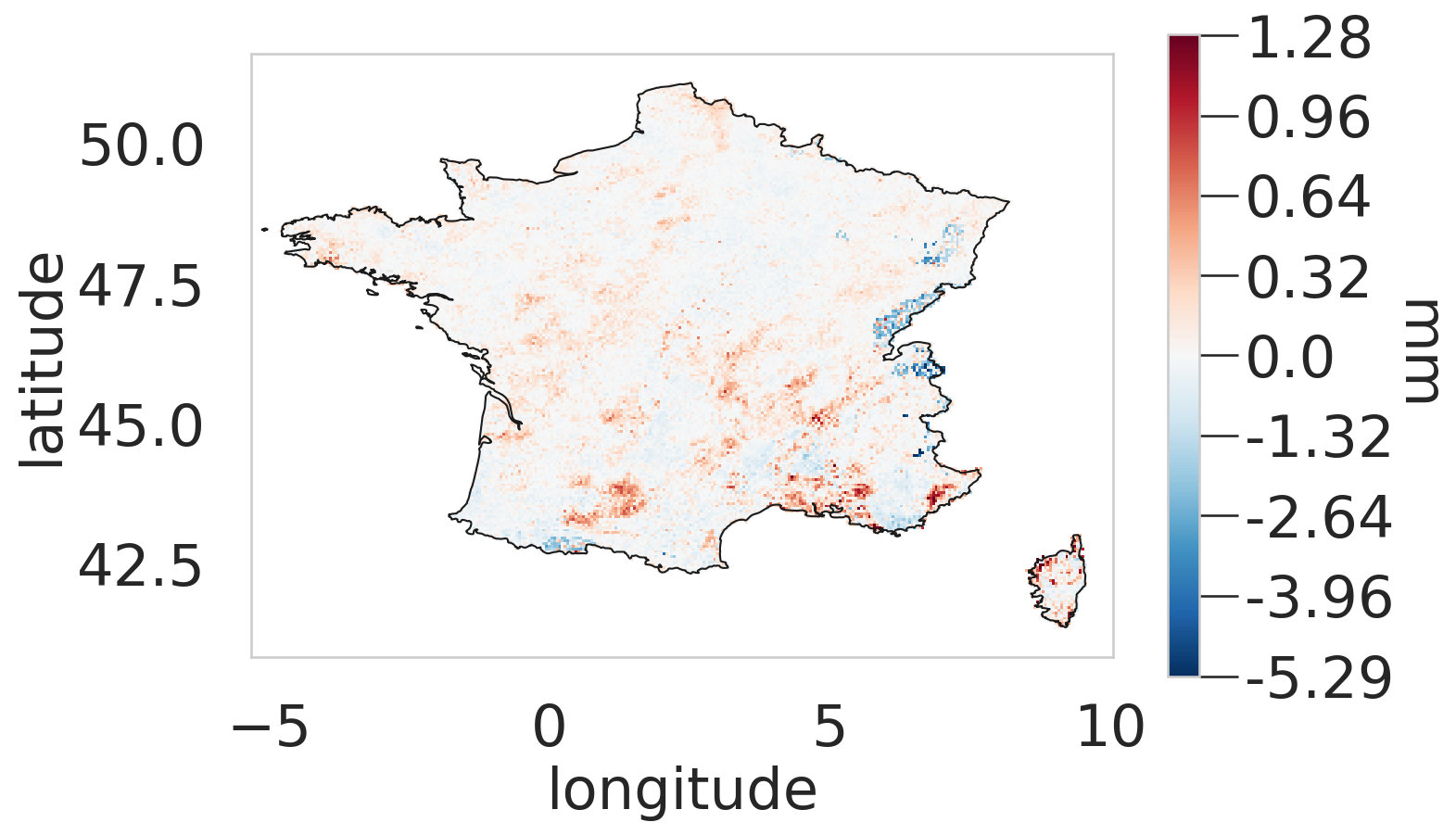}
    \caption{}
\end{subfigure}

\caption{Maps of $MAEdiff_{Stitch-BJ,G}$ for DJF (first row) and JJA (second row) with $G$ being the following models: EGP (a,d), \textit{emp} (b,e) and ExpW (c,f). Note that value ranges are specific to each map.} 
\label{fig:MAE_diff_DJF_JJA}
\end{figure}

Differences maps in Figure \ref{fig:MAE_diff_DJF_JJA} allow us to see how much the extreme errors are corrected with the Stitch-BJ method, compared to the EGP and ExpW. We refer to \textit{model}-\textit{season} the error maps corresponding to the differences between Stitch-BJ's error and \textit{model}'s error, for the given \textit{season}.

For the EGP-DJF (panel a), we see an important increase in performance around the Cévènnes region, with a difference of more than 10\textit{mm} in terms of MAE, while the EGP is at most 1.46\textit{mm} of MAE better than the Stitch-BJ. For EGP-JJA (panel d), performances are relatively similar between the Stitch-BJ and the EGP models.

For the ExpW-DJF (panel c), improvements cover a greater area but are less important with at most an improvement of the MAE of 5\textit{mm}.
As for the emp-DJF and emp-JJA (panels b and e), improvements are much less important and performance is relatively equal with no models improving significantly over the other. The only exception may be with the ExpW-JJA (panel f) where the Stitch-BJ was able to improve the MAE of more than 5\textit{mm} in some areas.

\begin{figure}[!h]
\centering
\begin{subfigure}[b]{0.45\textwidth} 
    \includegraphics[width=\textwidth]{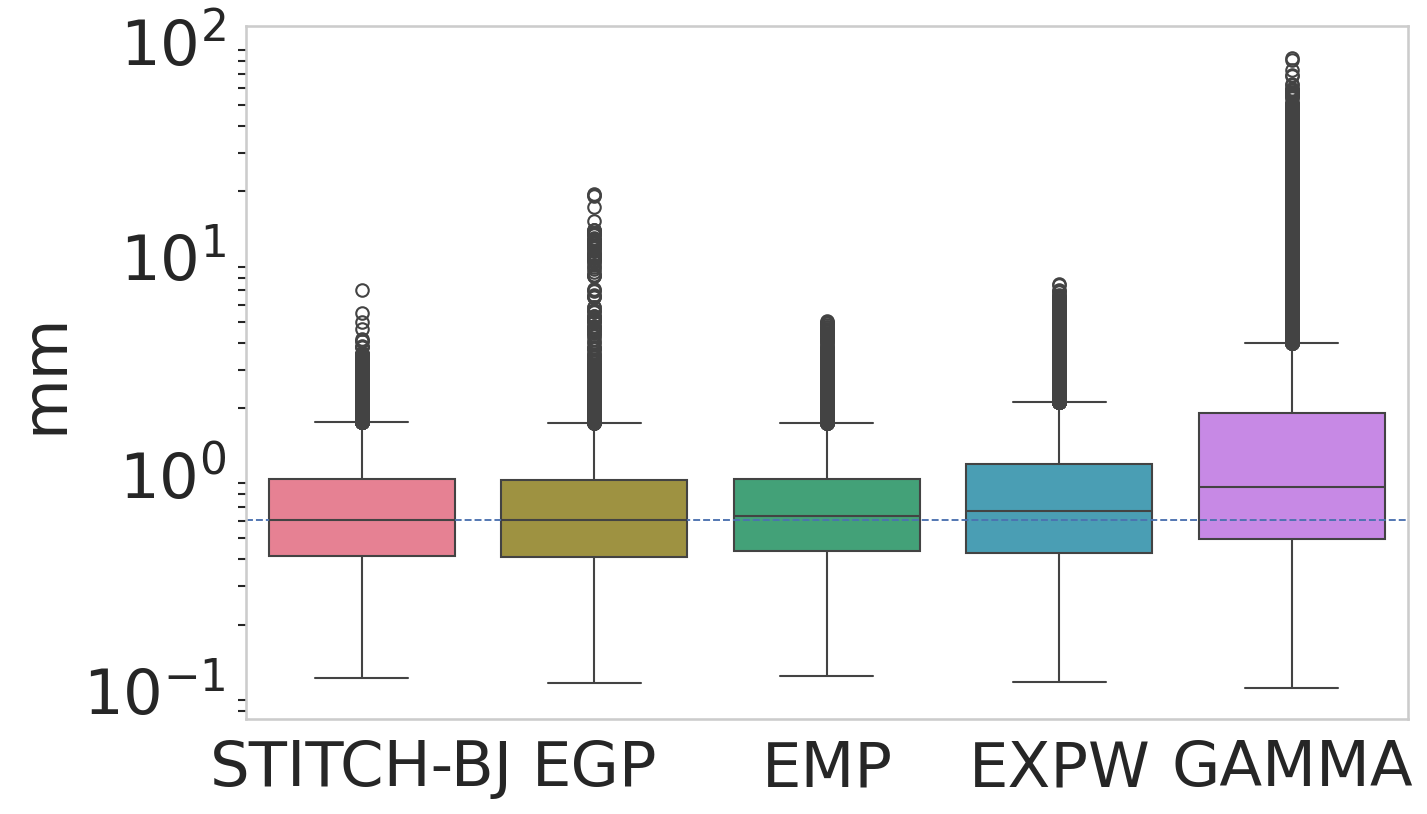}
\end{subfigure}
\hfill
\begin{subfigure}[b]{0.45\textwidth} 
    
    \includegraphics[width=\textwidth]{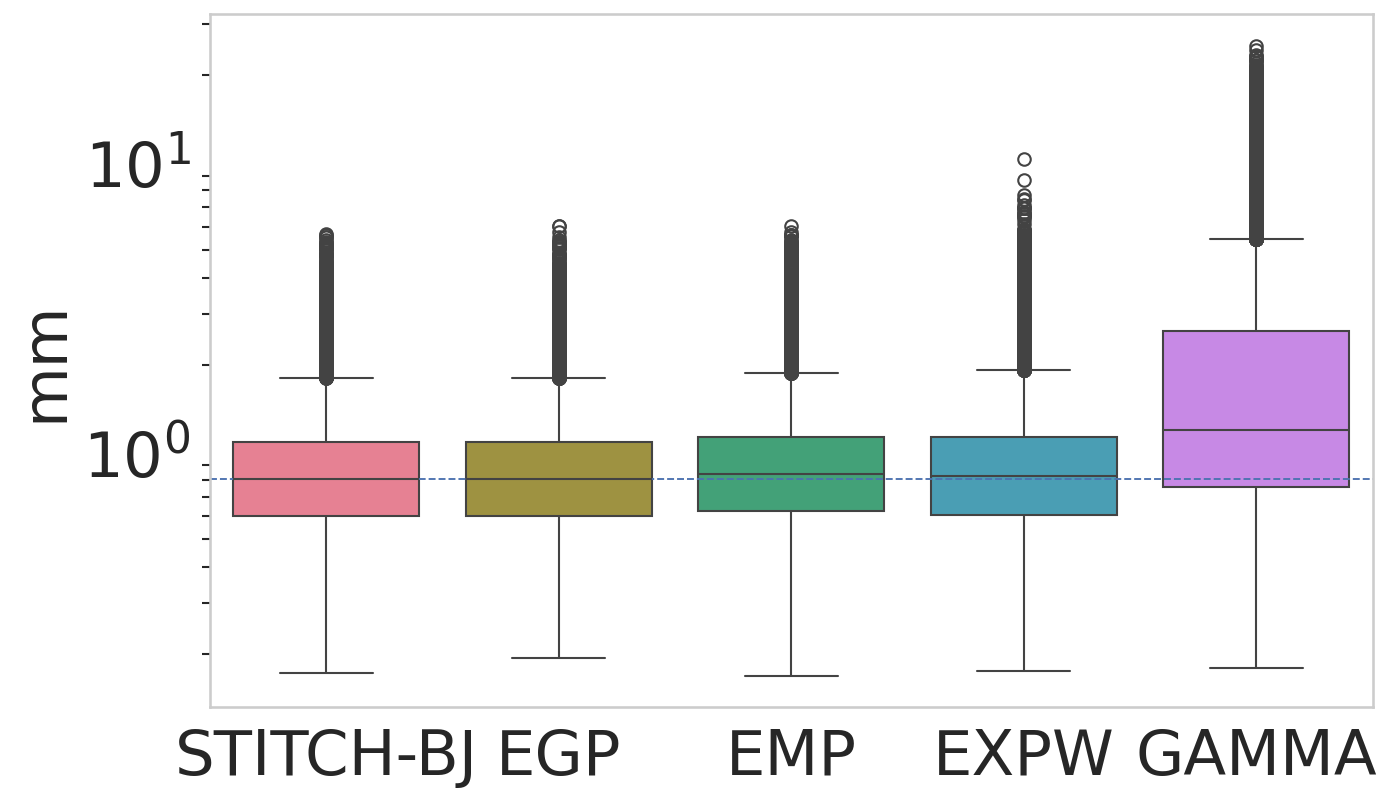}
\end{subfigure}
\caption{Boxplots of MAE for DJF (a) and JJA (b) for all models. Note that the $y$-axis is in log10-scale.} 
\label{fig:BP_MAE_DJF_JJA}
\end{figure}

Boxplots of Figure \ref{fig:BP_MAE_DJF_JJA} show how extreme errors are reduced for both the DJF and JJA seasons {by using the Stitch-BJ model to replace misfitted portions of the EGP and ExpW distributions}. For the DJF season (panel a), the Stitch-BJ was able to improve on the extremes error of both the EGP and ExpW distributions, but {fail to reduce the maximum error against the empirical distribution}.  However, the bulk of the outliers are essentially identical for the Stitch-BJ and empirical distribution. Moreover, in terms of median MAE, the Stitch-BJ and EGP are improving against both the ExpW and empirical models {(with an MAE of  0.61, 0.61, 0.64 and 0.67 for the Stitch-BJ, EGP, \textit{emp} and ExpW respectively)}.
For the JJA season (panel b), conclusions are similar to the DJF season but at a lesser extent with much less extreme outliers produced by the EGP and ExpW distributions. All models are relatively competitive for the JJA season and the Stitch-BJ and EGP still produce very slightly lower median MAE than other models.

\subsection{Mean Absolute Error over the 95th percentile}
\label{sec:3_2}
\begin{figure}[!h]
\centering
 \begin{subfigure}[b]{0.24\textwidth}   
    \centering
    \includegraphics[width=\textwidth]{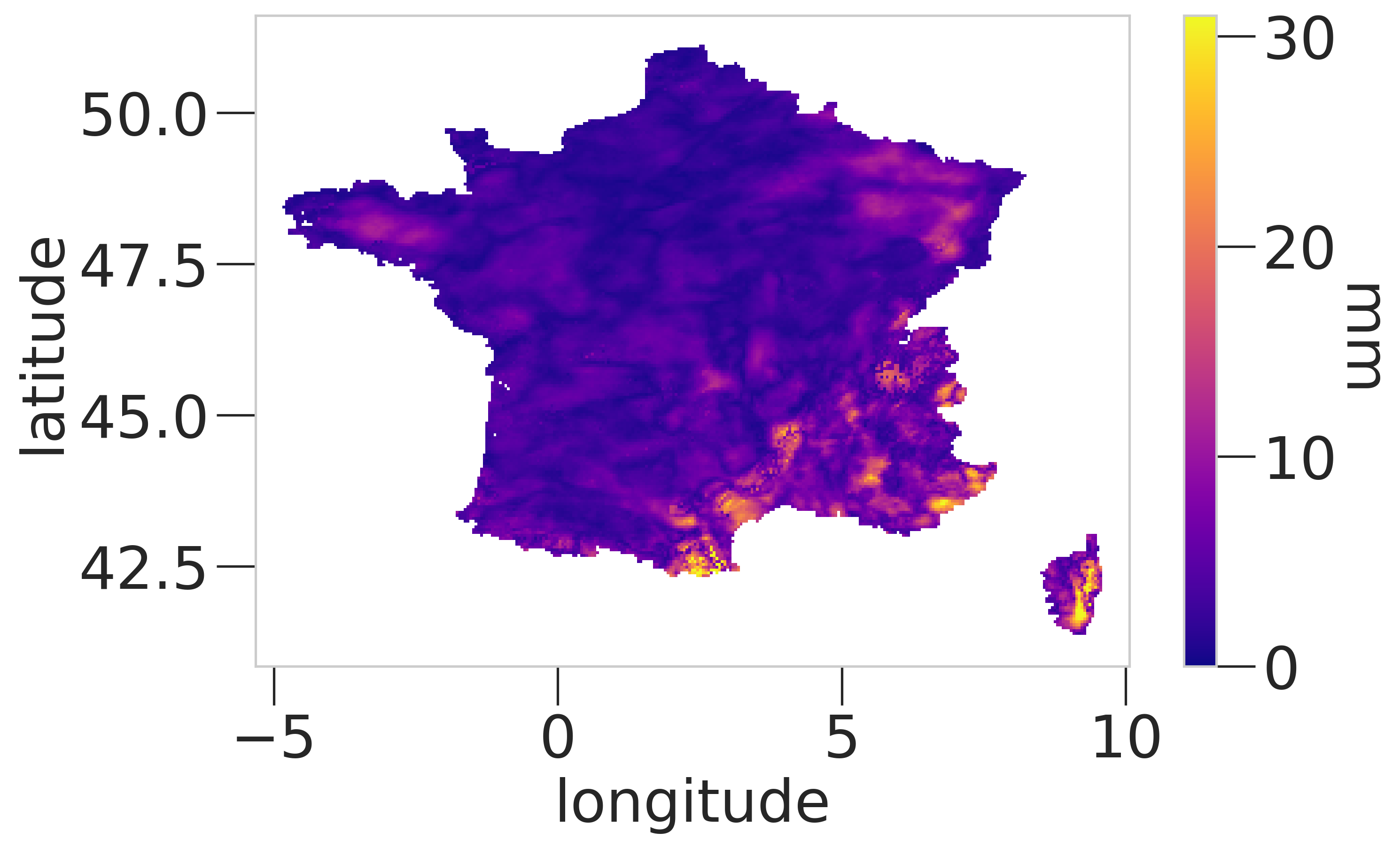}
    \caption{}
\end{subfigure}
\hfill
 \begin{subfigure}[b]{0.24\textwidth}   
    \centering
    \includegraphics[width=\textwidth]{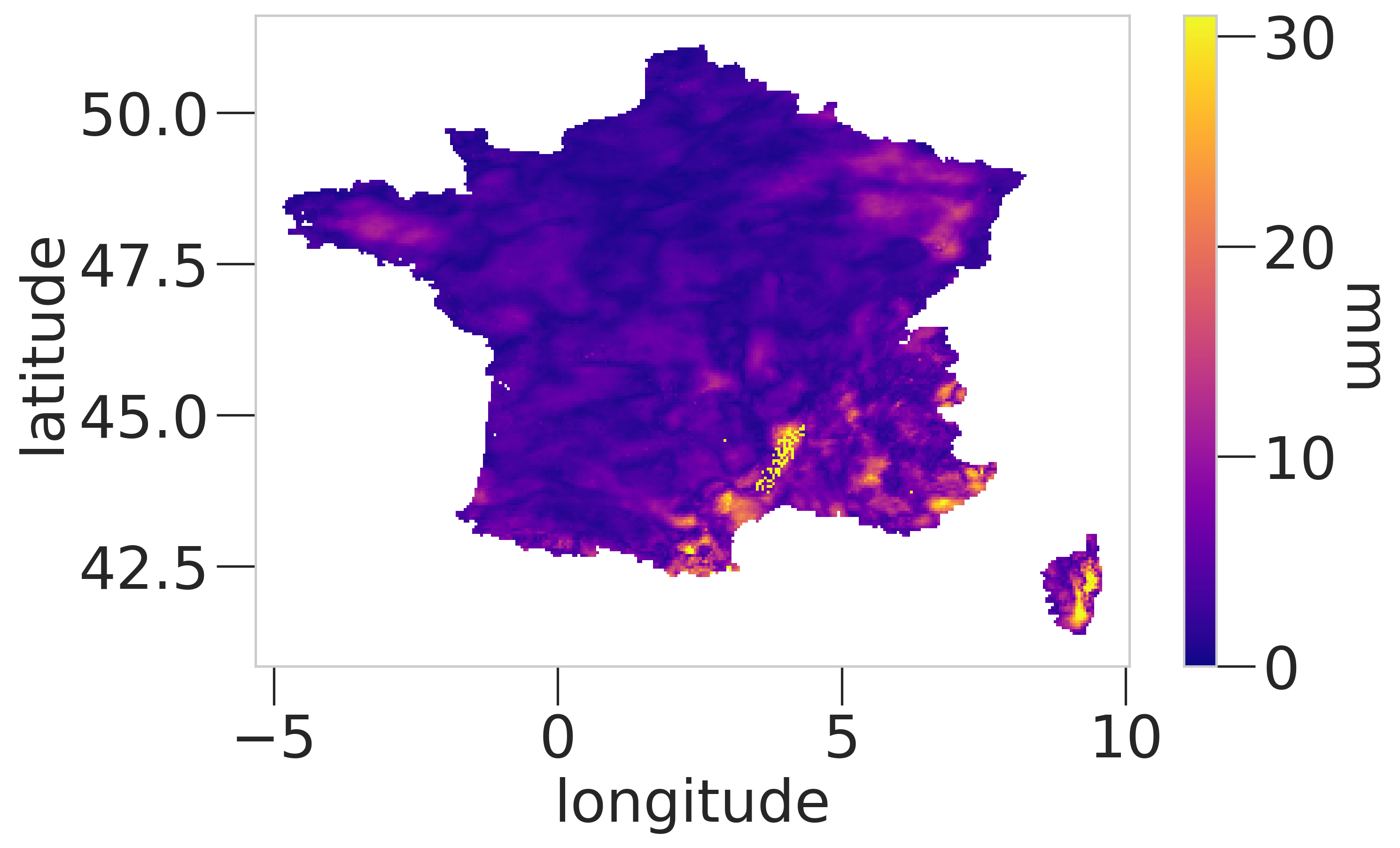}
    \caption{}
\end{subfigure}
\hfill
\begin{subfigure}[b]{0.24\textwidth}   
    \centering
    \includegraphics[width=\textwidth]{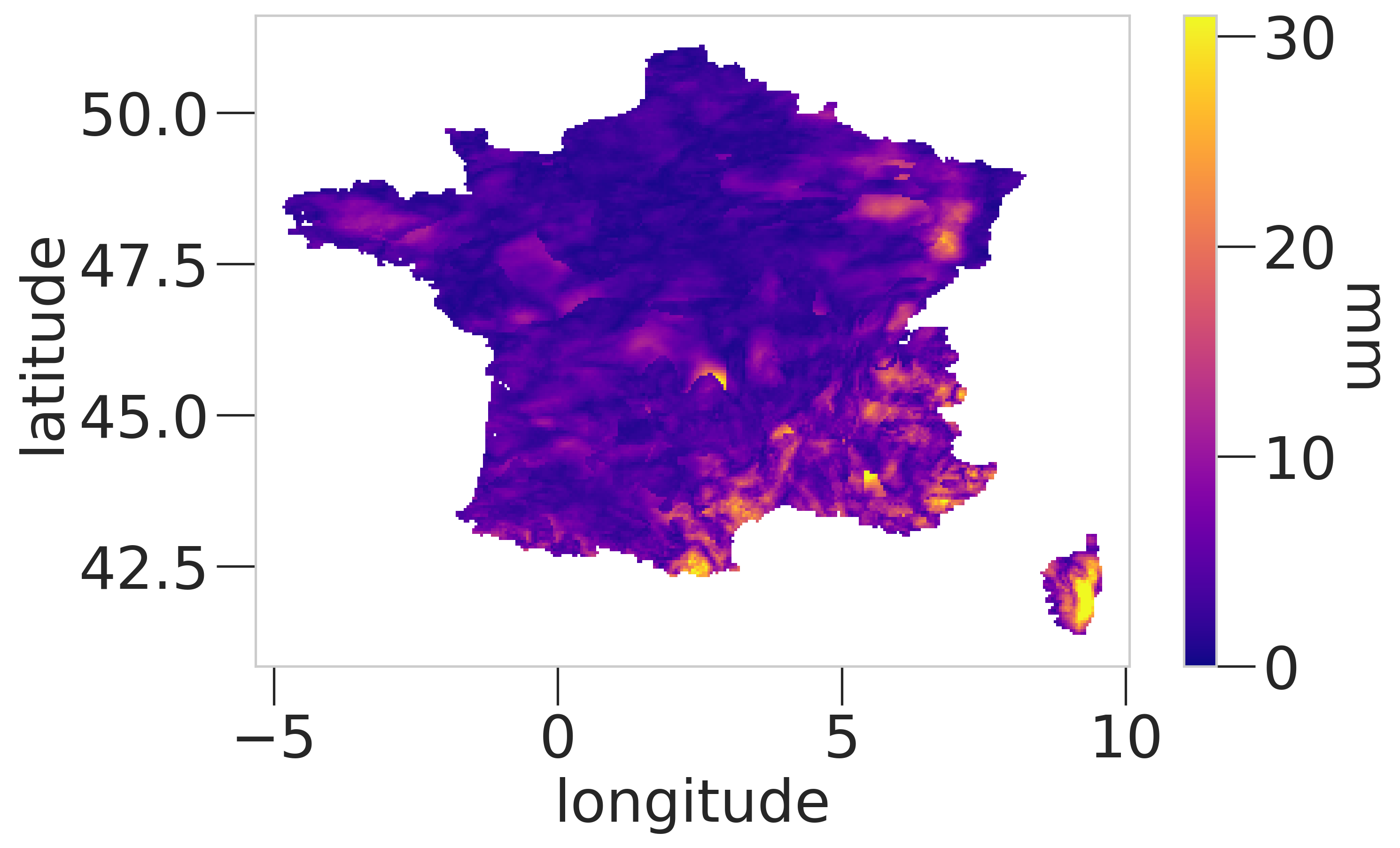}
    \caption{}
\end{subfigure}
\hfill
 \begin{subfigure}[b]{0.24\textwidth}   
    \centering
    \includegraphics[width=\textwidth]{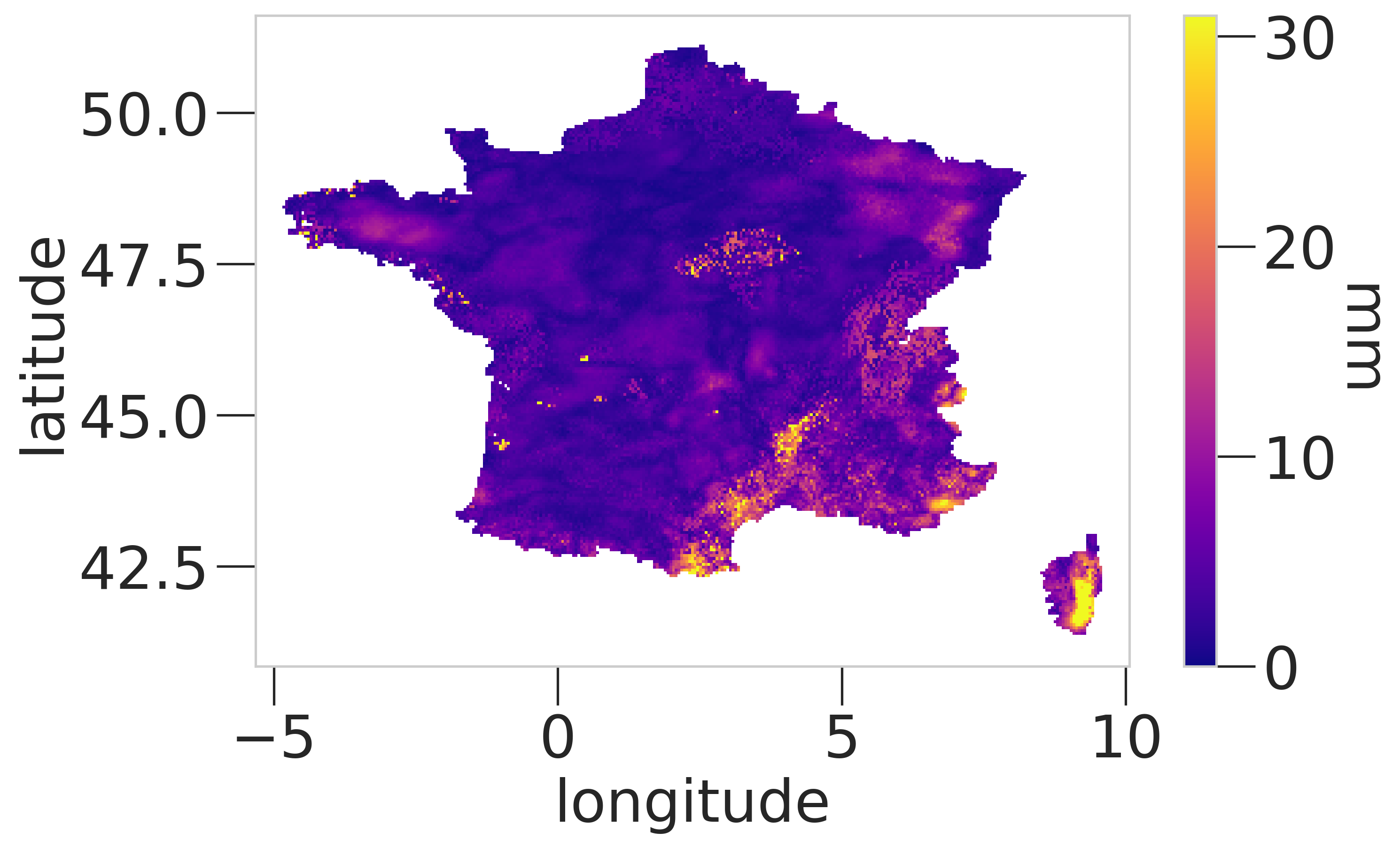}
    \caption{}
\end{subfigure}
\vskip\baselineskip
 \begin{subfigure}[b]{0.24\textwidth}   
    \centering
    \includegraphics[width=\textwidth]{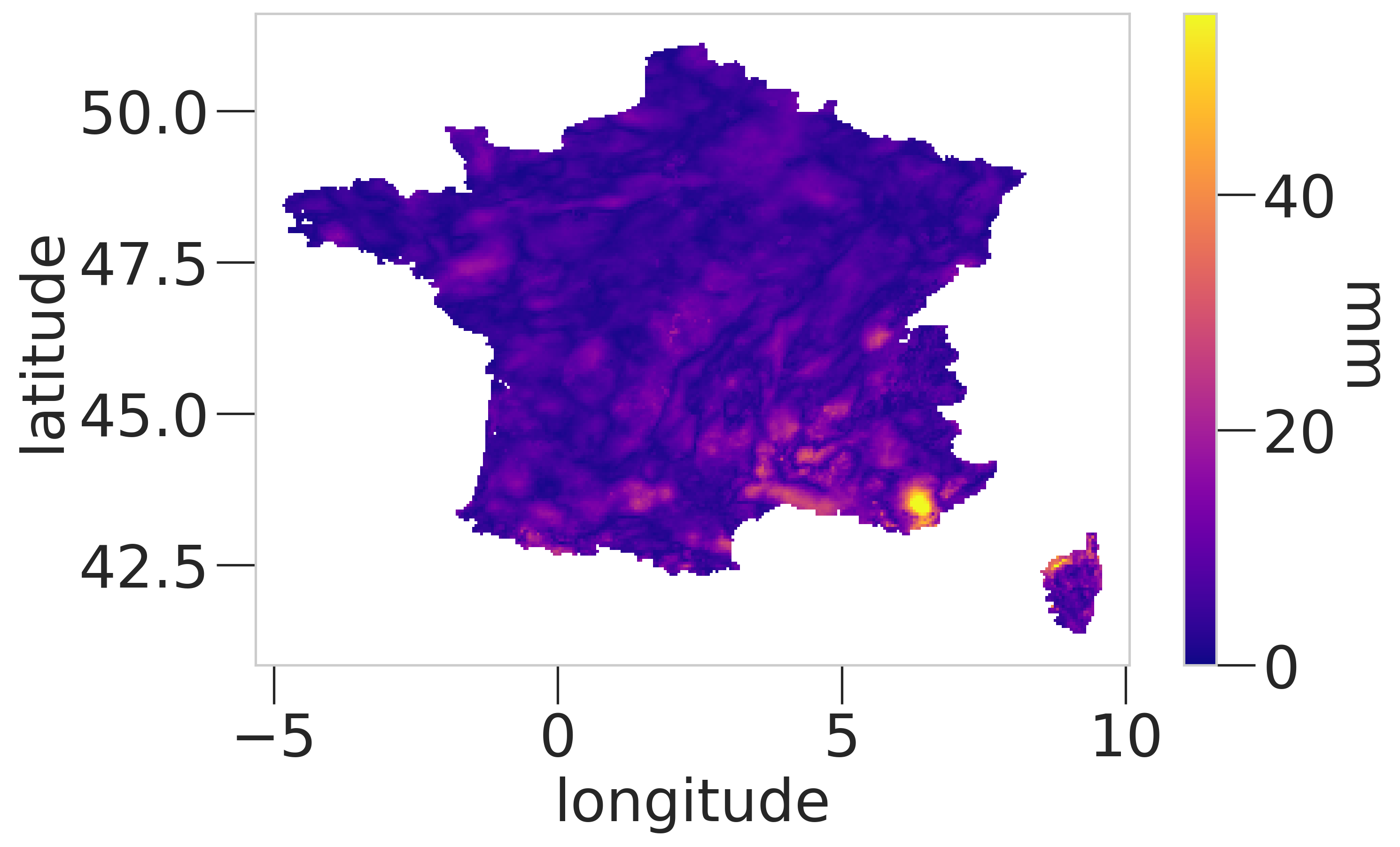}
    \caption{}
\end{subfigure}
\hfill
 \begin{subfigure}[b]{0.24\textwidth}   
    \centering
    \includegraphics[width=\textwidth]{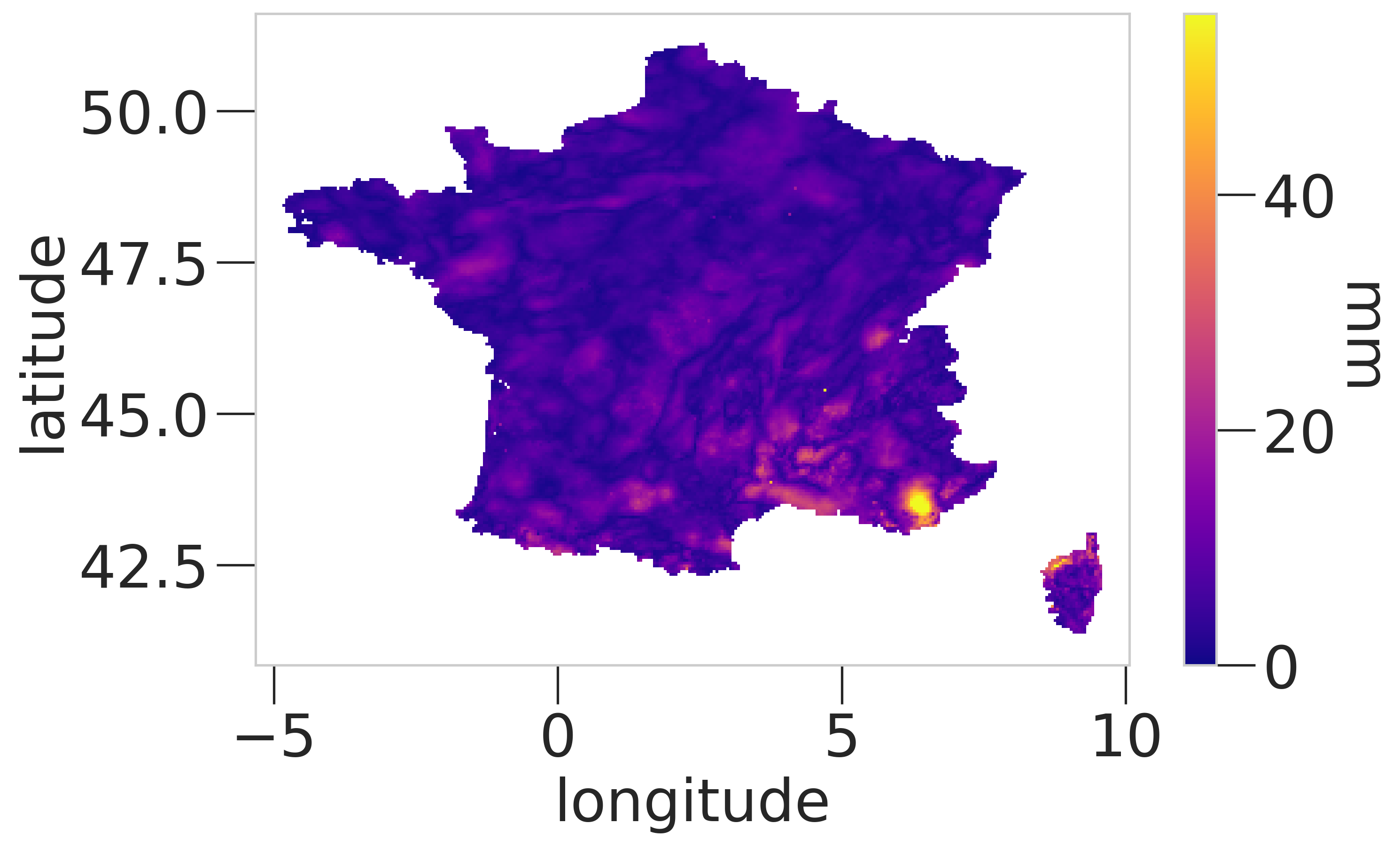}
    \caption{}
\end{subfigure}
\hfill
\begin{subfigure}[b]{0.24\textwidth}   
    \centering
    \includegraphics[width=\textwidth]{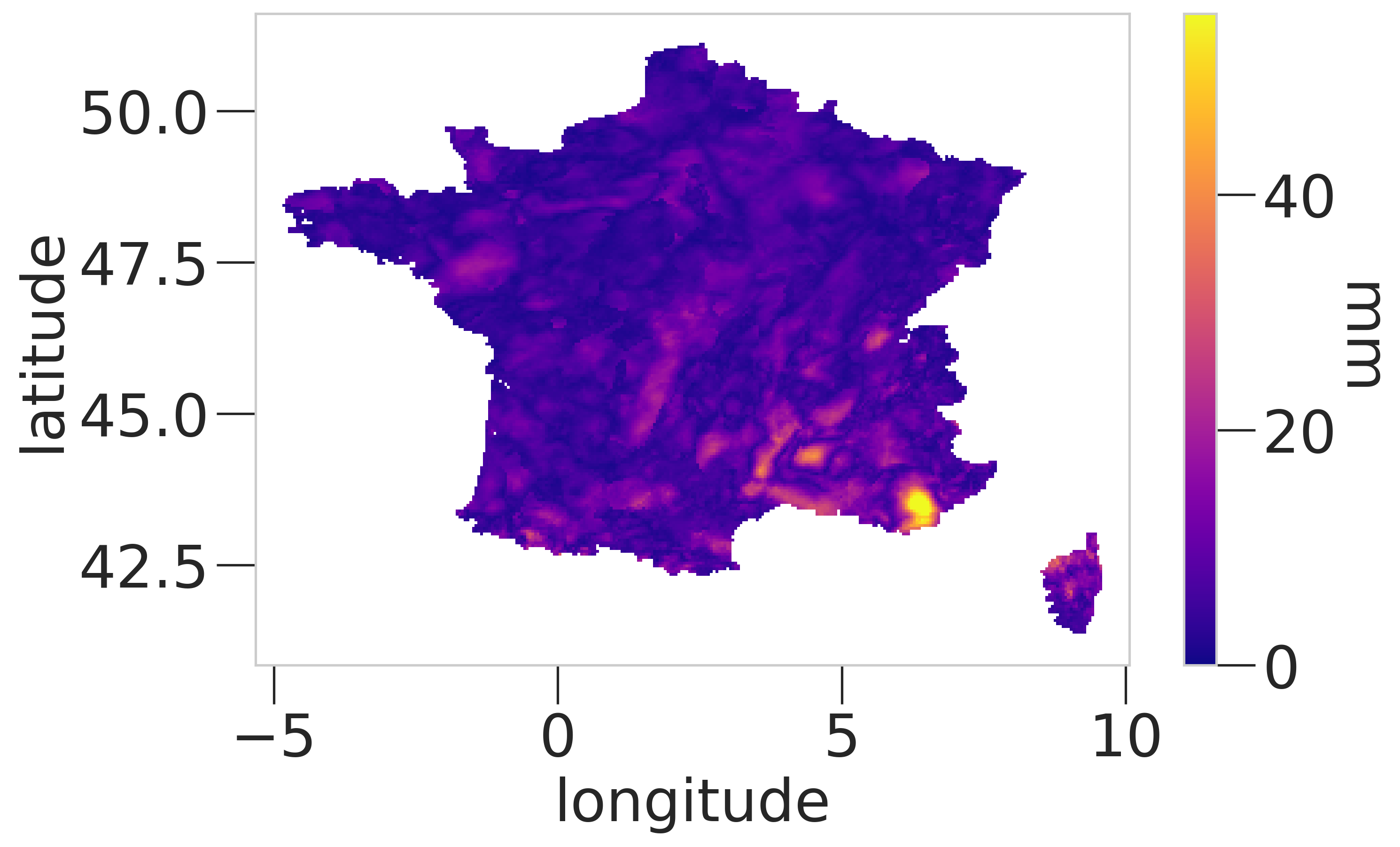}
    \caption{}
\end{subfigure}
\hfill
 \begin{subfigure}[b]{0.24\textwidth}   
    \centering
    \includegraphics[width=\textwidth]{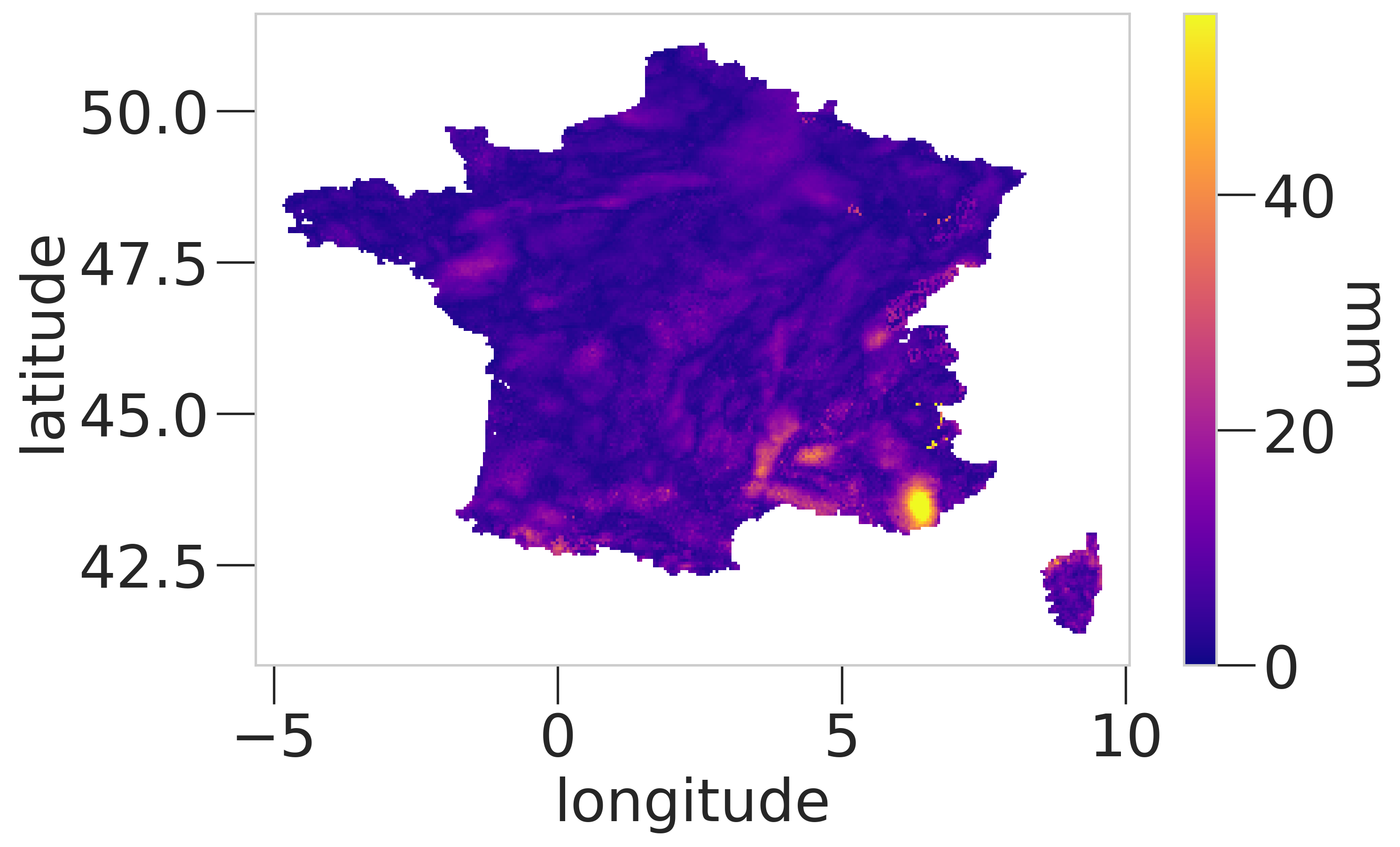}
    \caption{}
\end{subfigure}
\caption{Maps of MAE95sup for DJF (first row) and JJA (second row) for the following models: Stitch-BJ panels (panels a,e), EGP (panels b,f), \textit{emp} (panels c,g) and ExpW (panels d,h).} 
\label{fig:MAE95sup_DJF_JJA}
\end{figure}

MAE95sup maps in Figure \ref{fig:MAE95sup_DJF_JJA} are very similar to the MAE maps of Figure \ref{fig:MAE_DJF_JJA}. However, for the DJF season, the Stitch-BJ was able to improve against all other shown models significantly in some regions: Cévènnes and Corsica regions for both EGP and ExpW (panels a and c) and Alps and Corsica regions for the empirical model (panel b).
However, for the JJA season (panels d, e and f), no significant improvements are easily noticeable.

\begin{figure}[!h]
    \centering
     \begin{subfigure}[b]{0.32\textwidth}   
        \centering
        \includegraphics[width=\textwidth]{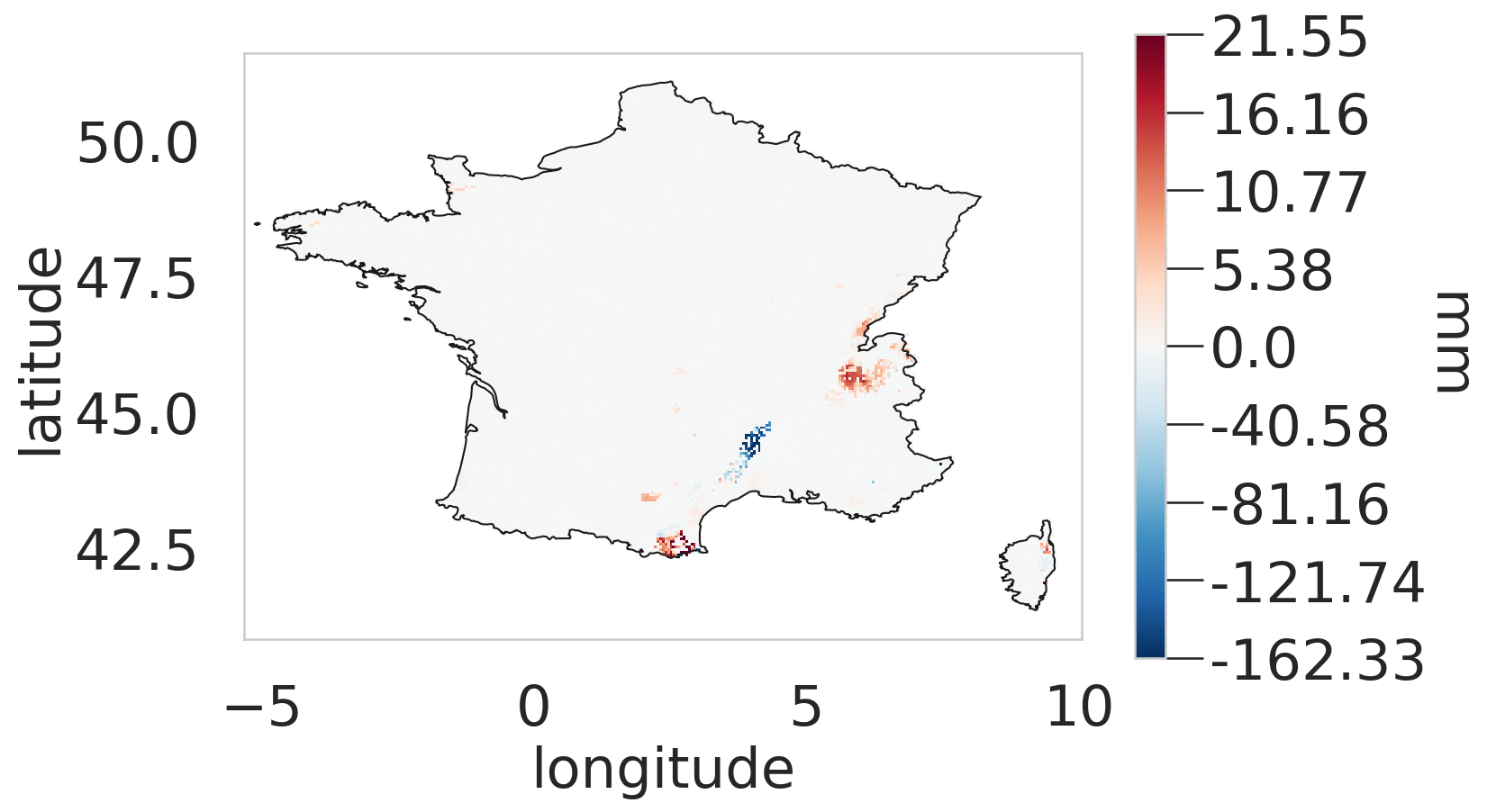}
        \caption{}
    \end{subfigure}
    \hfill
     \begin{subfigure}[b]{0.32\textwidth}   
        \centering
        \includegraphics[width=\textwidth]{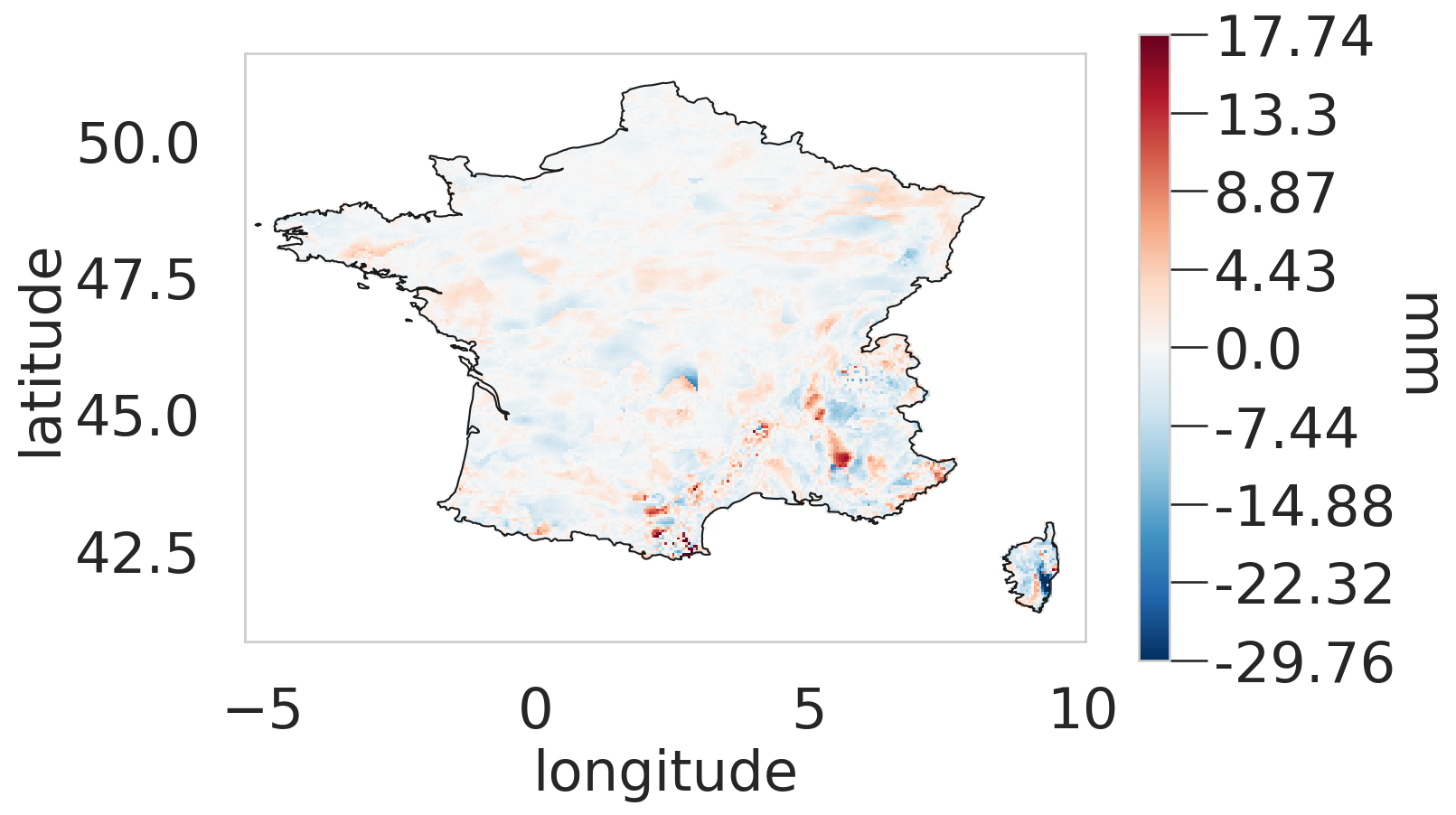}
        \caption{}
    \end{subfigure}
    \hfill
    \begin{subfigure}[b]{0.32\textwidth}   
        \centering
        \includegraphics[width=\textwidth]{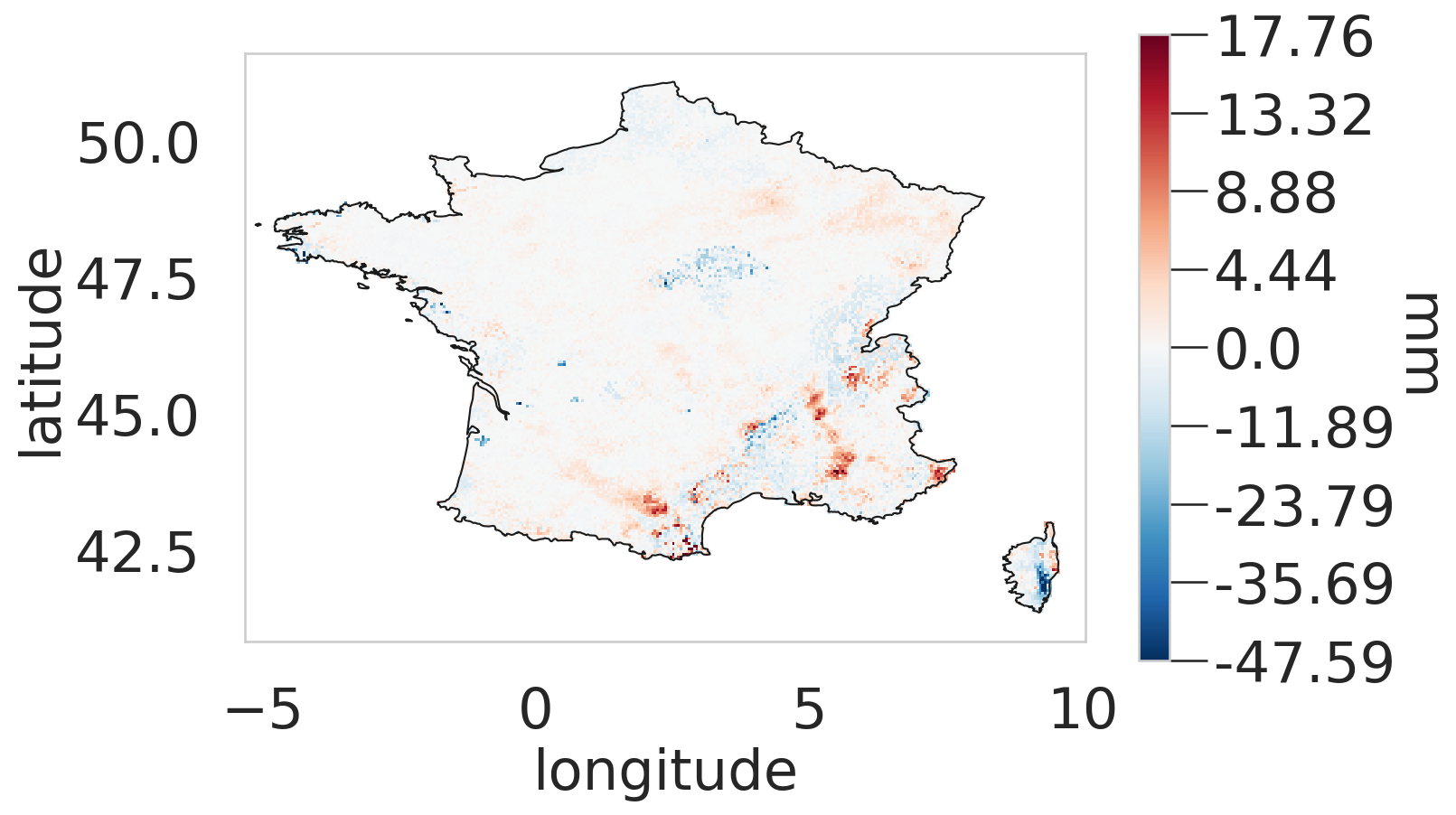}
        \caption{}
    \end{subfigure}
    
    \vskip\baselineskip
     \begin{subfigure}[b]{0.32\textwidth}   
        \centering
        \includegraphics[width=\textwidth]{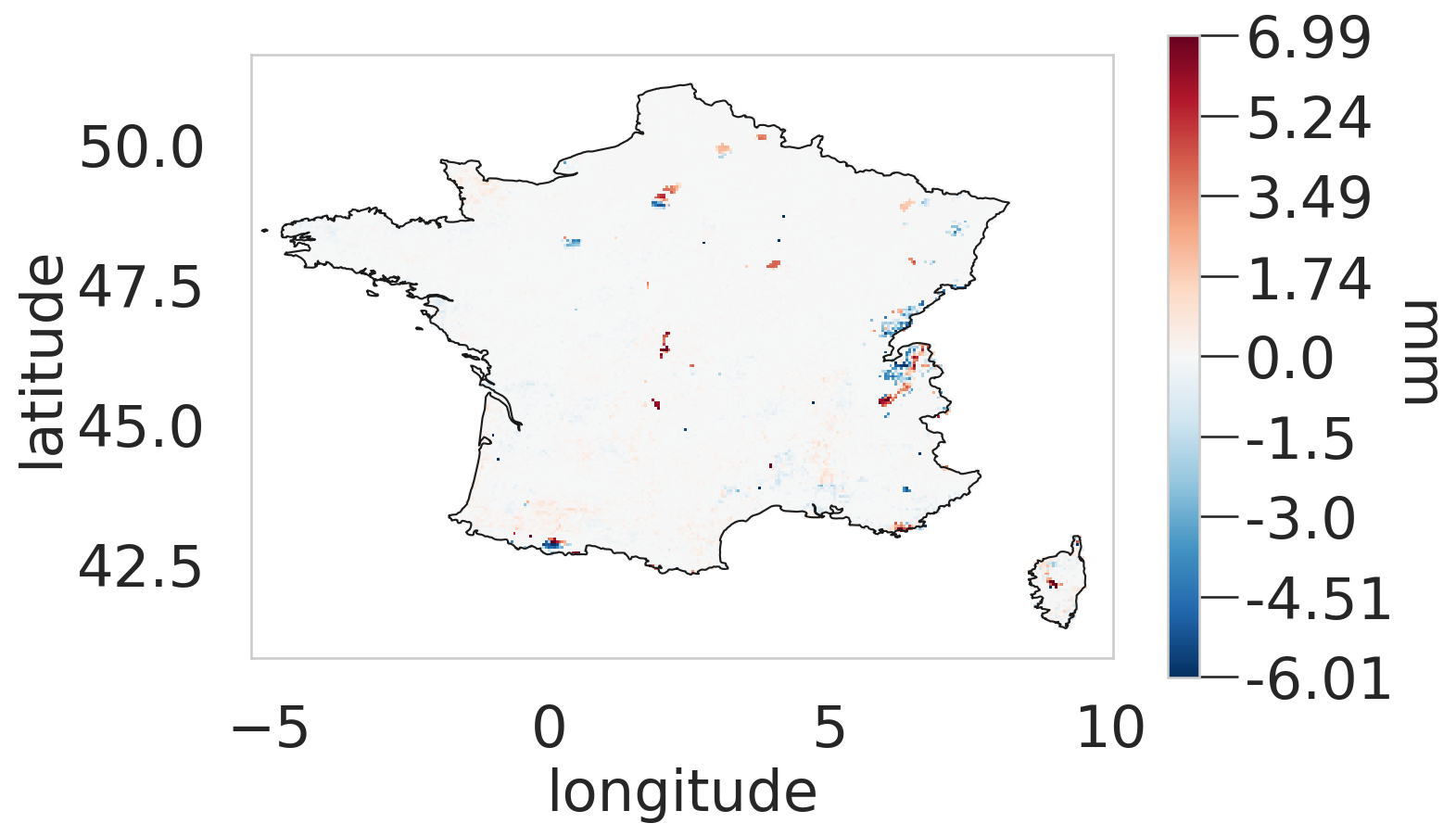}
        \caption{}
    \end{subfigure}
    \hfill
     \begin{subfigure}[b]{0.32\textwidth}   
        \centering
        \includegraphics[width=\textwidth]{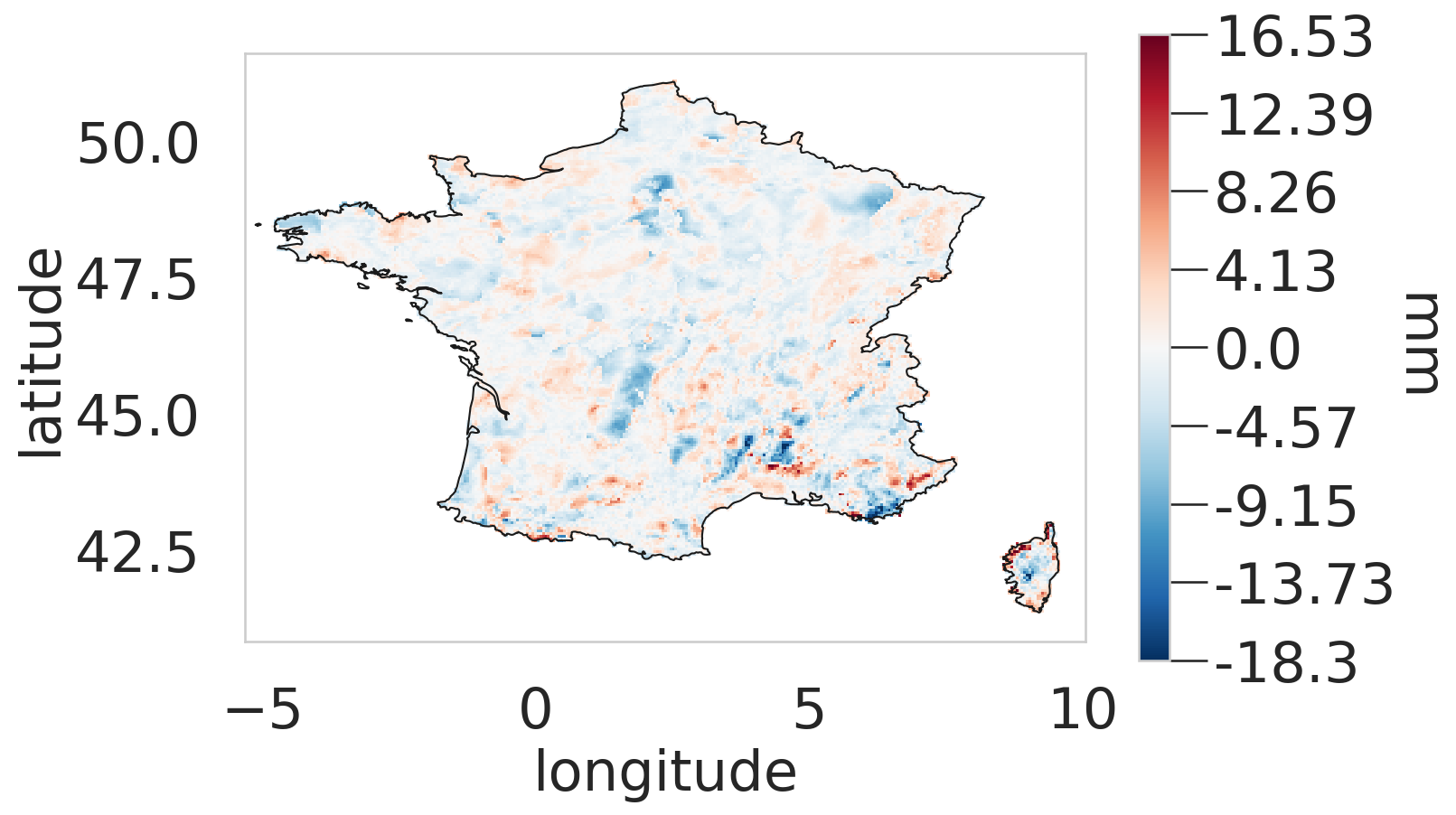}
        \caption{}
    \end{subfigure}
    \hfill
    \begin{subfigure}[b]{0.32\textwidth}   
        \centering
        \includegraphics[width=\textwidth]{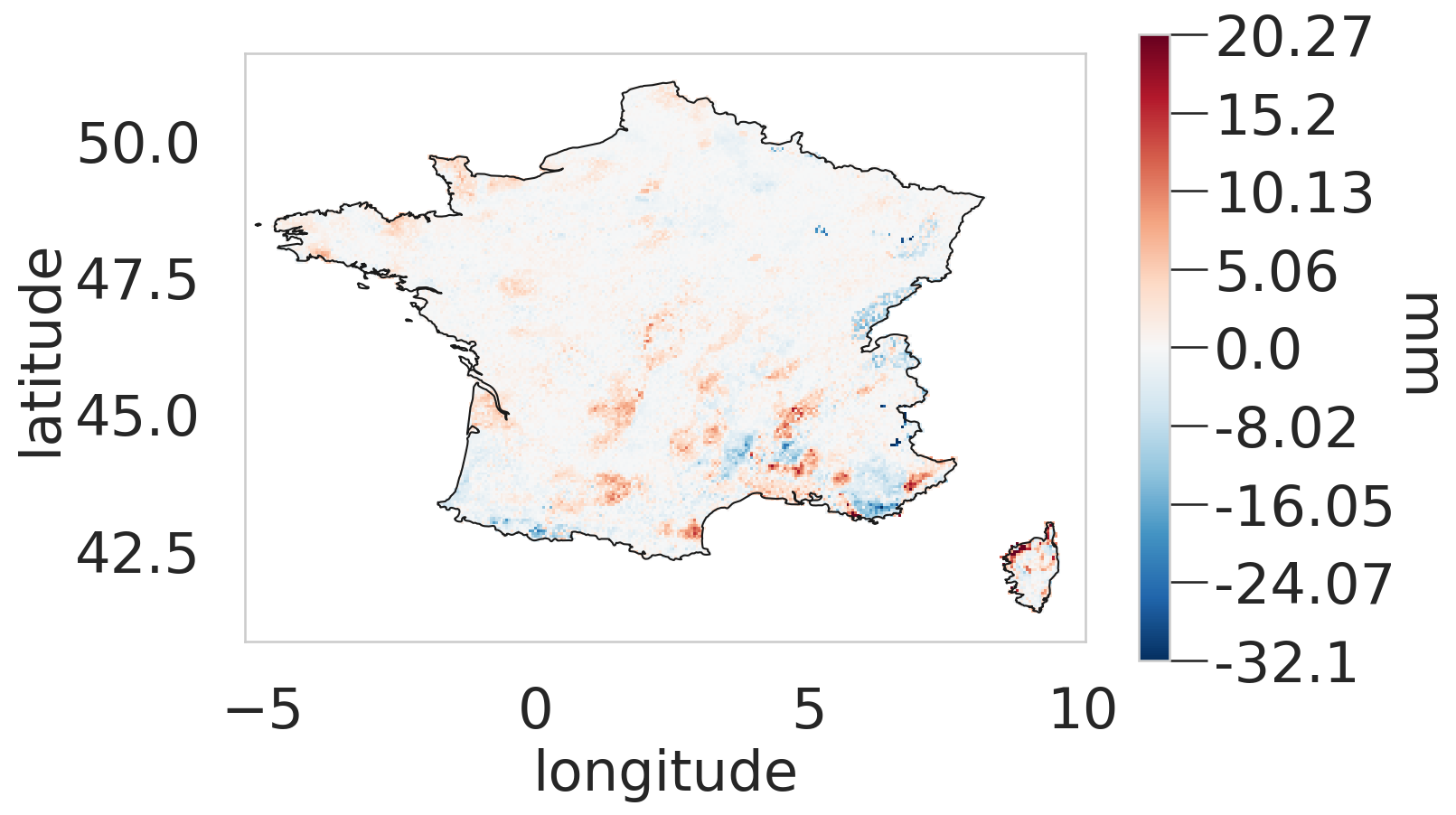}
        \caption{}
    \end{subfigure}
    
    \caption{Maps of $MAE95supdiff_{Stitch-BJ,G}$ for DJF (first row) and JJA (second row) with $G$ being the following models: EGP (a,d), \textit{emp} (b,e) and ExpW (c,f). Note that value ranges are specific to each map.} 
    \label{fig:MAE95sup_diff_DJF_JJA}
    \end{figure}

Differences of MAE95sup in Figure \ref{fig:MAE95sup_diff_DJF_JJA} clearly show the advantages of the Stitch-BJ against all other methods for the DJF season, mainly focused on the Cévènnes and Corsica region, with errors differences ranging from 30\textit{mm} to 162\textit{mm} (panels a, b and c). However, some locations show poor performance from the Stitch-BJ compared to other models, especially in the Eastern Pyrenees where other models produce MAE95sup 17\textit{mm} to 21\textit{mm} lower.
For the JJA season, performance is equivalent for most models (panels d and e), with only against the ExpW (panel f) where the Stitch-BJ can show higher improvements than the opposite with a maximum improvement of 32\textit{mm} against 20\textit{mm} for the ExpW.

\begin{figure}[H]
\centering
\begin{subfigure}[b]{0.45\textwidth}   
\includegraphics[width=\linewidth]{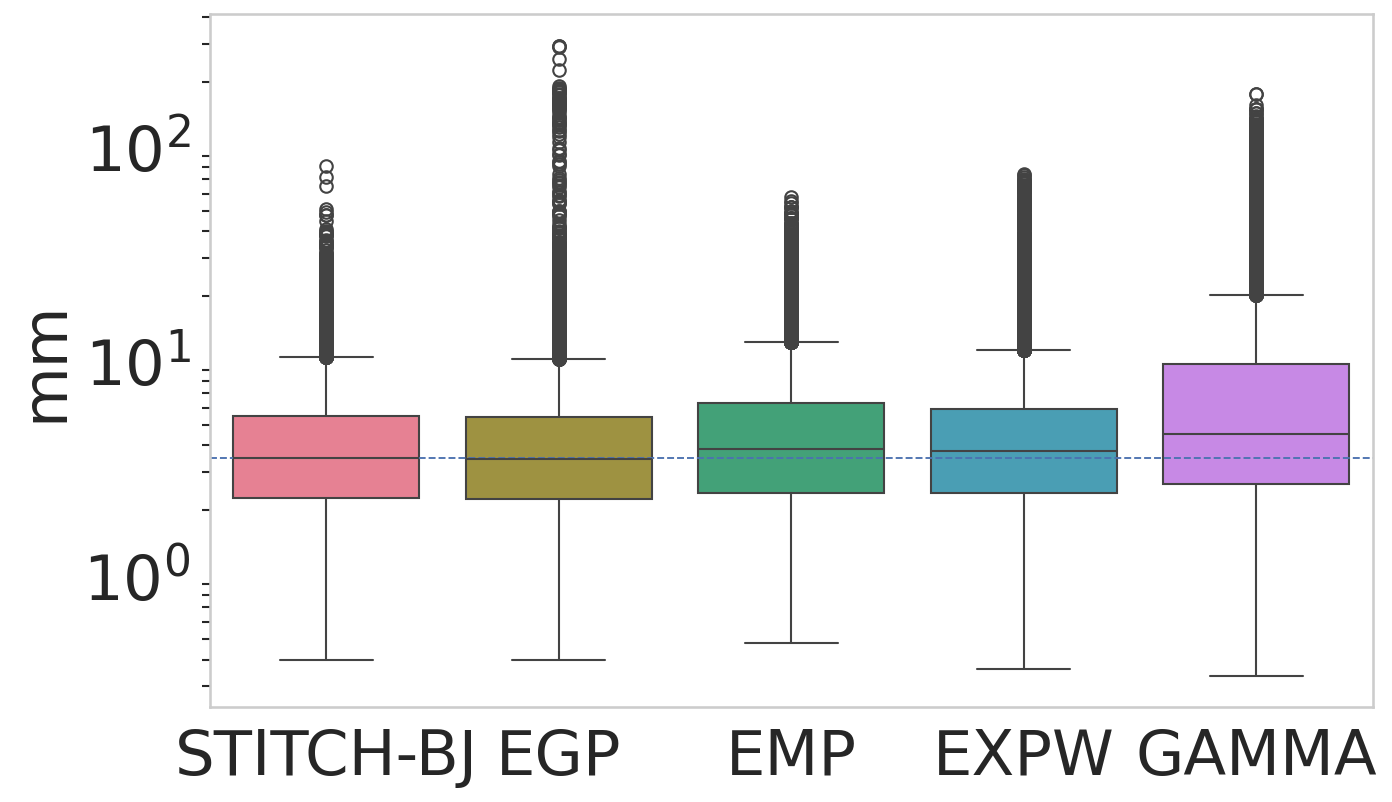}\end{subfigure}
\hfill
\begin{subfigure}[b]{0.45\textwidth}   
\includegraphics[width=\linewidth]{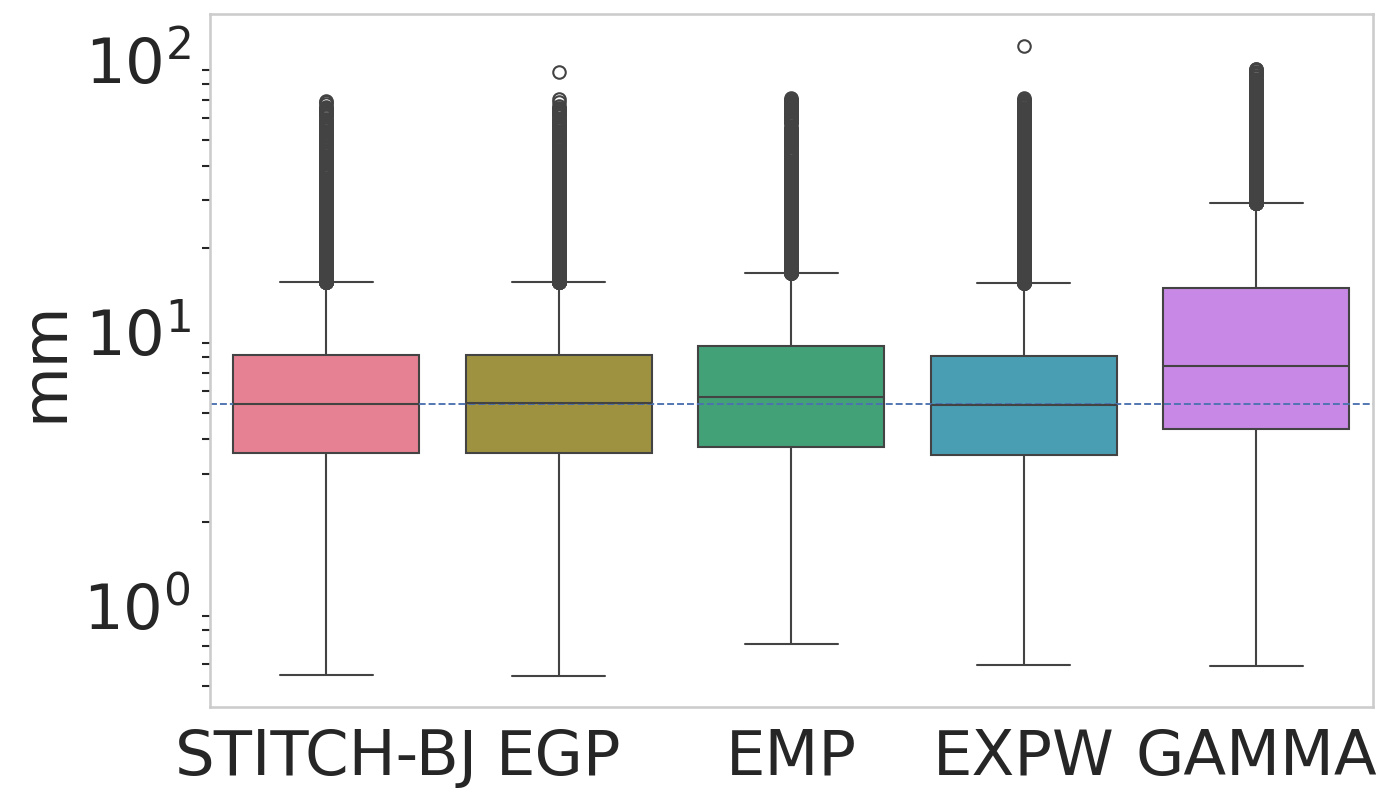}\end{subfigure}
\caption{Boxplots of MAE95sup for DJF (a) and JJA (b) for all models. Note that the $y$-axis is in log10-scale.} 
\label{fig:BP_MAE95sup_DJF_JJA}
\end{figure}

In Figure \ref{fig:BP_MAE95sup_DJF_JJA}, boxplots for the DJF season 
(panel a) show the clear improvement of the Stitch-BJ against the EGP and ExpW models. At first glance, the ExpW model produces lower outliers than the Stitch-BJ but this is only true for the three most extreme outliers produced by the Stitch-BJ. The rest of the outliers are much more condensed and generally lower than both the ExpW and empirical models' outliers. Median errors are also slightly lower for the EGP and Stitch-BJ models compared to all other models for the DJF season.
Models over the JJA season (panel b) produce very similar extreme outliers, and median errors are also very close and higher than for DJF.

Additional figures for the RMSE results are available in Appendix \ref{app:rmse}.

\subsection{Impact of seasonality on performance}
\label{sec:3_4}

\begin{figure}[H]
    \centering
    \includegraphics[width=\linewidth]{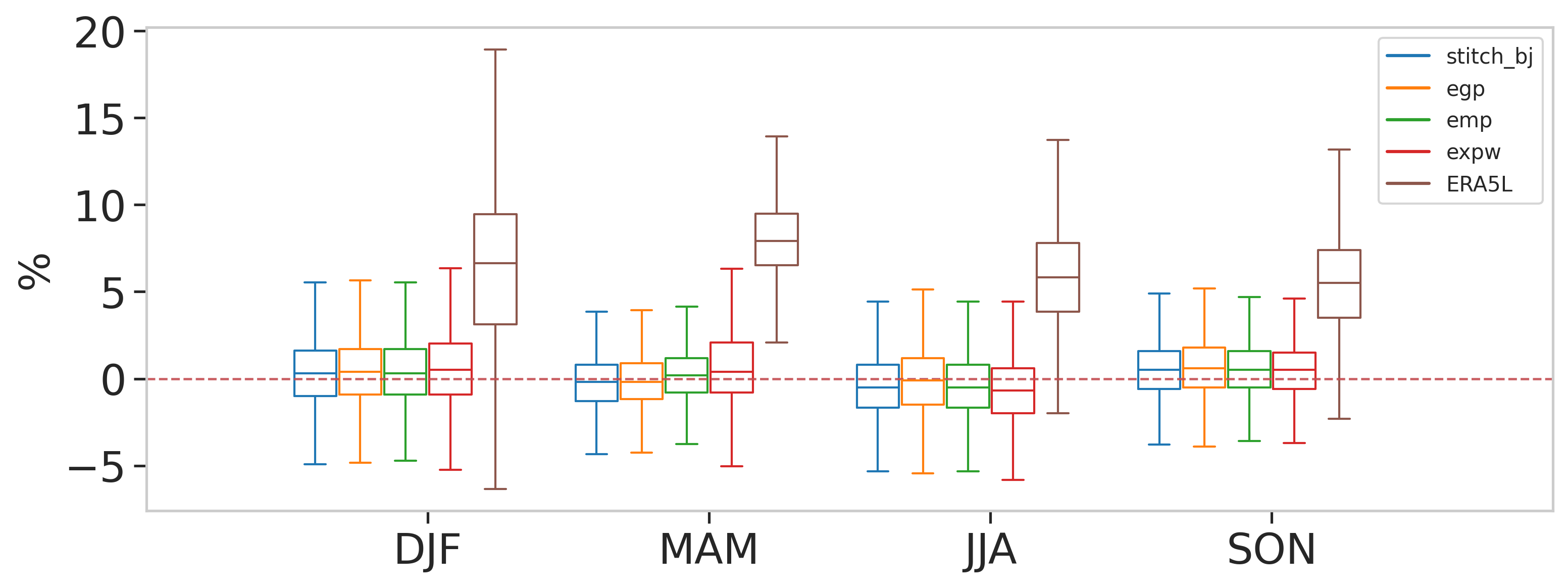}
    \caption{Multi-season dry days probability difference between CERRAL and Stitch-BJ, EGP, \textit{emp}, ExpW and uncorrected model ERA5L data.}
    \label{fig:BP_mult_p0}
\end{figure}

In the following figures, all seasons are shown on the same boxplot to assess the performance of the different methods season-wise. For the dry days' probability in Figure \ref{fig:BP_mult_p0}, we show the difference between the target dry days probability and the modeled one: a positive value means a dry bias, while a negative one implies a wet bias. 

In Figure \ref{fig:BP_mult_p0}, the impact of the SSR method is very noticeable for all seasons, with a large reduction in differences in dry days probability for all methods using the SSR correction. The method performed best for the DJF and MAM season, while a small wet bias remains in JJA season and a dry bias in the SON season. Since the underlying model has almost no impact on the SSR correction, all models corrected with the SSR produce similar dry days probability.

\begin{figure}[!h]
\centering

\begin{subfigure}[b]{0.7\textwidth}   
\includegraphics[width=\linewidth]{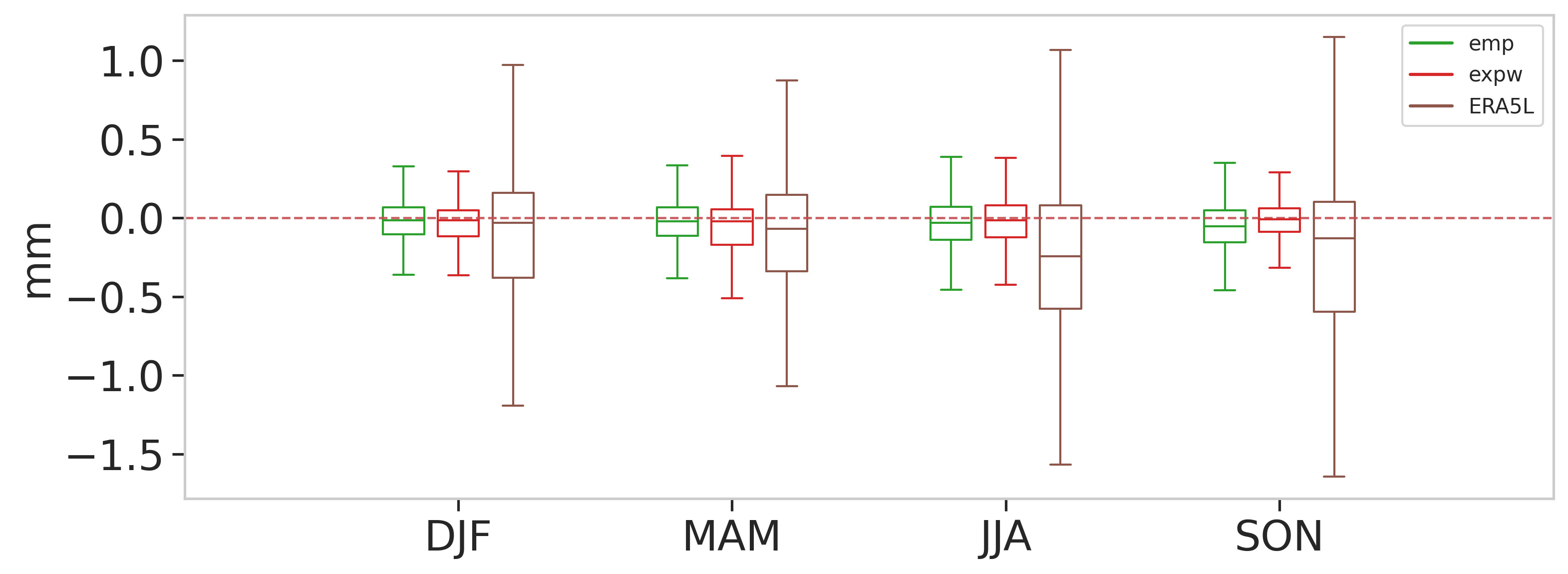}
\end{subfigure}
\begin{subfigure}[b]{0.7\textwidth}   
\includegraphics[width=\linewidth]{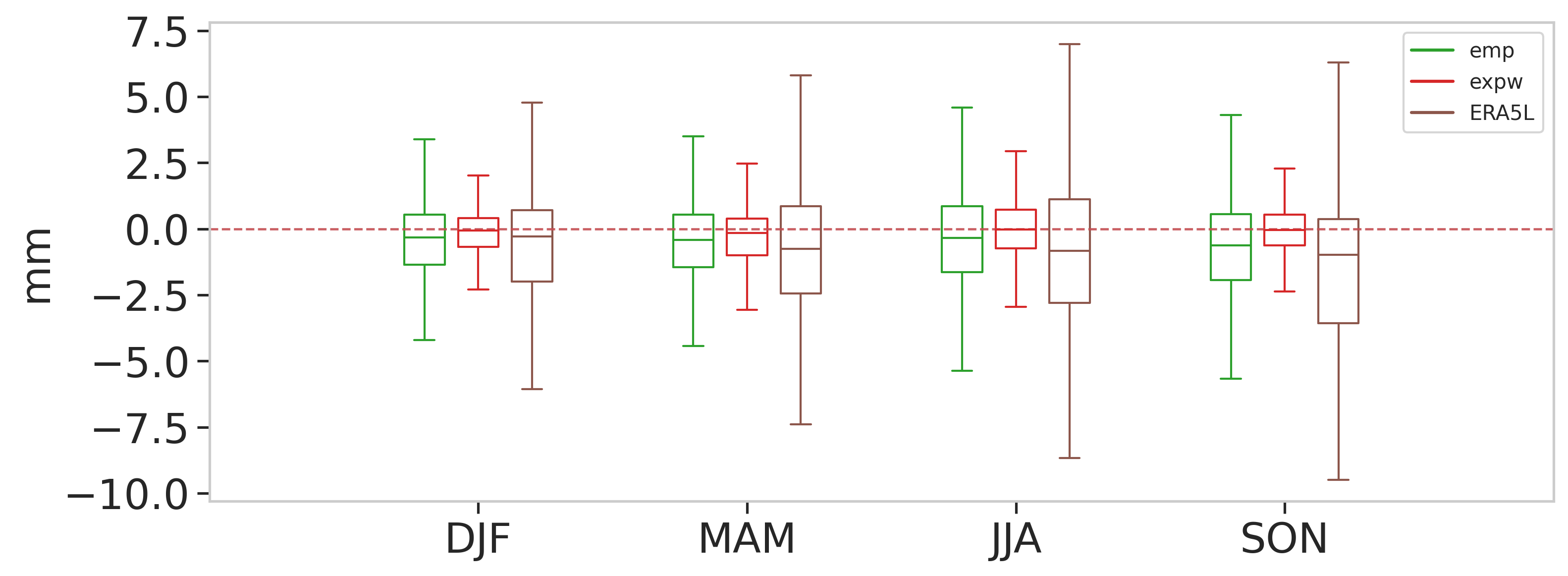}
\end{subfigure}  

\begin{subfigure}[b]{0.7\textwidth}   
\includegraphics[width=\linewidth]{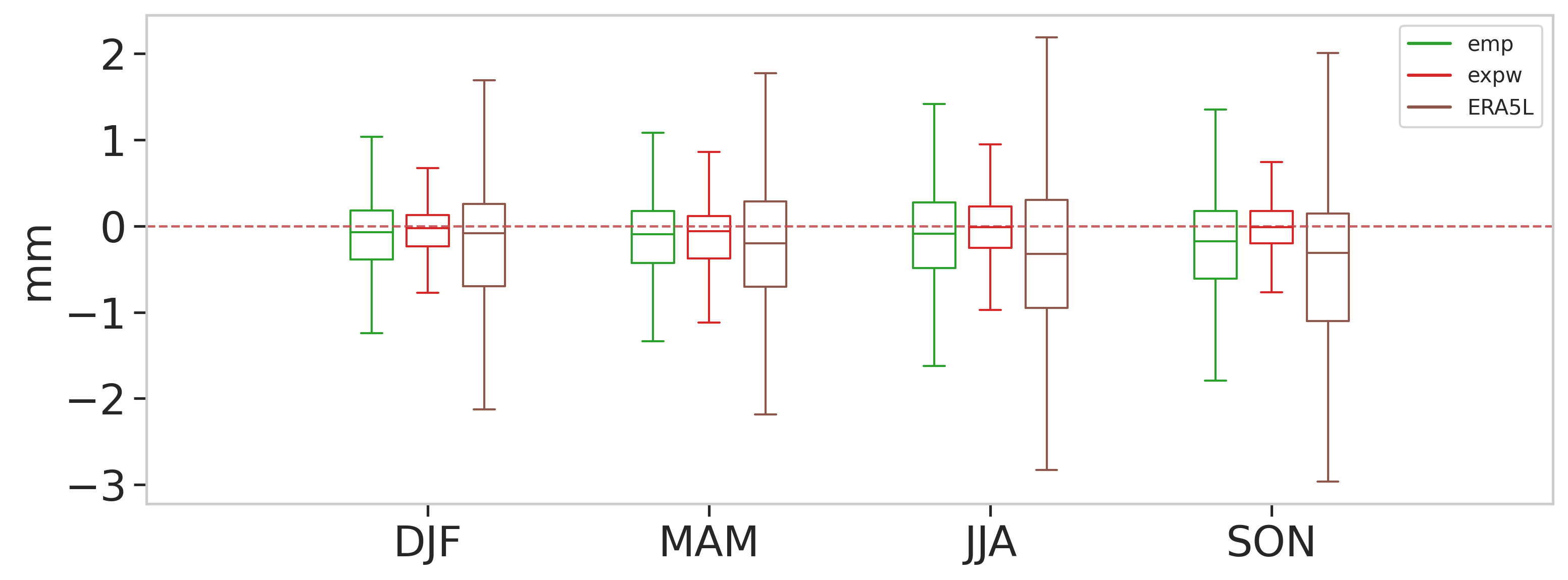}
\end{subfigure}
\caption{Multi-season MAE (a), MAE95sup (b) and RMSE (c) difference of Stitch-BJ, against \textit{emp}, ExpW and uncorrected ERA5L. Note that outliers are not shown in these boxplots for readability.}  
\label{fig:BP_mult}
\end{figure}

\begin{figure}[H]
\centering
\includegraphics[width=\textwidth]{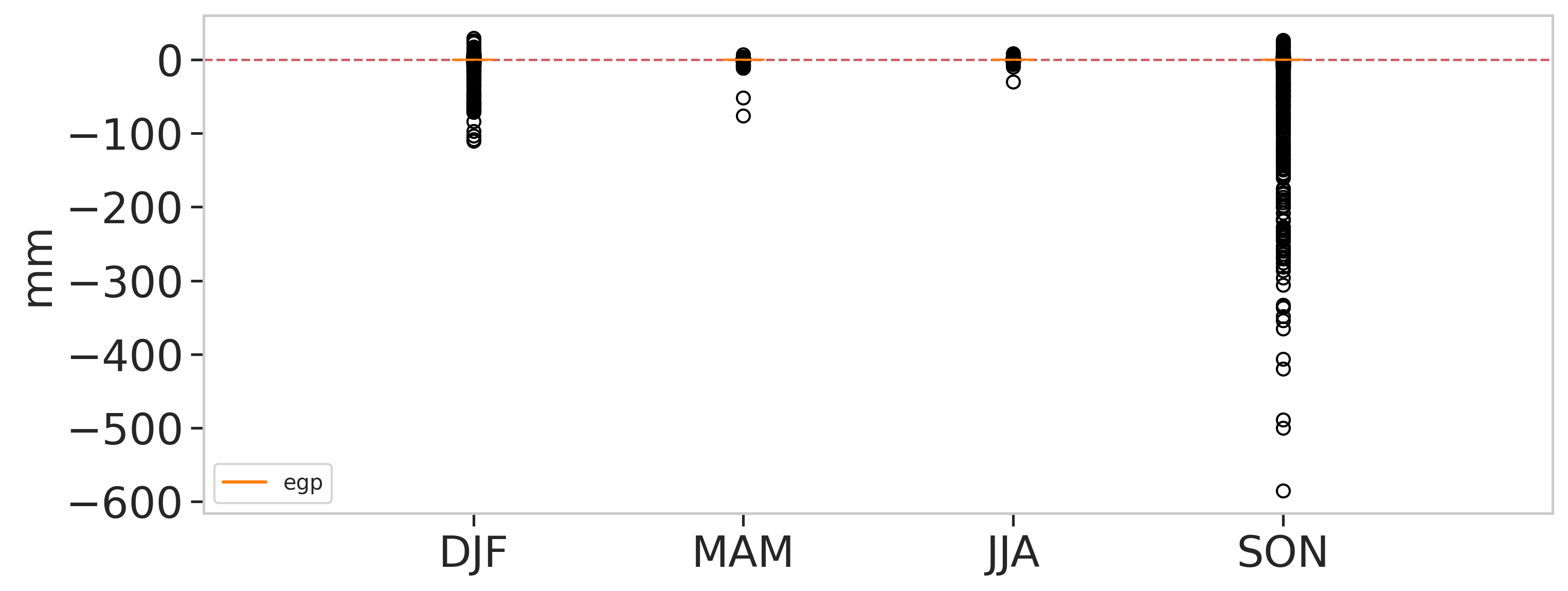}
\caption{Multi-season RMSE difference between Stitch-BJ and EGP.} 
\label{fig:BP_RMSE_diff_EGP}
\end{figure}

\begin{sloppypar}For all  metrics boxplots in Figures \ref{fig:BP_mult} and \ref{fig:BP_RMSE_diff_EGP}, the $MAEdiff_{Stitch-BJ,mod}$, 
$MAE95supdiff_{Stitch-BJ,mod}$ and $RMSEdiff_{Stitch-BJ,mod}$ (see Section \ref{sec:3}) are shown with the EGP, ExpW and empirical distributions as competing models. A positive value means the competing model performed better while a negative one means the Stitch-BJ performed better.\end{sloppypar}

{In Figure \ref{fig:BP_mult}, we have seasonal boxplots of respectively the MAE, MAE95sup and RMSE differences in panels a, b and c.} {For the MAE differences in panel (a), the median differences are always negative and the first quartile is consistently lower than the third quartile for all seasons for the empirical and ExpW models. This means that the Stitch-BJ resulted in a lower median MAE for all seasons compared to all the other competing distributions.} However, improvements on MAE are marginal with the highest median improvement being for the SON season over the empirical distribution with an improvement of approximately 0.05\textit{mm}. As for the uncorrected model ERA5L, improvements are more noticeable with a median MAE differences being much lower than the others.
We have similar observations in panel (b), with an improvement over the MAE95sup for all seasons over the empirical distribution. The ExpW model produces a median MAE95sup almost identical to the Stitch-BJ model for all seasons. This may be due to the ExpW being a flexible model as well, being able to model both light and heavy tails distributions. 
Results are constant throughout the seasons, with JJA and SON producing the most variations when comparing with the empirical distribution. The empirical and uncorrected model were both outperformed in terms of median MAE95sup by the proposed model.

Similarly to Figure \ref{fig:BP_mult} (panels a and b), an improvement can be seen over the empirical method for the median differences in panel c. A small improvement on the MAM season is noticeable against the ExpW, while performances are similar for all other seasons. The first and third quartiles are once again asymmetric, favouring the Stitch-BJ method slightly over the two others.

In Figure \ref{fig:BP_RMSE_diff_EGP}, we show the RMSE difference between the Stitch-BJ and EGP models, including outliers in the boxplot. {We separated the EGP from the other distributions from Figure \ref{fig:BP_RMSE_DJF_JJA}} to show the extent of the errors of the EGP without impacting the $y$-axis scale. For the MAM and JJA season, differences are quite symmetrical. This is expected due to the low replacement rate of the EGP for these seasons (see Figure \ref{fig:replacement_ERA5L_CERRAL}). However, for DJF and SON seasons, improvements provided by the  Stitch-BJ model over the EGP model are much more noticeable. Many locations produced RMSE 100 to 600\textit{mm} higher than the Stitch-BJ respective location. This shows that even though most of the locations used the EGP to model the upper tail, the model cannot be used on all the study area at the risk of producing extremely high errors. This emphasize the importance of the considered stitch model.
\subsection{Local analysis on a selected location}
\label{sec:3_5}
\begin{figure}[H]
\centering
\begin{subfigure}[b]{0.7\textwidth}   
\includegraphics[width=\linewidth]{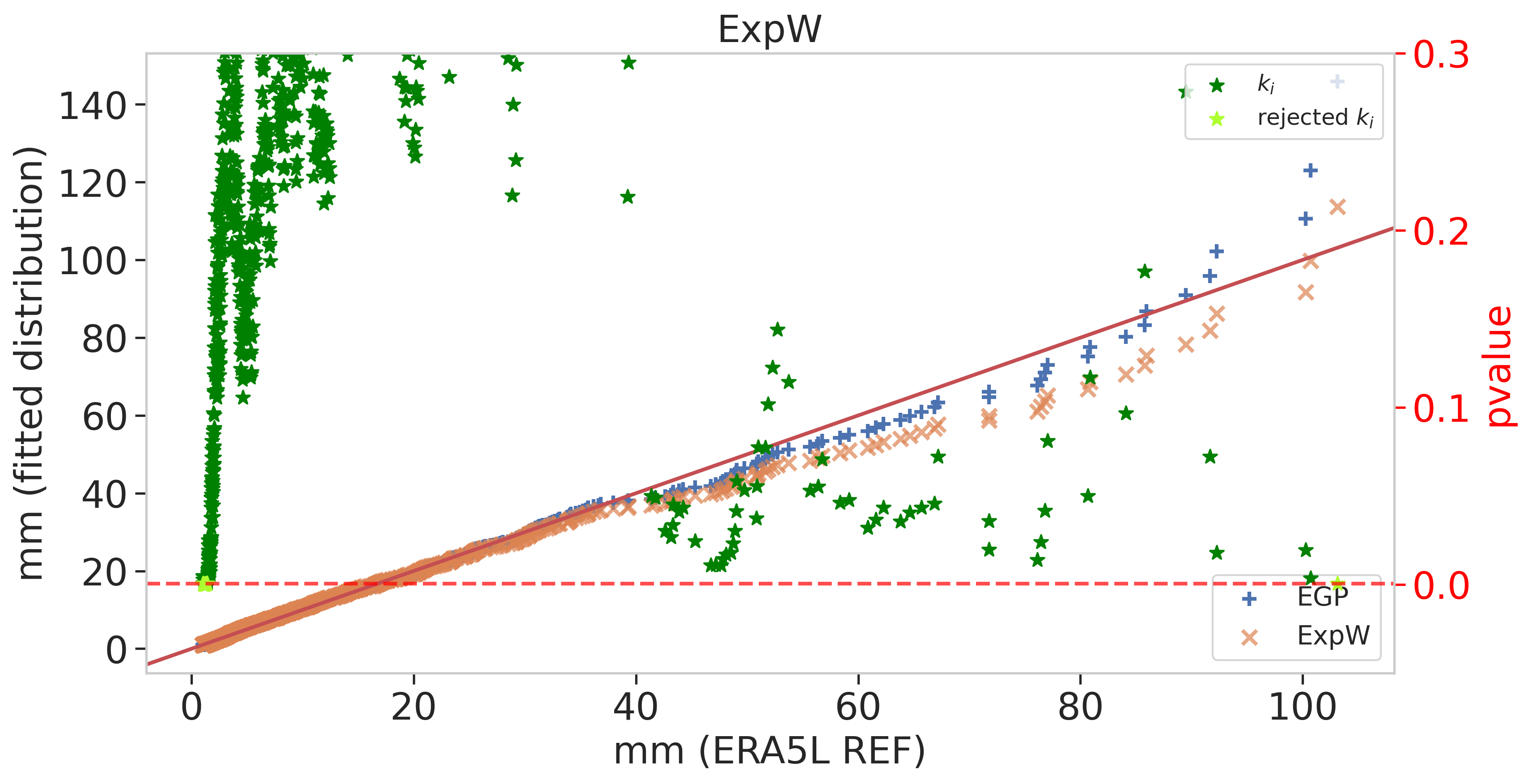}
\end{subfigure}
\begin{subfigure}[b]{0.7\textwidth}   
\includegraphics[width=\linewidth]{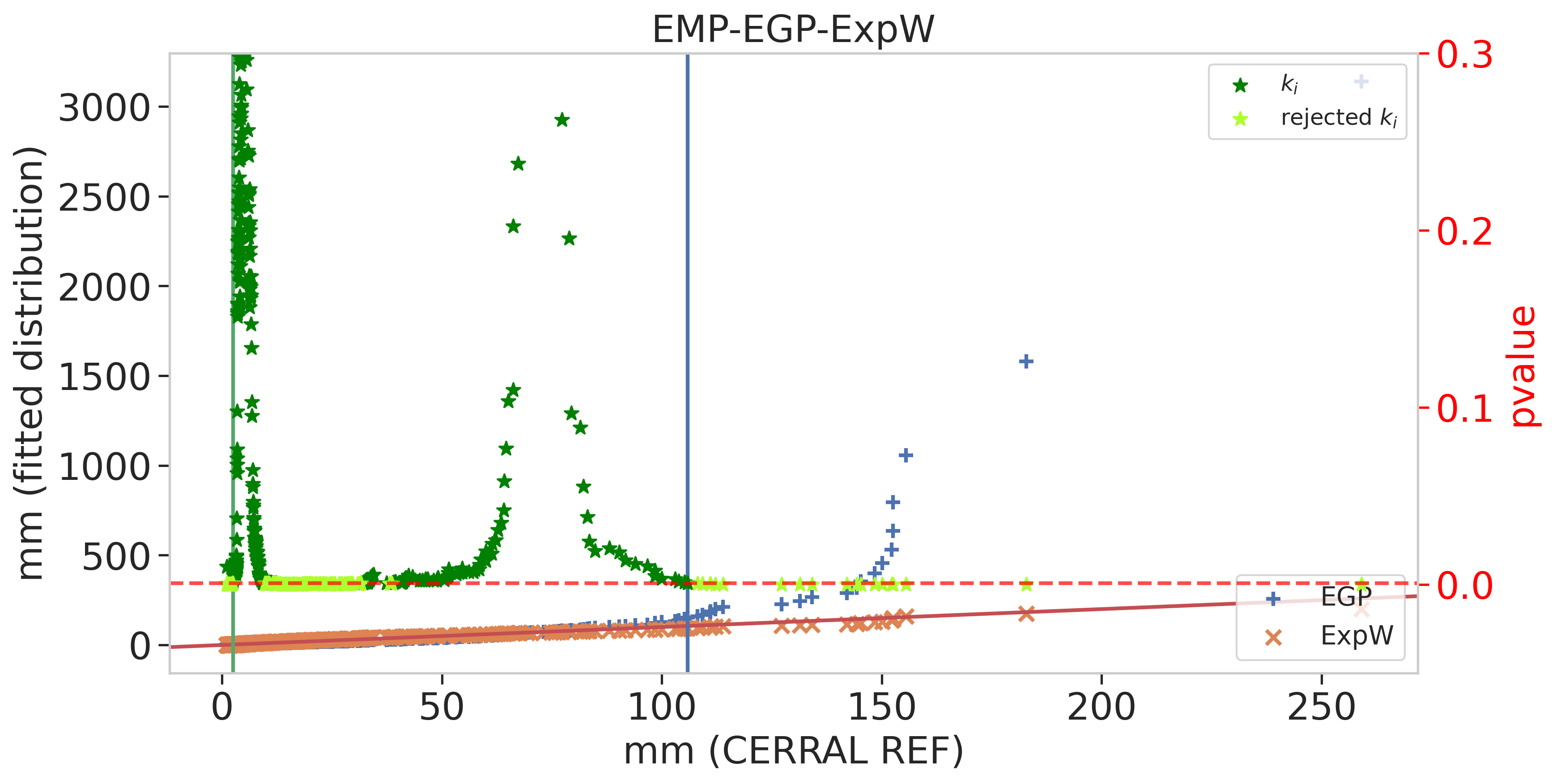}
\end{subfigure}
\begin{subfigure}[b]{0.7\textwidth}   
\includegraphics[width=\linewidth]{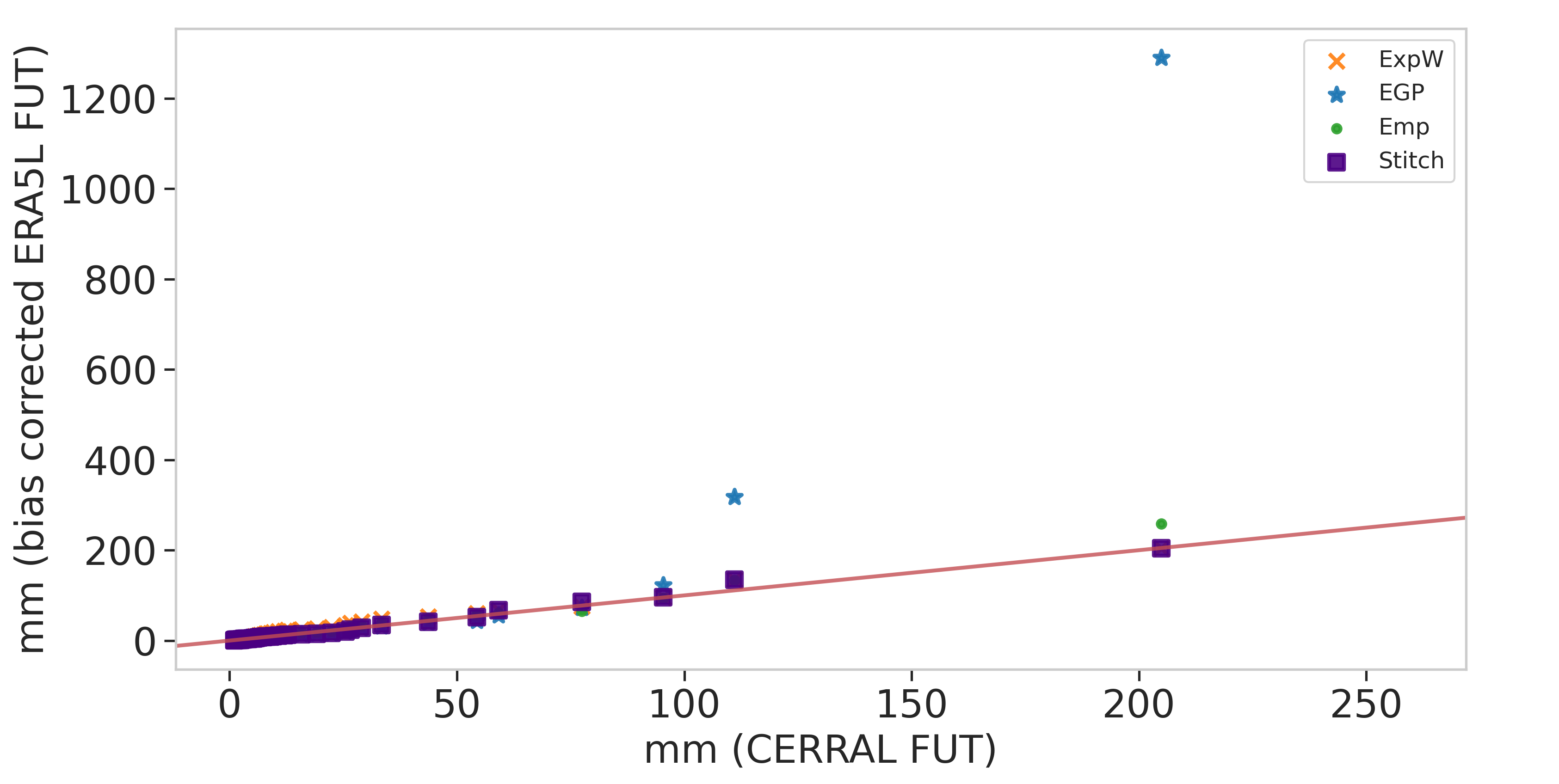}
\end{subfigure}

\caption{Local analysis for location (Lon 9.3 , Lat 41.96). Panel (a): Fitting performance of the EGP and ExpW on ERA5L. Panel (b): Fitting performance of the EGP and ExpW on CERRAL. Panel (c): Bias correction performance for Stitch-BJ, EGP, ExpW and empirical model.}  
\label{fig:egp_weird}
\end{figure}

While Figure \ref{fig:mecha_stitch} gave an idea of how the stitching procedure worked, let's take a corrected location for a finer analysis. 
In Figure \ref{fig:egp_weird}, we have quantile-quantile plots of the fitting procedure on both ERA5L and CERRAL data in panels (a, b), and the resulting bias-corrected data in panel (c).
In the PBJ test, p-values $k_i$ are produced and are used in the Stitch-BJ model to detect misfitted quantiles. When they cross a given threshold (red dashed line), we consider the corresponding quantile misfitted and we reject it. 
In the first two panels, dark green and light green stars correspond to the $k_i$ and rejected $k_i$ respectively. The green and blue vertical lines correspond to the lower and upper rejection index $i_l$ and $i_u$ introduced in Section \ref{sec:2_3}, and any quantiles lower (resp. higher) than the index are replaced with ones from another distribution.
 
In Figure \ref{fig:egp_weird} (panel a),  we have a local analysis where a full ExpW model has been used on ERA5L data. While the ExpW performed globally worse on high quantiles, the extreme errors of the EGP distribution (50\textit{mm} on the last quantile) resulted in the selection of the ExpW distribution as the final distributional model. Note that while rejected $k_i$ can be seen in the figure, since we completely replaced the distribution with an ExpW, no lower and upper stitching indexes are shown with vertical lines, {meaning that the adopted model is a pure ExpW for this ERA5L's location}.
In Figure \ref{fig:egp_weird} (panel b),  a stitch distribution has been used for this location for CERRAL data. This specific stitch model is composed by an empirical distribution in the lower tail, an EGP distribution in the centre of the law and an ExpW distribution in the upper tail. The notation associated is EMP-EGP-ExpW (see Figure \ref{fig:replacement_ERA5L_CERRAL}).

This resulted in the bias-correction qqplot in panel (c), where the Stitch-BJ is the best fitting model over all contenders. Moreover, we can see the impact of the stitch: if left as is, the EGP produced a large overestimation of almost 1000\textit{mm} for the highest quantile.

To put into perspective, for this location, the MAE is at respectively 1.1, 3.3, 4.1 and 25.6\textit{mm} for the Stitch-BJ, empirical, ExpW and EGP models. When considering the MAE95sup which is more focused on the extremes, the errors are 6.3, 29.5, 17 and 440 \textit{mm} respectively.
For differences metrics as shown in Figures \ref{fig:MAE_diff_DJF_JJA}, \ref{fig:MAE95sup_diff_DJF_JJA} and \ref{fig:RMSE_diff_DJF_JJA}, the MAE differences are then respectively  -2.2, -3 and -26.5\textit{mm} against the empirical, ExpW and EGP models. MAE95sup differences exhibit similar results with -23.2, -10.6 and -433.5 \textit{mm} respectively.

The analysis of Figure \ref{fig:egp_weird} shows how the stitching procedure can prevent extremely large errors from a misfitted EGP, and even improve upon the bias-corrected data using and ExpW or the empirical distribution.

\section{Conclusion and discussion}
\label{sec:4}

In this study, our objective was to seasonally bias correct daily precipitation data from ERA5-Land with CERRA-Land data using out-of-sample validation. We applied the Stitch-BJ \citep{ear_semi-parametric_2025} model along with other distributions (see Section \ref{sec:2_3}) combined with the Singularity Stochastic Removal method from \cite{vrac_bias_2016} for dry days probability correction. This allowed to show the potential of parametric and semi-parametric bias correction methods against the classical empirical model.

We first showed the necessity of introducing a seasonal separation of our datasets in Section \ref{sec:2_1} and correcting the probability of dry days in Section \ref{sec:2_2}. If left as is, residual bias would be left in the final result, which may worsen the performance of all models. To correct the dry-day probabilities of our models, we adapted the SSR methodto apply it to semi and fully-parametric models. This allowed us to obtain well-calibrated dry-day probabilities compared to uncorrected models, with medians of differences with the target dry-day probabilities for the validation period of less than 0.01 across all seasons.

In Section \ref{sec:3}, we show the bias correction performance of all models for the DJF and JJA season across multiple metrics. We showed that the Stitch-BJ model produced performance similar to the empirical model in terms of MAE, with a slight reduction in the median error across the study area.
For more extreme focused metrics such as the MAE95sup or the RMSE, the Stitch-BJ model was able to show a reduction in extreme errors against all models for the DJF season, while performance were mostly equivalent for the JJA season. Again, median errors were lower for the Stitch-BJ compared to the empirical model, and error outliers were globally lower, or similar to the empirical model's.

Comparing all seasons together, we do not see an impact on the relative performance of models against the Stitch-BJ, with the latter outperforming the ExpW and empirical models for all seasons.

While this study shows promising results for the Stitch-BJ and demonstrates its flexibility compared to other models,   performances could be improved by considering other distributions.  All metrics from this study were also solely based on quantiles of wet days intensity distribution, and on the proportion of dry days. Interesting results may arise from studying the impact of such models on climate indicators, which often consider the temporality of such variables. 

Moreover, the present analysis shows a high sensitivity to extreme events. Indeed, depending on the user, one could want to produce the worst-case scenario and would rather produce extremes too high than potentially miss them, and in this case, the presented models can be used as such. However, some users might want to be conservative and avoid sending alarming signals and results as much as possible. In this case, multiple methods exist to reduce the sensitivity of the fitting procedure to the few highest data:

\begin{itemize}
    \item The few highest data points may be removed, either completely from the dataset (right-truncation) or only the information on their intensity (right-censoring), since outliers are known to cause misfitting in some cases as in \cite{berg_robust_2024}. Some variation can be found where extreme outliers are not corrected and left as is in \citep{gutjahr_comparing_2013};
    \item Adding a more robust model for the upper tail modelisation such as the EQM-LIN from \citep{holthuijzen_robust_2022} which uses a linear correction for extremes above a selected threshold. This allows the smoothening of the upper tail and ignores very high outliers produced by the very few last data quantiles.
\end{itemize}

Finally, this study's validation and training periods were all historical periods that could be compared with station data. However, climate projections such as data from CMIP6 also need to be bias-corrected. Such data are influenced by green gas emission scenarios which may greatly impact the probability of more extreme events. Such settings may be better suited for parametric distributions for their ability to extrapolate to new extremes and would be an interesting future study to pursue, as in \cite{andrade-velazquez_statistical_2023} or \cite{enyew_performance_2024}.

\appendix
\section[\appendixname~\thesection]{RMSE results for the bias correction of ERA5L on the 2010-2020 period using CERRAL data}
\label{app:rmse}
 
Maps from Figure \ref{fig:RMSE_DJF_JJA} are very similar to the one for the MAE95sup in Figure \ref{fig:MAE95sup_DJF_JJA}, with mostly similar results for the DJF season except for the Cévènnes, Corsica and Alps region where the Stitch-BJ may perform better than the other models.
For the JJA season, performance are very similar among all models. 


\begin{figure}[!h]
    \centering
     \begin{subfigure}[b]{0.32\textwidth}   
        \centering
        \includegraphics[width=\textwidth]{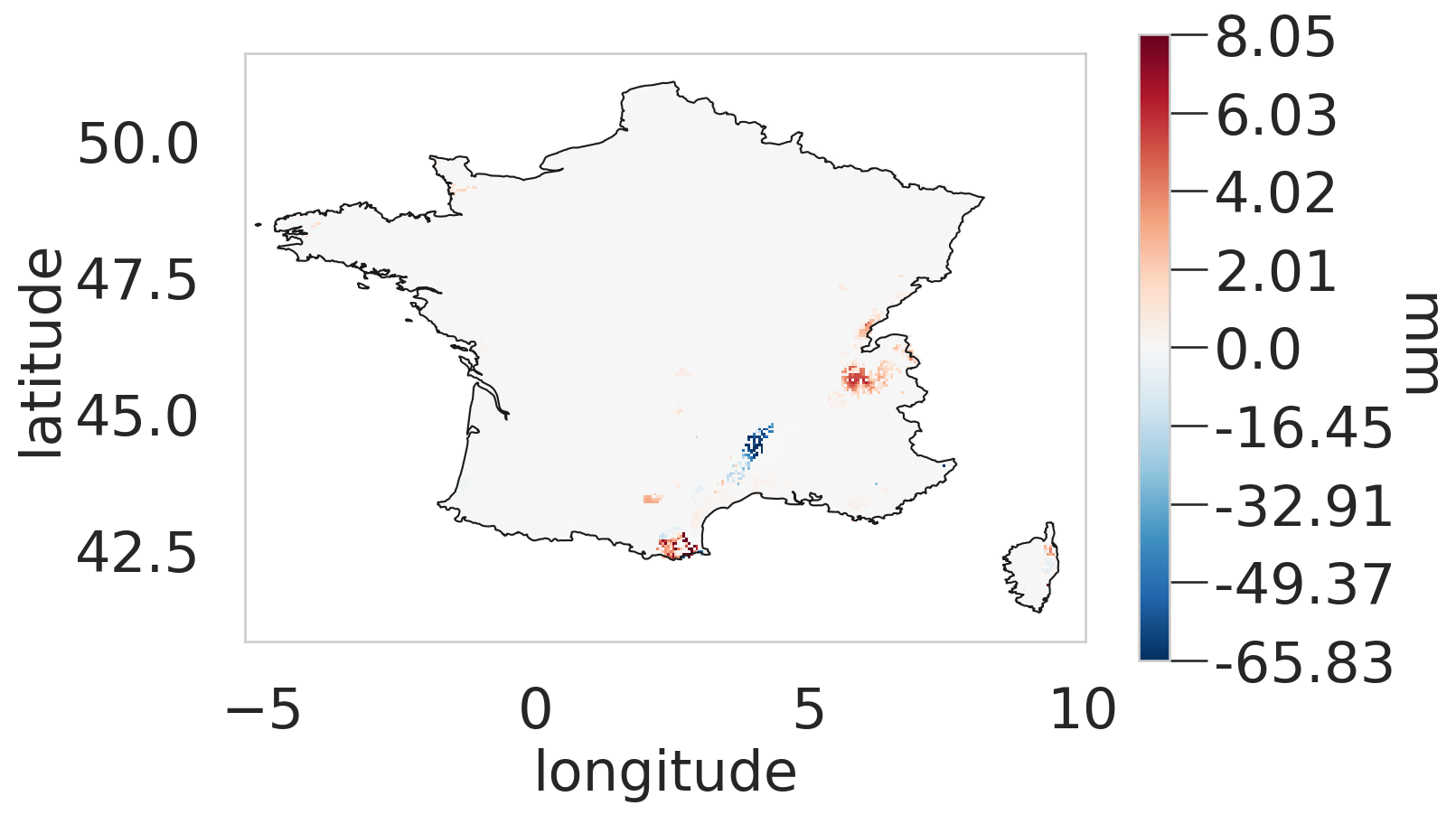}
        \caption{}
    \end{subfigure}
    \hfill
     \begin{subfigure}[b]{0.32\textwidth}   
        \centering
        \includegraphics[width=\textwidth]{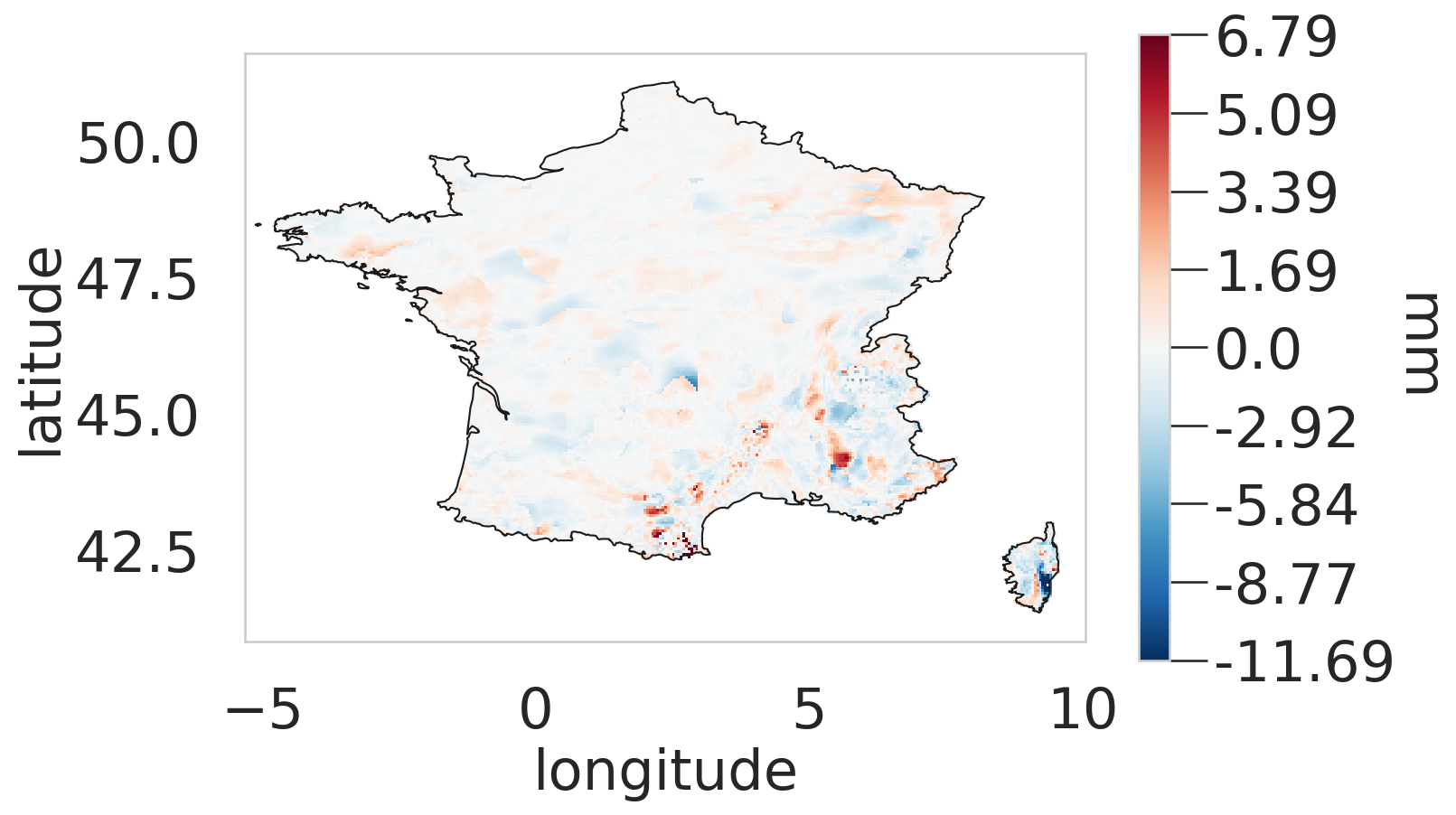}
        \caption{}
    \end{subfigure}
    \hfill
    \begin{subfigure}[b]{0.32\textwidth}   
        \centering
        \includegraphics[width=\textwidth]{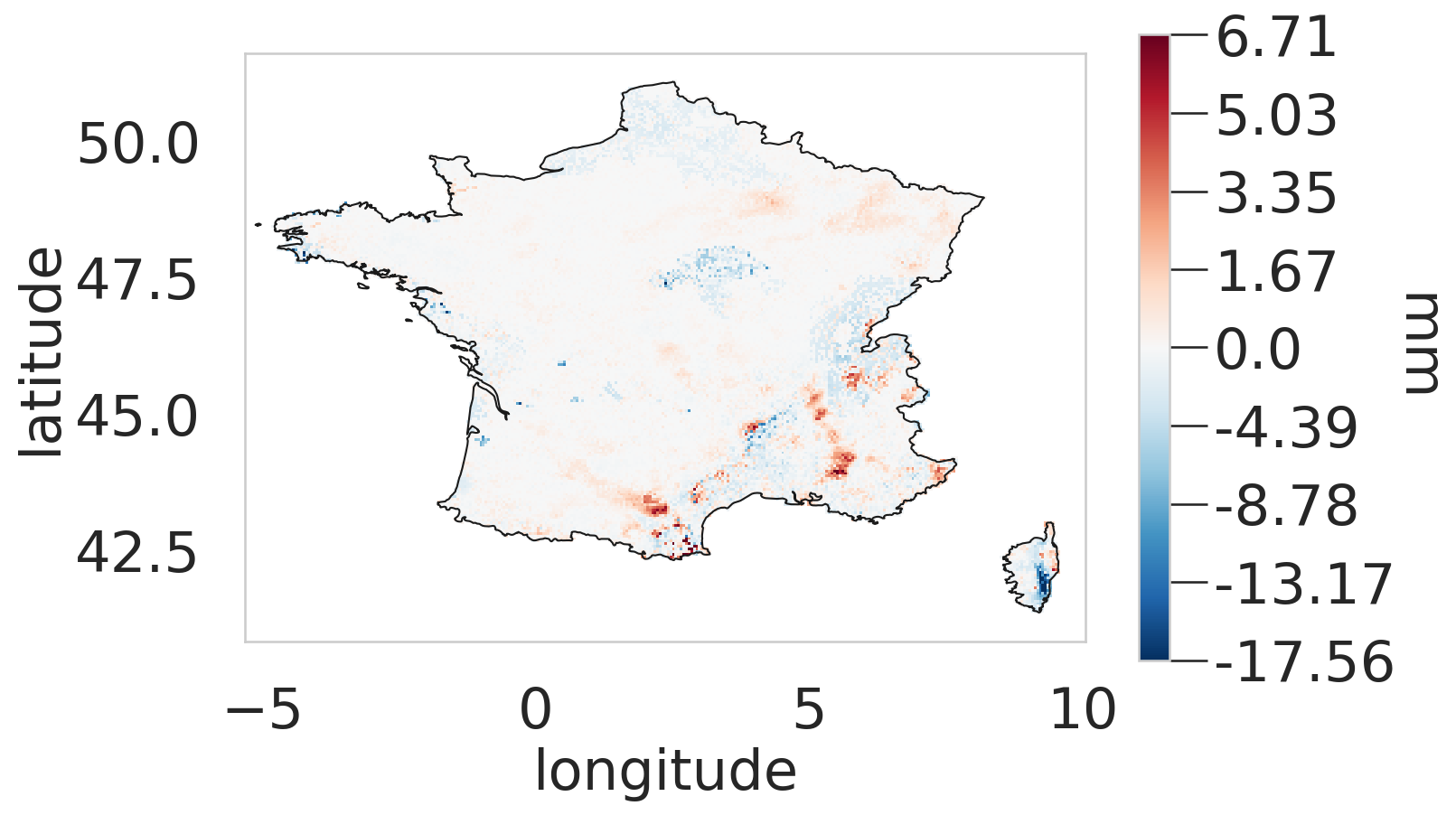}
        \caption{}
    \end{subfigure}
    
    \vskip\baselineskip
     \begin{subfigure}[b]{0.32\textwidth}   
        \centering
        \includegraphics[width=\textwidth]{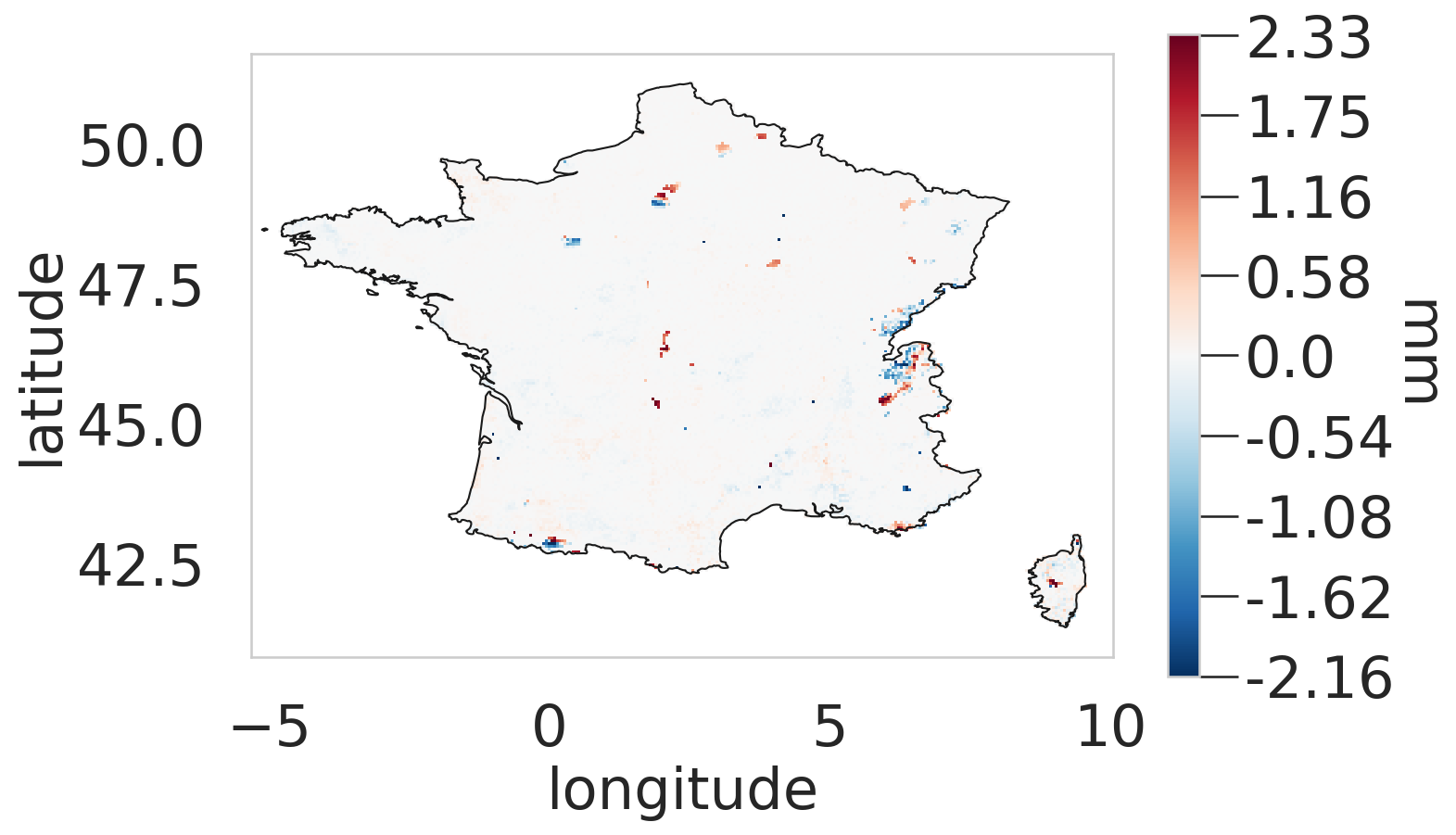}
        \caption{}
    \end{subfigure}
    \hfill
     \begin{subfigure}[b]{0.32\textwidth}   
        \centering
        \includegraphics[width=\textwidth]{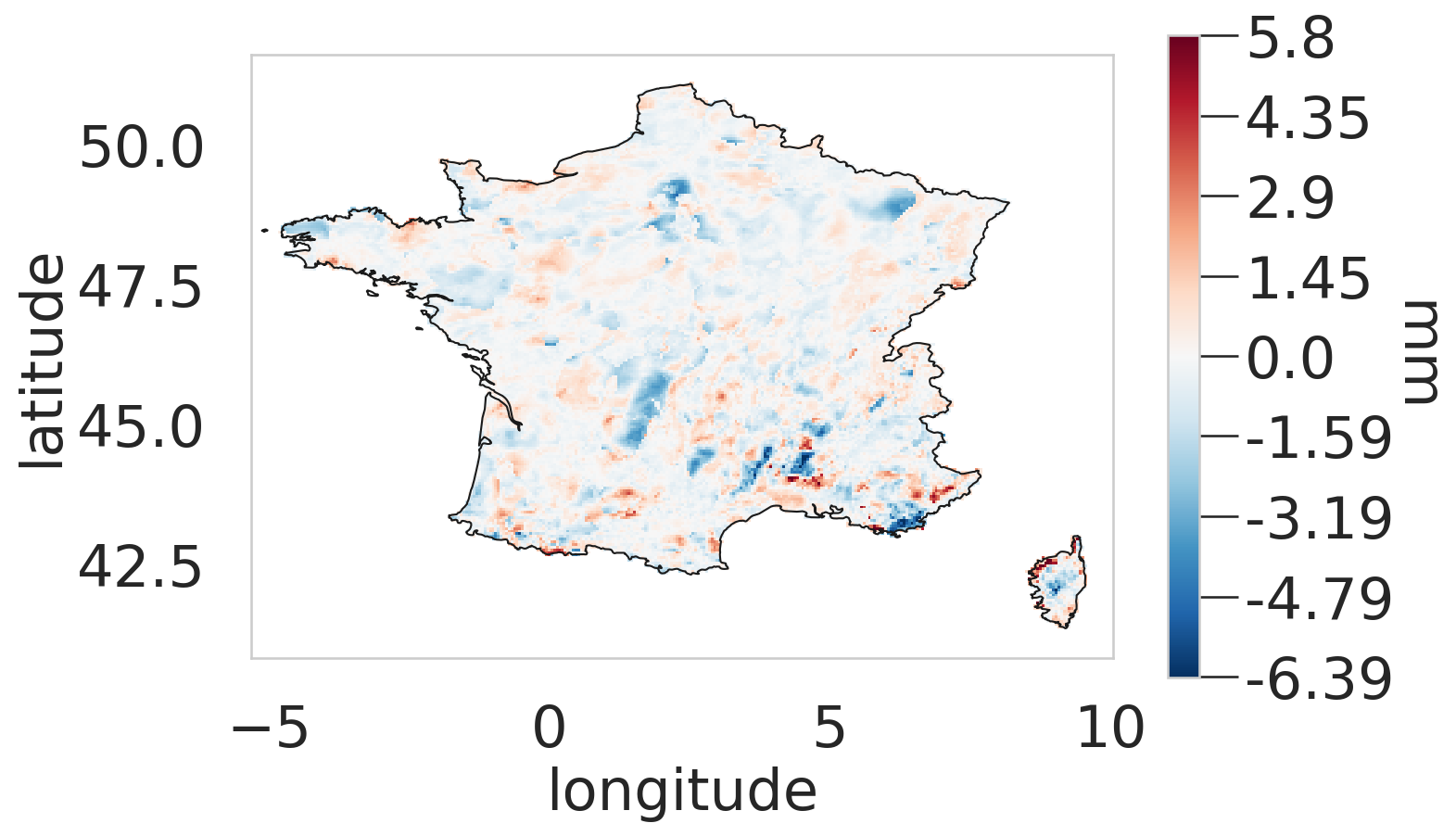}
        \caption{}
    \end{subfigure}
    \hfill
    \begin{subfigure}[b]{0.32\textwidth}   
        \centering
        \includegraphics[width=\textwidth]{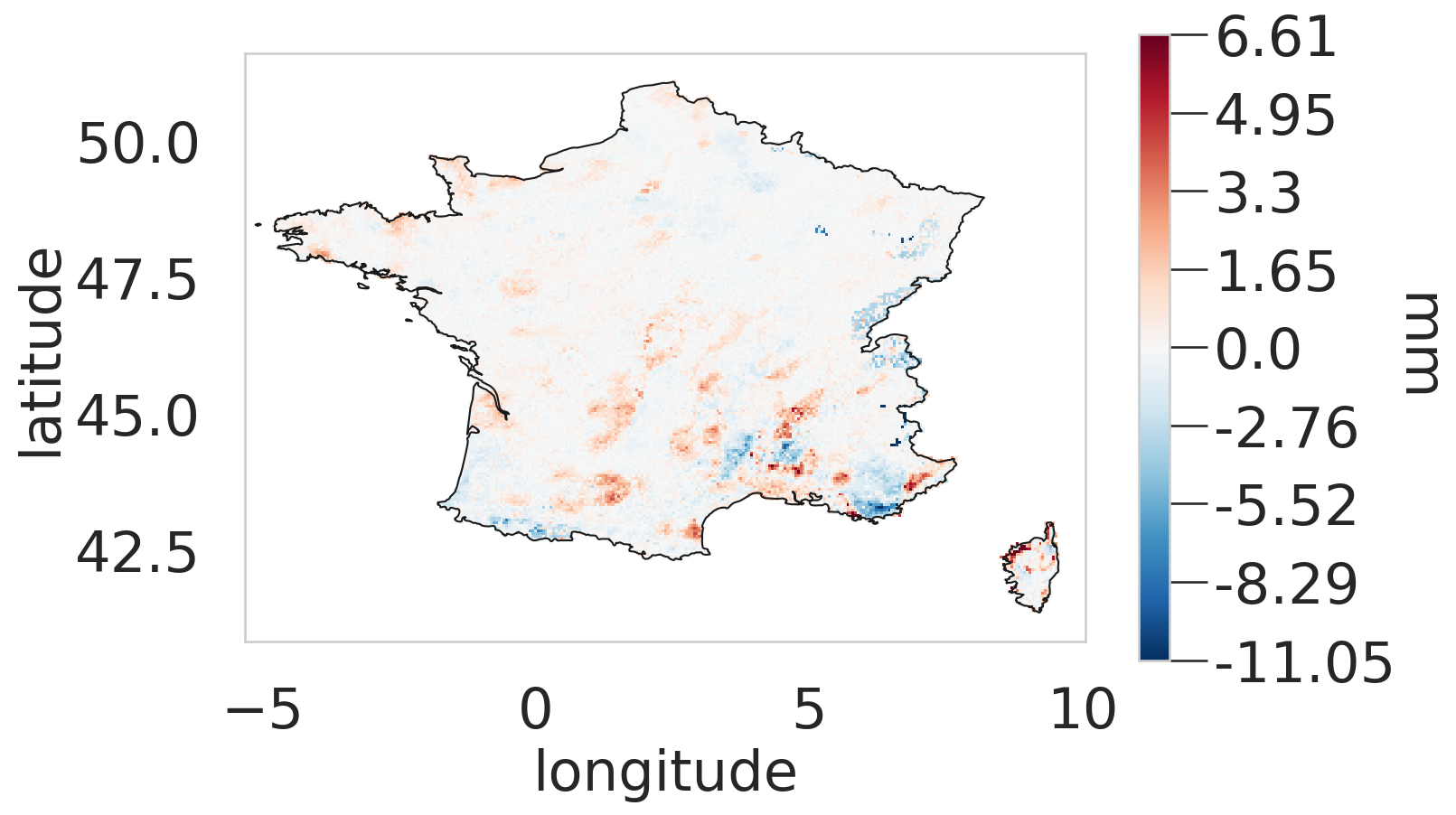}
        \caption{}
    \end{subfigure}
    
    \caption{Maps of $RMSEdiff_{Stitch-BJ,G}$ for DJF (first row) and JJA (second row) with $G$ being the following models: EGP (a,d), \textit{emp} (b,e) and ExpW (c,f). Note that value ranges are specific to each map.} 
    \label{fig:RMSE_diff_DJF_JJA}
    \end{figure}
\begin{figure}[H]
\centering
    \begin{subfigure}[b]{0.45\textwidth}   
        \centering
        \includegraphics[width=0.45\textwidth]
        {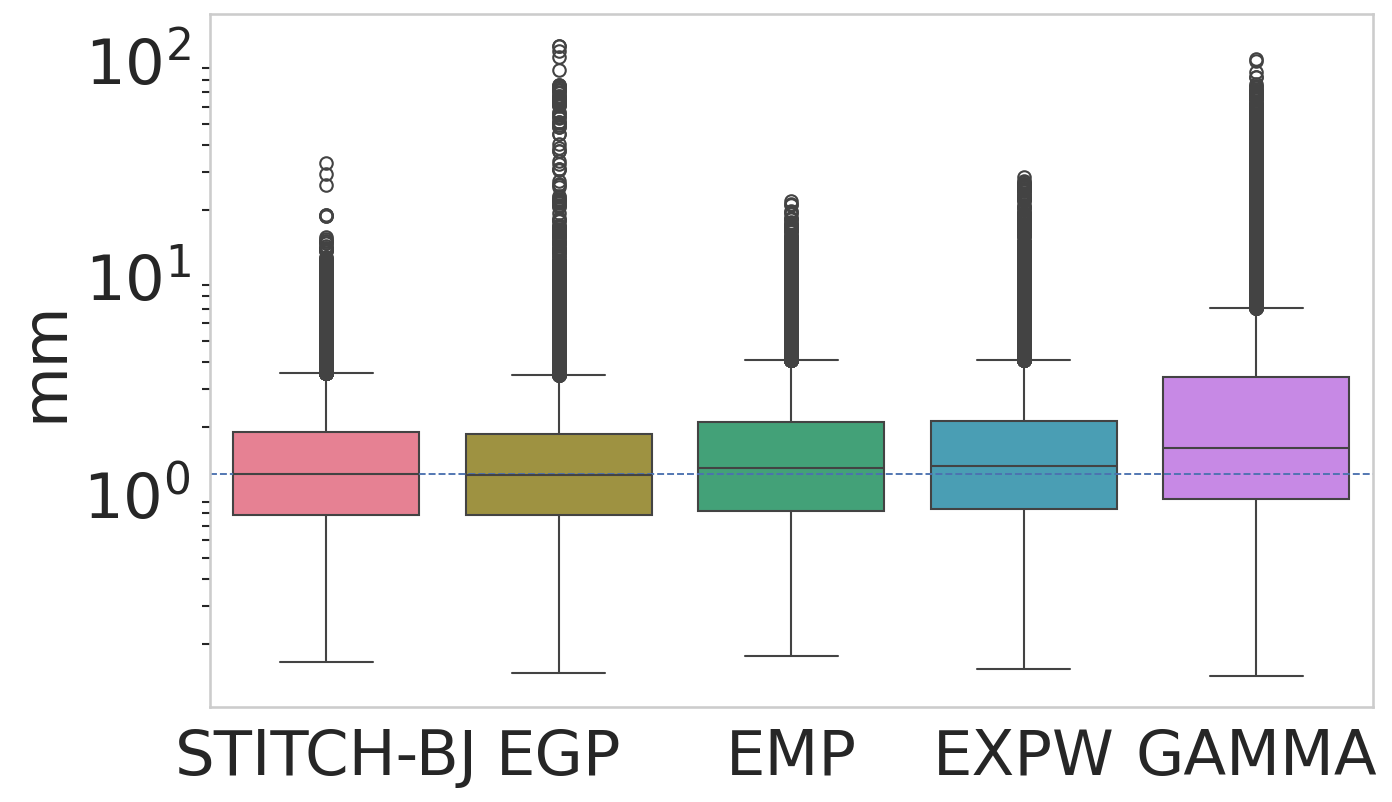}
        \caption{}
    \end{subfigure}
    \hfill
        \begin{subfigure}[b]{0.45\textwidth}   
        \centering
        \includegraphics[width=0.45\textwidth]{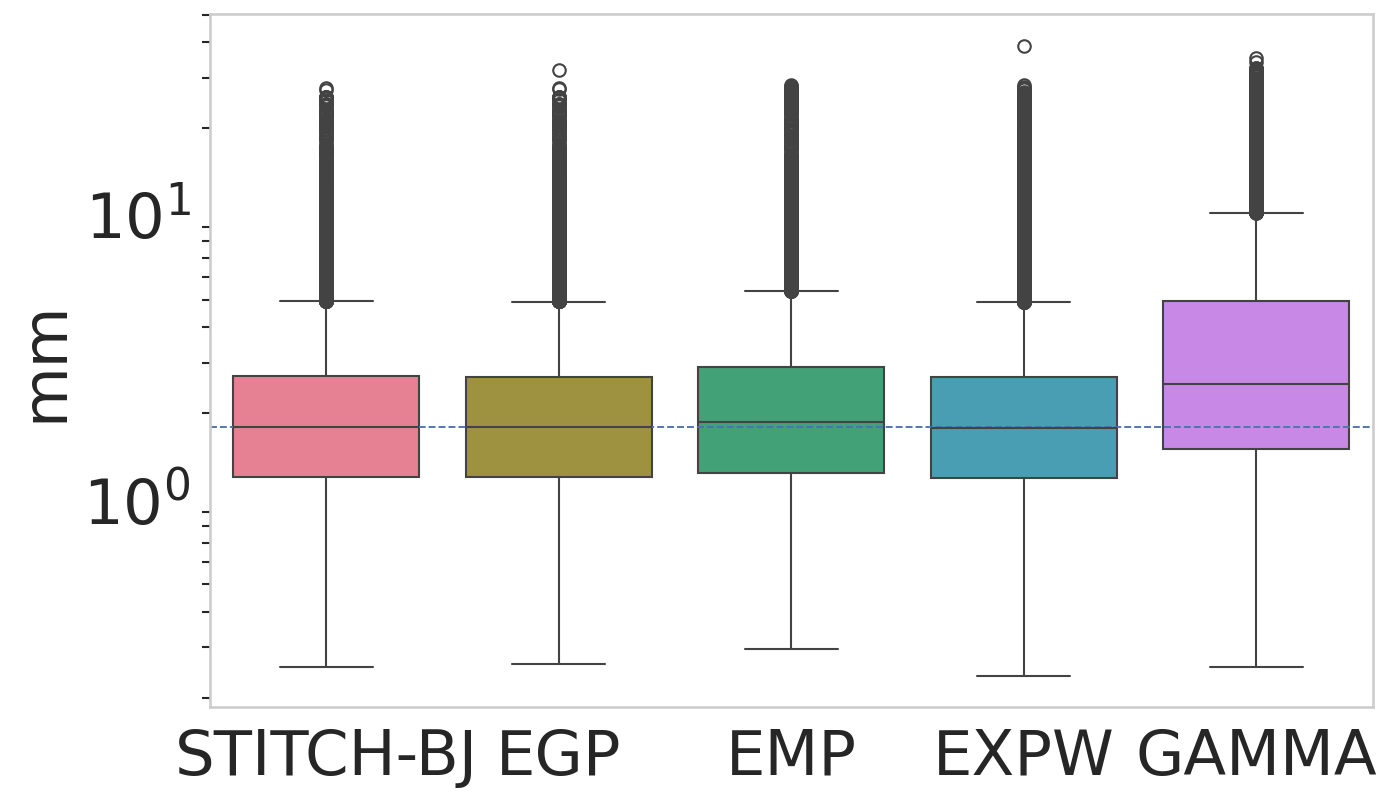}
        \caption{}
    \end{subfigure}

\caption{Boxplots of RMSE for DJF (a) and JJA (b) for all models. Note that the $y$-axis is in log10-scale.} 
\label{fig:BP_RMSE_DJF_JJA}
\end{figure}

{In Figure \ref{fig:RMSE_diff_DJF_JJA},} spatial patterns and intensity of errors match the ones from Figure \ref{fig:MAE95sup_diff_DJF_JJA} for both seasons. The Stitch-BJ method offers an advantage over the DJF season for all methods, while the JJA season shows very close performance for all models.
\begin{figure}[!h]
    \centering
     \begin{subfigure}[b]{0.32\textwidth}   
        \centering
        \includegraphics[width=\textwidth]{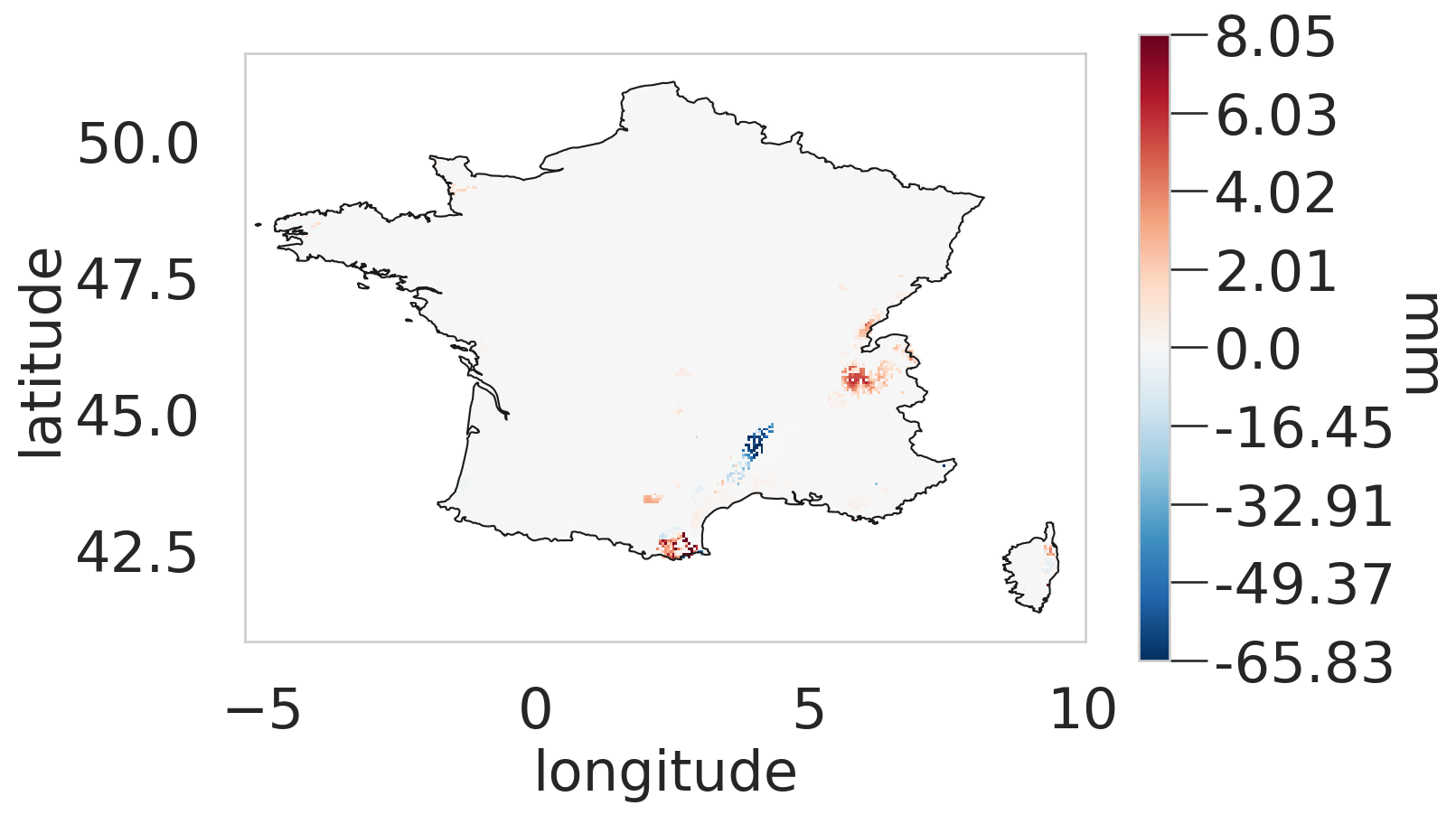}
        \caption{}
    \end{subfigure}
    \hfill
     \begin{subfigure}[b]{0.32\textwidth}   
        \centering
        \includegraphics[width=\textwidth]{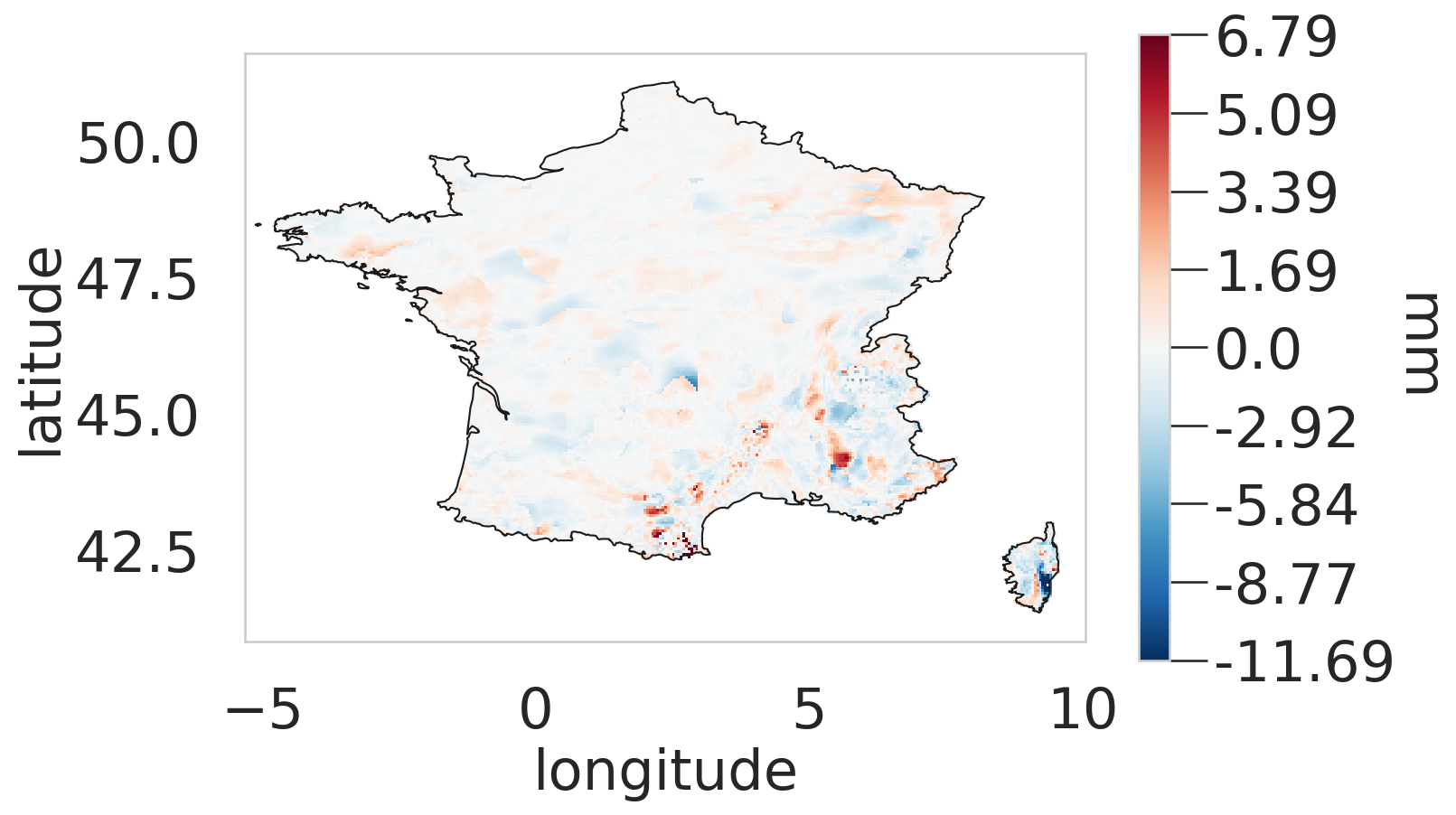}
        \caption{}
    \end{subfigure}
    \hfill
    \begin{subfigure}[b]{0.32\textwidth}   
        \centering
        \includegraphics[width=\textwidth]{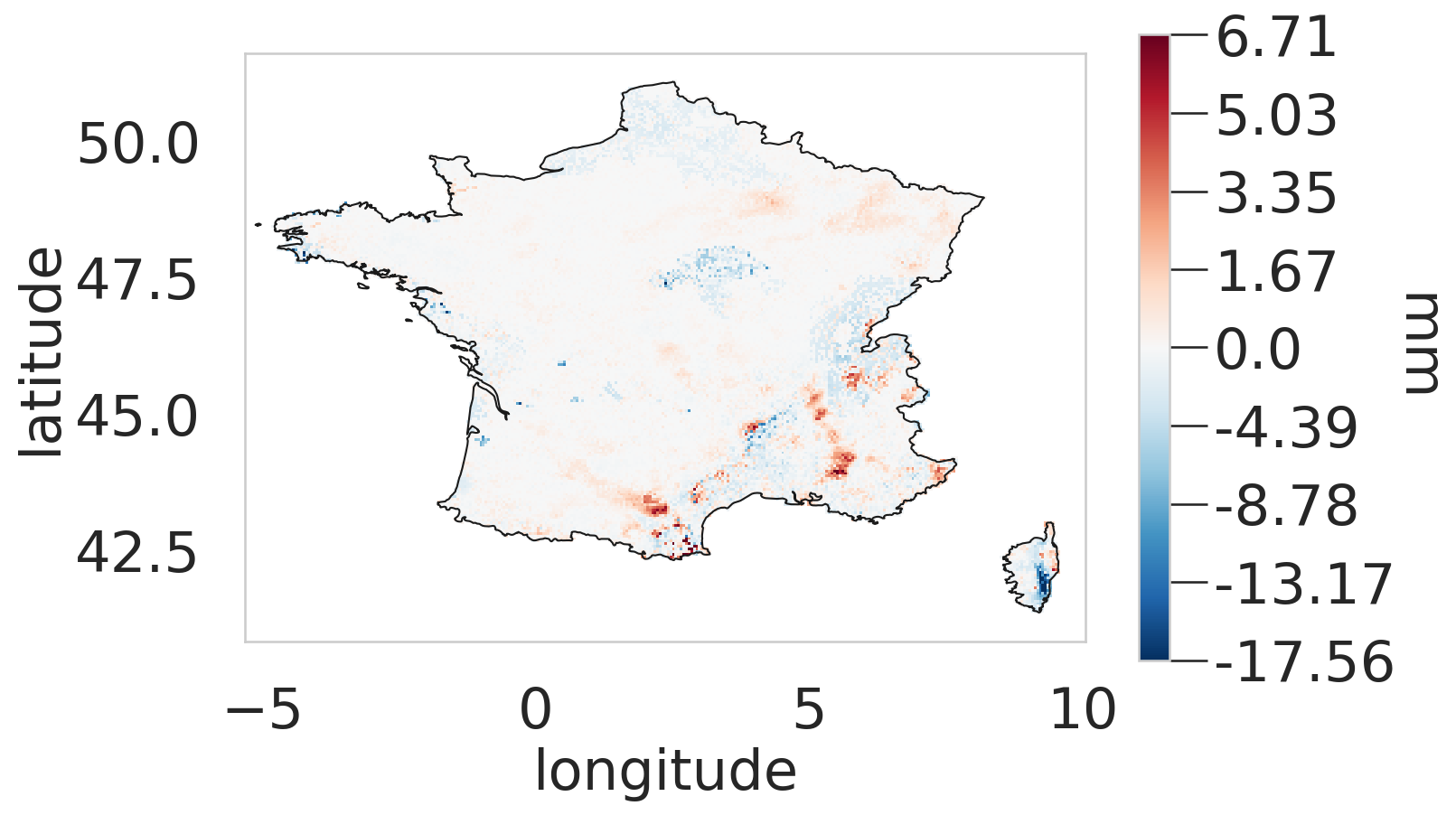}
        \caption{}
    \end{subfigure}
    
    \vskip\baselineskip
     \begin{subfigure}[b]{0.32\textwidth}   
        \centering
        \includegraphics[width=\textwidth]{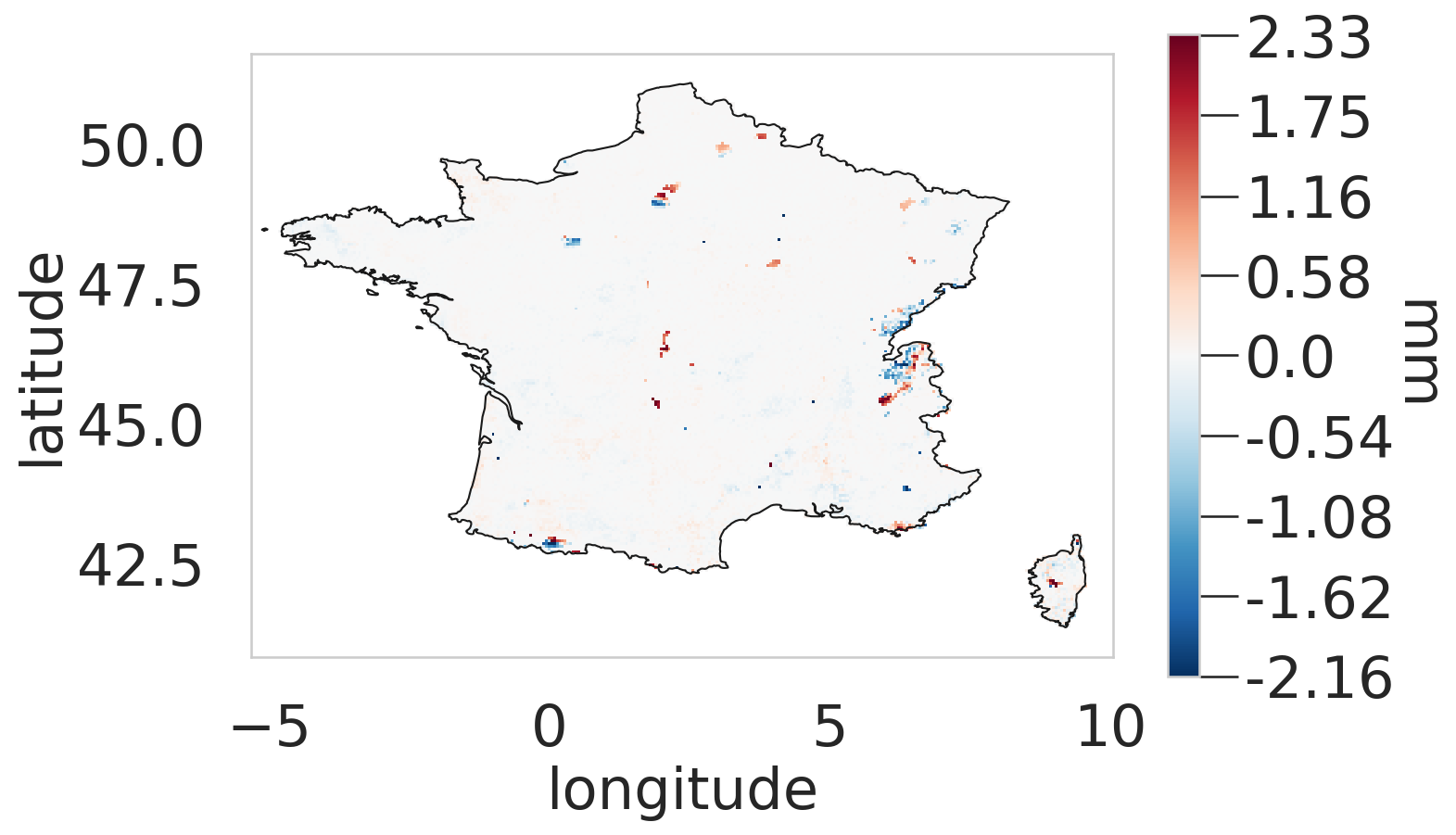}
        \caption{}
    \end{subfigure}
    \hfill
     \begin{subfigure}[b]{0.32\textwidth}   
        \centering
        \includegraphics[width=\textwidth]{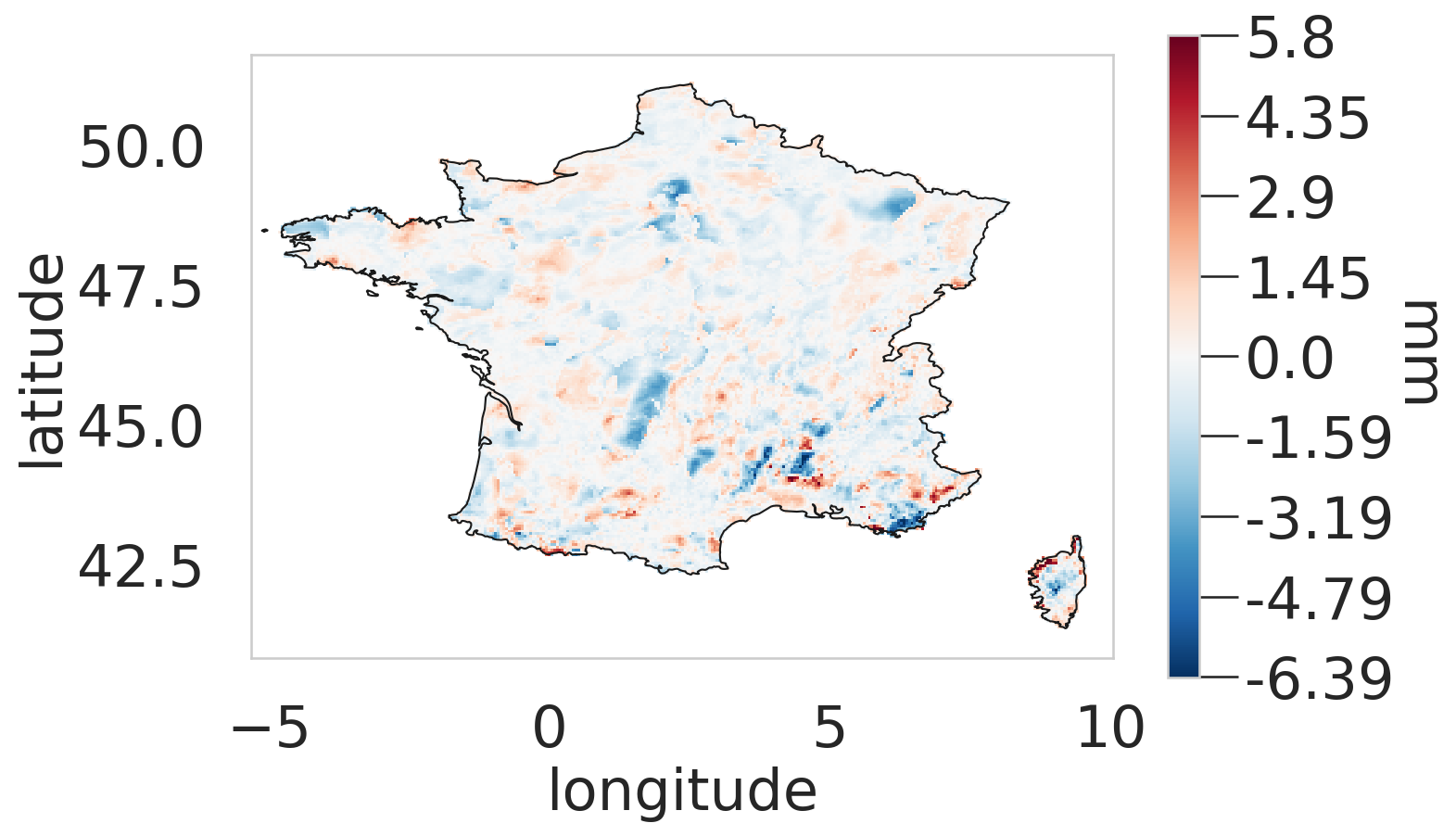}
        \caption{}
    \end{subfigure}
    \hfill
    \begin{subfigure}[b]{0.32\textwidth}   
        \centering
        \includegraphics[width=\textwidth]{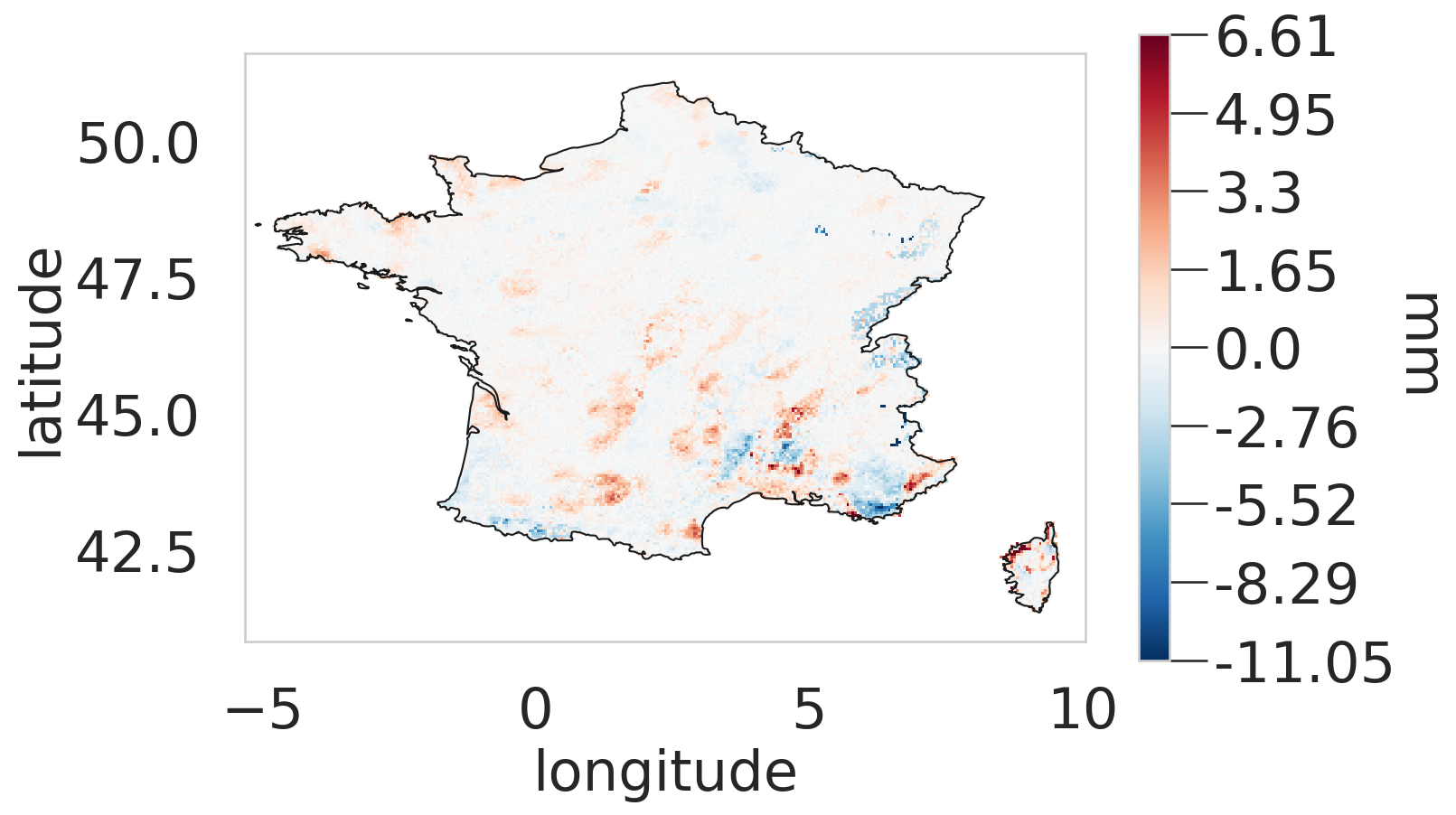}
        \caption{}
    \end{subfigure}
    
    \caption{Maps of $RMSEdiff_{Stitch-BJ,G}$ for DJF (first row) and JJA (second row) with $G$ being the following models: EGP (a,d), \textit{emp} (b,e) and ExpW (c,f). Note that value ranges are specific to each map.} 
    \label{fig:RMSE_diff_DJF_JJA}
    \end{figure}
With similar observations to previous figures, conclusion from RMSE boxplots of Figure \ref{fig:BP_RMSE_DJF_JJA} are almost identical to Figure \ref{fig:BP_MAE95sup_DJF_JJA} with a slight improvement over median error for the Stitch-BJ and EGP models, and an improvement on extremes for the Stitch-BJ compared to the EGP and ExpW models for both seasons.

\textbf{Author contributions:} {Conceptualization, P.E., E.D., T.L., A.L. and M.T.; methodology, P.E., E.D. and T.L.; software, P.E.; validation, P.E., E.D. and T.L.; formal analysis, P.E.; investigation, P.E.; resources,
P.E.; data curation, P.E.; writing--original draft preparation, P.E.; writing--review and editing,
P.E., E.D. and T.L.; visualization, P.E.; supervision, E.D., T.L. and A.L.; project
administration, E.D., T.L., A.L. and M.T.; funding acquisition, E.D., T.L., A.L. and M.T. All authors have read and agreed to
the published version of the manuscript.
}   

\textbf{Funding: }This work has been partially supported by the French government through the CIFRE funding (CIFRE n$\circ$ 2022/0519) managed by the National Association of Research and Technology and the 3IA Côte d'Azur Investments in the Future project managed by the National Research Agency (ANR-19-P3IA-0002).

\textbf{Data availability:} All data, material, and programming codes used in this study are available upon request. ERA5-Land and CERRA-Land datasets analyzed in the current study are available on the 
\hypertarget{link:data} \href{}{}
\href{https://cds.climate.copernicus.eu/}{Copernicus Climate Change Service (C3S) Climate Data Store}.

\textbf{Conflict of interests:} The authors declare no conflicts of interest. The funders had no role in the design of the study; in the collection, analyses, or interpretation of data; in the writing of the manuscript; or in the decision to publish the results.

\bibliography{These}


\begin{thebibliography}{71}
\ifx \bisbn   \undefined \def \bisbn  #1{ISBN #1}\fi
\ifx \binits  \undefined \def \binits#1{#1}\fi
\ifx \bauthor  \undefined \def \bauthor#1{#1}\fi
\ifx \batitle  \undefined \def \batitle#1{#1}\fi
\ifx \bjtitle  \undefined \def \bjtitle#1{#1}\fi
\ifx \bvolume  \undefined \def \bvolume#1{\textbf{#1}}\fi
\ifx \byear  \undefined \def \byear#1{#1}\fi
\ifx \bissue  \undefined \def \bissue#1{#1}\fi
\ifx \bfpage  \undefined \def \bfpage#1{#1}\fi
\ifx \blpage  \undefined \def \blpage #1{#1}\fi
\ifx \burl  \undefined \def \burl#1{\textsf{#1}}\fi
\ifx \doiurl  \undefined \def \doiurl#1{\url{https://doi.org/#1}}\fi
\ifx \betal  \undefined \def \betal{\textit{et al.}}\fi
\ifx \binstitute  \undefined \def \binstitute#1{#1}\fi
\ifx \binstitutionaled  \undefined \def \binstitutionaled#1{#1}\fi
\ifx \bctitle  \undefined \def \bctitle#1{#1}\fi
\ifx \beditor  \undefined \def \beditor#1{#1}\fi
\ifx \bpublisher  \undefined \def \bpublisher#1{#1}\fi
\ifx \bbtitle  \undefined \def \bbtitle#1{#1}\fi
\ifx \bedition  \undefined \def \bedition#1{#1}\fi
\ifx \bseriesno  \undefined \def \bseriesno#1{#1}\fi
\ifx \blocation  \undefined \def \blocation#1{#1}\fi
\ifx \bsertitle  \undefined \def \bsertitle#1{#1}\fi
\ifx \bsnm \undefined \def \bsnm#1{#1}\fi
\ifx \bsuffix \undefined \def \bsuffix#1{#1}\fi
\ifx \bparticle \undefined \def \bparticle#1{#1}\fi
\ifx \barticle \undefined \def \barticle#1{#1}\fi
\bibcommenthead
\ifx \bconfdate \undefined \def \bconfdate #1{#1}\fi
\ifx \botherref \undefined \def \botherref #1{#1}\fi
\ifx \url \undefined \def \url#1{\textsf{#1}}\fi
\ifx \bchapter \undefined \def \bchapter#1{#1}\fi
\ifx \bbook \undefined \def \bbook#1{#1}\fi
\ifx \bcomment \undefined \def \bcomment#1{#1}\fi
\ifx \oauthor \undefined \def \oauthor#1{#1}\fi
\ifx \citeauthoryear \undefined \def \citeauthoryear#1{#1}\fi
\ifx \endbibitem  \undefined \def \endbibitem {}\fi
\ifx \bconflocation  \undefined \def \bconflocation#1{#1}\fi
\ifx \arxivurl  \undefined \def \arxivurl#1{\textsf{#1}}\fi
\csname PreBibitemsHook\endcsname

\bibitem[\protect\citeauthoryear{Ambrosino
  et~al.}{2014}]{ambrosino_rainfall-derived_2014}
\begin{barticle}
\bauthor{\bsnm{Ambrosino}, \binits{C.}},
\bauthor{\bsnm{Chandler}, \binits{R.E.}},
\bauthor{\bsnm{Todd}, \binits{M.C.}}:
\batitle{Rainfall-derived growing season characteristics for agricultural
  impact assessments in {South} {Africa}}.
\bjtitle{Theoretical and Applied Climatology}
\bvolume{115}(\bissue{3}),
\bfpage{411}--\blpage{426}
(\byear{2014})
\doiurl{10.1007/s00704-013-0896-y} .
Accessed 2025-01-21
\end{barticle}
\endbibitem

\bibitem[\protect\citeauthoryear{Argüeso
  et~al.}{2013}]{argueso_precipitation_2013}
\begin{barticle}
\bauthor{\bsnm{Argüeso}, \binits{D.}},
\bauthor{\bsnm{Evans}, \binits{J.P.}},
\bauthor{\bsnm{Fita}, \binits{L.}}:
\batitle{Precipitation bias correction of very high resolution regional climate
  models}.
\bjtitle{Hydrology and Earth System Sciences}
\bvolume{17}(\bissue{11}),
\bfpage{4379}--\blpage{4388}
(\byear{2013})
\doiurl{10.5194/hess-17-4379-2013} .
Accessed 2024-10-29
\end{barticle}
\endbibitem

\bibitem[\protect\citeauthoryear{Ajaaj et~al.}{2016}]{ajaaj_comparison_2016}
\begin{barticle}
\bauthor{\bsnm{Ajaaj}, \binits{A.A.}},
\bauthor{\bsnm{Mishra}, \binits{A.K.}},
\bauthor{\bsnm{Khan}, \binits{A.A.}}:
\batitle{Comparison of {BIAS} correction techniques for {GPCC} rainfall data in
  semi-arid climate}.
\bjtitle{Stochastic Environmental Research and Risk Assessment}
\bvolume{30}(\bissue{6}),
\bfpage{1659}--\blpage{1675}
(\byear{2016})
\doiurl{10.1007/s00477-015-1155-9} .
Accessed 2024-04-25
\end{barticle}
\endbibitem

\bibitem[\protect\citeauthoryear{Abbott et~al.}{2016}]{abbott_long_2016}
\begin{barticle}
\bauthor{\bsnm{Abbott}, \binits{T.H.}},
\bauthor{\bsnm{Stechmann}, \binits{S.N.}},
\bauthor{\bsnm{Neelin}, \binits{J.D.}}:
\batitle{Long temporal autocorrelations in tropical precipitation data and
  spike train prototypes}.
\bjtitle{Geophysical Research Letters}
\bvolume{43}(\bissue{21}),
\bfpage{11472}--\blpage{11480}
(\byear{2016})
\doiurl{10.1002/2016GL071282} .
Accessed 2024-10-29
\end{barticle}
\endbibitem

\bibitem[\protect\citeauthoryear{Alfieri and
  Thielen}{2015}]{alfieri_european_2015}
\begin{barticle}
\bauthor{\bsnm{Alfieri}, \binits{L.}},
\bauthor{\bsnm{Thielen}, \binits{J.}}:
\batitle{A {European} precipitation index for extreme rain-storm and flash
  flood early warning}.
\bjtitle{Meteorological Applications}
\bvolume{22}(\bissue{1}),
\bfpage{3}--\blpage{13}
(\byear{2015})
\doiurl{10.1002/met.1328} .
Accessed 2023-05-17
\end{barticle}
\endbibitem

\bibitem[\protect\citeauthoryear{Andrade-Velázquez and
  Montero-Martínez}{2023}]{andrade-velazquez_statistical_2023}
\begin{barticle}
\bauthor{\bsnm{Andrade-Velázquez}, \binits{M.}},
\bauthor{\bsnm{Montero-Martínez}, \binits{M.J.}}:
\batitle{Statistical {Downscaling} of {Precipitation} in the {South} and
  {Southeast} of {Mexico}}.
\bjtitle{Climate}
\bvolume{11}(\bissue{9}),
\bfpage{186}
(\byear{2023})
\doiurl{10.3390/cli11090186} .
Accessed 2025-01-13
\end{barticle}
\endbibitem

\bibitem[\protect\citeauthoryear{Berg et~al.}{2024}]{berg_robust_2024}
\begin{barticle}
\bauthor{\bsnm{Berg}, \binits{P.}},
\bauthor{\bsnm{Bosshard}, \binits{T.}},
\bauthor{\bsnm{Bozhinova}, \binits{D.}},
\bauthor{\bsnm{Bärring}, \binits{L.}},
\bauthor{\bsnm{Löw}, \binits{J.}},
\bauthor{\bsnm{Nilsson}, \binits{C.}},
\bauthor{\bsnm{Strandberg}, \binits{G.}},
\bauthor{\bsnm{Södling}, \binits{J.}},
\bauthor{\bsnm{Thuresson}, \binits{J.}},
\bauthor{\bsnm{Wilcke}, \binits{R.}},
\bauthor{\bsnm{Yang}, \binits{W.}}:
\batitle{Robust handling of extremes in quantile mapping – “{Murder} your
  darlings”}.
\bjtitle{Geoscientific Model Development}
\bvolume{17}(\bissue{22}),
\bfpage{8173}--\blpage{8179}
(\byear{2024})
\doiurl{10.5194/gmd-17-8173-2024} .
Accessed 2025-01-21
\end{barticle}
\endbibitem

\bibitem[\protect\citeauthoryear{Bador et~al.}{2020}]{bador_impact_2020}
\begin{barticle}
\bauthor{\bsnm{Bador}, \binits{M.}},
\bauthor{\bsnm{Boé}, \binits{J.}},
\bauthor{\bsnm{Terray}, \binits{L.}},
\bauthor{\bsnm{Alexander}, \binits{L.V.}},
\bauthor{\bsnm{Baker}, \binits{A.}},
\bauthor{\bsnm{Bellucci}, \binits{A.}},
\bauthor{\bsnm{Haarsma}, \binits{R.}},
\bauthor{\bsnm{Koenigk}, \binits{T.}},
\bauthor{\bsnm{Moine}, \binits{M.-P.}},
\bauthor{\bsnm{Lohmann}, \binits{K.}},
\bauthor{\bsnm{Putrasahan}, \binits{D.A.}},
\bauthor{\bsnm{Roberts}, \binits{C.}},
\bauthor{\bsnm{Roberts}, \binits{M.}},
\bauthor{\bsnm{Scoccimarro}, \binits{E.}},
\bauthor{\bsnm{Schiemann}, \binits{R.}},
\bauthor{\bsnm{Seddon}, \binits{J.}},
\bauthor{\bsnm{Senan}, \binits{R.}},
\bauthor{\bsnm{Valcke}, \binits{S.}},
\bauthor{\bsnm{Vanniere}, \binits{B.}}:
\batitle{Impact of {Higher} {Spatial} {Atmospheric} {Resolution} on
  {Precipitation} {Extremes} {Over} {Land} in {Global} {Climate} {Models}}.
\bjtitle{Journal of Geophysical Research: Atmospheres}
\bvolume{125}(\bissue{13}),
\bfpage{2019}--\blpage{032184}
(\byear{2020})
\doiurl{10.1029/2019JD032184} .
Accessed 2025-01-21
\end{barticle}
\endbibitem

\bibitem[\protect\citeauthoryear{Bouvier
  et~al.}{2003}]{bouvier_generating_2003}
\begin{barticle}
\bauthor{\bsnm{Bouvier}, \binits{C.}},
\bauthor{\bsnm{Cisneros}, \binits{L.}},
\bauthor{\bsnm{Dominguez}, \binits{R.}},
\bauthor{\bsnm{Laborde}, \binits{J.-P.}},
\bauthor{\bsnm{Lebel}, \binits{T.}}:
\batitle{Generating rainfall fields using principal components ({PC})
  decomposition of the covariance matrix: a case study in {Mexico} {City}}.
\bjtitle{Journal of Hydrology}
\bvolume{278}(\bissue{1}),
\bfpage{107}--\blpage{120}
(\byear{2003})
\doiurl{10.1016/S0022-1694(03)00122-7} .
Accessed 2025-01-21
\end{barticle}
\endbibitem

\bibitem[\protect\citeauthoryear{Byun and Hamlet}{2024}]{byun_improved_2024}
\begin{barticle}
\bauthor{\bsnm{Byun}, \binits{K.}},
\bauthor{\bsnm{Hamlet}, \binits{A.F.}}:
\batitle{An improved empirical quantile mapping approach for bias correction of
  extreme values in climate model simulations}.
\bjtitle{Environmental Research Letters}
\bvolume{20}(\bissue{1}),
\bfpage{014041}
(\byear{2024})
\doiurl{10.1088/1748-9326/ad9b3d} .
Accessed 2025-01-21
\end{barticle}
\endbibitem

\bibitem[\protect\citeauthoryear{Braunstein}{1992}]{braunstein_how_1992}
\begin{barticle}
\bauthor{\bsnm{Braunstein}, \binits{S.L.}}:
\batitle{How large a sample is needed for the maximum likelihood estimator to
  be approximately {Gaussian}?}
\bjtitle{Journal of Physics A: Mathematical and General}
\bvolume{25}(\bissue{13}),
\bfpage{3813}
(\byear{1992})
\doiurl{10.1088/0305-4470/25/13/027} .
Accessed 2025-01-21
\end{barticle}
\endbibitem

\bibitem[\protect\citeauthoryear{Boé et~al.}{2007}]{boe_statistical_2007}
\begin{barticle}
\bauthor{\bsnm{Boé}, \binits{J.}},
\bauthor{\bsnm{Terray}, \binits{L.}},
\bauthor{\bsnm{Habets}, \binits{F.}},
\bauthor{\bsnm{Martin}, \binits{E.}}:
\batitle{Statistical and dynamical downscaling of the {Seine} basin climate for
  hydro-meteorological studies}.
\bjtitle{International Journal of Climatology}
\bvolume{27}(\bissue{12}),
\bfpage{1643}--\blpage{1655}
(\byear{2007})
\doiurl{10.1002/joc.1602} .
Accessed 2023-09-18
\end{barticle}
\endbibitem

\bibitem[\protect\citeauthoryear{Chen
  et~al.}{2021}]{chen_convective--total_2021}
\begin{barticle}
\bauthor{\bsnm{Chen}, \binits{D.}},
\bauthor{\bsnm{Dai}, \binits{A.}},
\bauthor{\bsnm{Hall}, \binits{A.}}:
\batitle{The {Convective}-{To}-{Total} {Precipitation} {Ratio} and the
  “{Drizzling}” {Bias} in {Climate} {Models}}.
\bjtitle{Journal of Geophysical Research: Atmospheres}
\bvolume{126}(\bissue{16}),
\bfpage{2020}--\blpage{034198}
(\byear{2021})
\doiurl{10.1029/2020JD034198} .
Accessed 2024-07-08
\end{barticle}
\endbibitem

\bibitem[\protect\citeauthoryear{Cortés-Hernández
  et~al.}{2024}]{cortes-hernandez_evaluation_2024}
\begin{barticle}
\bauthor{\bsnm{Cortés-Hernández}, \binits{V.E.}},
\bauthor{\bsnm{Caillaud}, \binits{C.}},
\bauthor{\bsnm{Bellon}, \binits{G.}},
\bauthor{\bsnm{Brisson}, \binits{E.}},
\bauthor{\bsnm{Alias}, \binits{A.}},
\bauthor{\bsnm{Lucas-Picher}, \binits{P.}}:
\batitle{Evaluation of the convection permitting regional climate model
  {CNRM}-{AROME} on the orographically complex island of {Corsica}}.
\bjtitle{Climate Dynamics}
\bvolume{62}(\bissue{6}),
\bfpage{4673}--\blpage{4696}
(\byear{2024})
\doiurl{10.1007/s00382-024-07232-z} .
Accessed 2025-01-21
\end{barticle}
\endbibitem

\bibitem[\protect\citeauthoryear{Casanueva et~al.}{2016}]{casanueva_daily_2016}
\begin{barticle}
\bauthor{\bsnm{Casanueva}, \binits{A.}},
\bauthor{\bsnm{Kotlarski}, \binits{S.}},
\bauthor{\bsnm{Herrera}, \binits{S.}},
\bauthor{\bsnm{Fernández}, \binits{J.}},
\bauthor{\bsnm{Gutiérrez}, \binits{J.M.}},
\bauthor{\bsnm{Boberg}, \binits{F.}},
\bauthor{\bsnm{Colette}, \binits{A.}},
\bauthor{\bsnm{Christensen}, \binits{O.B.}},
\bauthor{\bsnm{Goergen}, \binits{K.}},
\bauthor{\bsnm{Jacob}, \binits{D.}},
\bauthor{\bsnm{Keuler}, \binits{K.}},
\bauthor{\bsnm{Nikulin}, \binits{G.}},
\bauthor{\bsnm{Teichmann}, \binits{C.}},
\bauthor{\bsnm{Vautard}, \binits{R.}}:
\batitle{Daily precipitation statistics in a {EURO}-{CORDEX} {RCM} ensemble:
  added value of raw and bias-corrected high-resolution simulations}.
\bjtitle{Climate Dynamics}
\bvolume{47}(\bissue{3}),
\bfpage{719}--\blpage{737}
(\byear{2016})
\doiurl{10.1007/s00382-015-2865-x} .
Accessed 2024-12-03
\end{barticle}
\endbibitem

\bibitem[\protect\citeauthoryear{Chaouche
  et~al.}{2010}]{chaouche_analyses_2010}
\begin{barticle}
\bauthor{\bsnm{Chaouche}, \binits{K.}},
\bauthor{\bsnm{Neppel}, \binits{L.}},
\bauthor{\bsnm{Dieulin}, \binits{C.}},
\bauthor{\bsnm{Pujol}, \binits{N.}},
\bauthor{\bsnm{Ladouche}, \binits{B.}},
\bauthor{\bsnm{Martin}, \binits{E.}},
\bauthor{\bsnm{Salas}, \binits{D.}},
\bauthor{\bsnm{Caballero}, \binits{Y.}}:
\batitle{Analyses of precipitation, temperature and evapotranspiration in a
  {French} {Mediterranean} region in the context of climate change}.
\bjtitle{Comptes Rendus Geoscience}
\bvolume{342}(\bissue{3}),
\bfpage{234}--\blpage{243}
(\byear{2010})
\doiurl{10.1016/j.crte.2010.02.001} .
Accessed 2025-01-21
\end{barticle}
\endbibitem

\bibitem[\protect\citeauthoryear{Derdour et~al.}{2022}]{derdour_bias_2022}
\begin{barticle}
\bauthor{\bsnm{Derdour}, \binits{S.}},
\bauthor{\bsnm{Ghenim}, \binits{A.N.}},
\bauthor{\bsnm{Megnounif}, \binits{A.}},
\bauthor{\bsnm{Tangang}, \binits{F.}},
\bauthor{\bsnm{Chung}, \binits{J.X.}},
\bauthor{\bsnm{Ayoub}, \binits{A.B.}}:
\batitle{Bias {Correction} and {Evaluation} of {Precipitation} {Data} from the
  {CORDEX} {Regional} {Climate} {Model} for {Monitoring} {Climate} {Change} in
  the {Wadi} {Chemora} {Basin} ({Northeastern} {Algeria})}.
\bjtitle{Atmosphere}
\bvolume{13}(\bissue{11}),
\bfpage{1876}
(\byear{2022})
\doiurl{10.3390/atmos13111876} .
Accessed 2025-01-13
\end{barticle}
\endbibitem

\bibitem[\protect\citeauthoryear{Déqué}{2007}]{deque_frequency_2007}
\begin{barticle}
\bauthor{\bsnm{Déqué}, \binits{M.}}:
\batitle{Frequency of precipitation and temperature extremes over {France} in
  an anthropogenic scenario: {Model} results and statistical correction
  according to observed values}.
\bjtitle{Global and Planetary Change}
\bvolume{57}(\bissue{1}),
\bfpage{16}--\blpage{26}
(\byear{2007})
\doiurl{10.1016/j.gloplacha.2006.11.030} .
Accessed 2023-02-06
\end{barticle}
\endbibitem

\bibitem[\protect\citeauthoryear{Enayati et~al.}{2021}]{enayati_bias_2021}
\begin{barticle}
\bauthor{\bsnm{Enayati}, \binits{M.}},
\bauthor{\bsnm{Bozorg-Haddad}, \binits{O.}},
\bauthor{\bsnm{Bazrafshan}, \binits{J.}},
\bauthor{\bsnm{Hejabi}, \binits{S.}},
\bauthor{\bsnm{Chu}, \binits{X.}}:
\batitle{Bias correction capabilities of quantile mapping methods for rainfall
  and temperature variables}.
\bjtitle{Journal of Water and Climate Change}
\bvolume{12}(\bissue{2}),
\bfpage{401}--\blpage{419}
(\byear{2021})
\doiurl{10.2166/wcc.2020.261} .
Accessed 2024-12-03
\end{barticle}
\endbibitem

\bibitem[\protect\citeauthoryear{Ear et~al.}{2025}]{ear_semi-parametric_2025}
\begin{botherref}
\oauthor{\bsnm{Ear}, \binits{P.}},
\oauthor{\bsnm{Di~Bernardino}, \binits{E.}},
\oauthor{\bsnm{Laloë}, \binits{T.}},
\oauthor{\bsnm{Troin}, \binits{M.}},
\oauthor{\bsnm{Lambert}, \binits{A.}}:
A semi-parametric distribution stitch based on the {Berk}-{Jones} test for
  {French} daily precipitation bias correction.
Stochastic Environmental Research and Risk Assessment
(2025).
to appear.
Accessed 2024-10-29
\end{botherref}
\endbibitem

\bibitem[\protect\citeauthoryear{Emmanuel et~al.}{2017}]{emmanuel_method_2017}
\begin{barticle}
\bauthor{\bsnm{Emmanuel}, \binits{I.}},
\bauthor{\bsnm{Payrastre}, \binits{O.}},
\bauthor{\bsnm{Andrieu}, \binits{H.}},
\bauthor{\bsnm{Zuber}, \binits{F.}}:
\batitle{A method for assessing the influence of rainfall spatial variability
  on hydrograph modeling. {First} case study in the {Cevennes} {Region},
  southern {France}}.
\bjtitle{Journal of Hydrology}
\bvolume{555},
\bfpage{314}--\blpage{322}
(\byear{2017})
\doiurl{10.1016/j.jhydrol.2017.10.011} .
Accessed 2025-01-21
\end{barticle}
\endbibitem

\bibitem[\protect\citeauthoryear{Estermann
  et~al.}{2025}]{estermann_projections_2025}
\begin{barticle}
\bauthor{\bsnm{Estermann}, \binits{R.}},
\bauthor{\bsnm{Rajczak}, \binits{J.}},
\bauthor{\bsnm{Velasquez}, \binits{P.}},
\bauthor{\bsnm{Lorenz}, \binits{R.}},
\bauthor{\bsnm{Schär}, \binits{C.}}:
\batitle{Projections of {Heavy} {Precipitation} {Characteristics} {Over} the
  {Greater} {Alpine} {Region} {Using} a {Kilometer}–{Scale} {Climate} {Model}
  {Ensemble}}.
\bjtitle{Journal of Geophysical Research: Atmospheres}
\bvolume{130}(\bissue{2}),
\bfpage{2024}--\blpage{040901}
(\byear{2025})
\doiurl{10.1029/2024JD040901} .
Accessed 2025-01-21
\end{barticle}
\endbibitem

\bibitem[\protect\citeauthoryear{Enyew et~al.}{2024}]{enyew_performance_2024}
\begin{barticle}
\bauthor{\bsnm{Enyew}, \binits{F.B.}},
\bauthor{\bsnm{Sahlu}, \binits{D.}},
\bauthor{\bsnm{Tarekegn}, \binits{G.B.}},
\bauthor{\bsnm{Hama}, \binits{S.}},
\bauthor{\bsnm{Debele}, \binits{S.E.}}:
\batitle{Performance {Evaluation} of {CMIP6} {Climate} {Model} {Projections}
  for {Precipitation} and {Temperature} in the {Upper} {Blue} {Nile} {Basin},
  {Ethiopia}}.
\bjtitle{Climate}
\bvolume{12}(\bissue{11}),
\bfpage{169}
(\byear{2024})
\doiurl{10.3390/cli12110169} .
Accessed 2025-01-13
\end{barticle}
\endbibitem

\bibitem[\protect\citeauthoryear{Fosser
  et~al.}{2024}]{fosser_convection-permitting_2024}
\begin{barticle}
\bauthor{\bsnm{Fosser}, \binits{G.}},
\bauthor{\bsnm{Gaetani}, \binits{M.}},
\bauthor{\bsnm{Kendon}, \binits{E.J.}},
\bauthor{\bsnm{Adinolfi}, \binits{M.}},
\bauthor{\bsnm{Ban}, \binits{N.}},
\bauthor{\bsnm{Belušić}, \binits{D.}},
\bauthor{\bsnm{Caillaud}, \binits{C.}},
\bauthor{\bsnm{Careto}, \binits{J.A.M.}},
\bauthor{\bsnm{Coppola}, \binits{E.}},
\bauthor{\bsnm{Demory}, \binits{M.-E.}},
\bauthor{\bsnm{Vries}, \binits{H.}},
\bauthor{\bsnm{Dobler}, \binits{A.}},
\bauthor{\bsnm{Feldmann}, \binits{H.}},
\bauthor{\bsnm{Goergen}, \binits{K.}},
\bauthor{\bsnm{Lenderink}, \binits{G.}},
\bauthor{\bsnm{Pichelli}, \binits{E.}},
\bauthor{\bsnm{Schär}, \binits{C.}},
\bauthor{\bsnm{Soares}, \binits{P.M.M.}},
\bauthor{\bsnm{Somot}, \binits{S.}},
\bauthor{\bsnm{Tölle}, \binits{M.H.}}:
\batitle{Convection-permitting climate models offer more certain extreme
  rainfall projections}.
\bjtitle{npj Climate and Atmospheric Science}
\bvolume{7}(\bissue{1}),
\bfpage{1}--\blpage{10}
(\byear{2024})
\doiurl{10.1038/s41612-024-00600-w} .
Accessed 2025-01-21
\end{barticle}
\endbibitem

\bibitem[\protect\citeauthoryear{Gutowski
  et~al.}{2003}]{gutowski_temporalspatial_2003}
\begin{botherref}
\oauthor{\bsnm{Gutowski}, \binits{W.J.}},
\oauthor{\bsnm{Decker}, \binits{S.G.}},
\oauthor{\bsnm{Donavon}, \binits{R.A.}},
\oauthor{\bsnm{Pan}, \binits{Z.}},
\oauthor{\bsnm{Arritt}, \binits{R.W.}},
\oauthor{\bsnm{Takle}, \binits{E.S.}}:
Temporal–{Spatial} {Scales} of {Observed} and {Simulated} {Precipitation} in
  {Central} {U}.{S}. {Climate}.
Journal of Climate
(2003).
Accessed 2025-02-28
\end{botherref}
\endbibitem

\bibitem[\protect\citeauthoryear{Gutjahr and
  Heinemann}{2013}]{gutjahr_comparing_2013}
\begin{barticle}
\bauthor{\bsnm{Gutjahr}, \binits{O.}},
\bauthor{\bsnm{Heinemann}, \binits{G.}}:
\batitle{Comparing precipitation bias correction methods for high-resolution
  regional climate simulations using {COSMO}-{CLM}}.
\bjtitle{Theoretical and Applied Climatology}
\bvolume{114}(\bissue{3}),
\bfpage{511}--\blpage{529}
(\byear{2013})
\doiurl{10.1007/s00704-013-0834-z} .
Accessed 2023-07-18
\end{barticle}
\endbibitem

\bibitem[\protect\citeauthoryear{Gutiérrez
  et~al.}{2019}]{gutierrez_intercomparison_2019}
\begin{barticle}
\bauthor{\bsnm{Gutiérrez}, \binits{J.M.}},
\bauthor{\bsnm{Maraun}, \binits{D.}},
\bauthor{\bsnm{Widmann}, \binits{M.}},
\bauthor{\bsnm{Huth}, \binits{R.}},
\bauthor{\bsnm{Hertig}, \binits{E.}},
\bauthor{\bsnm{Benestad}, \binits{R.}},
\bauthor{\bsnm{Roessler}, \binits{O.}},
\bauthor{\bsnm{Wibig}, \binits{J.}},
\bauthor{\bsnm{Wilcke}, \binits{R.}},
\bauthor{\bsnm{Kotlarski}, \binits{S.}},
\bauthor{\bsnm{San~Martín}, \binits{D.}},
\bauthor{\bsnm{Herrera}, \binits{S.}},
\bauthor{\bsnm{Bedia}, \binits{J.}},
\bauthor{\bsnm{Casanueva}, \binits{A.}},
\bauthor{\bsnm{Manzanas}, \binits{R.}},
\bauthor{\bsnm{Iturbide}, \binits{M.}},
\bauthor{\bsnm{Vrac}, \binits{M.}},
\bauthor{\bsnm{Dubrovsky}, \binits{M.}},
\bauthor{\bsnm{Ribalaygua}, \binits{J.}},
\bauthor{\bsnm{Pórtoles}, \binits{J.}},
\bauthor{\bsnm{Räty}, \binits{O.}},
\bauthor{\bsnm{Räisänen}, \binits{J.}},
\bauthor{\bsnm{Hingray}, \binits{B.}},
\bauthor{\bsnm{Raynaud}, \binits{D.}},
\bauthor{\bsnm{Casado}, \binits{M.J.}},
\bauthor{\bsnm{Ramos}, \binits{P.}},
\bauthor{\bsnm{Zerenner}, \binits{T.}},
\bauthor{\bsnm{Turco}, \binits{M.}},
\bauthor{\bsnm{Bosshard}, \binits{T.}},
\bauthor{\bsnm{Štěpánek}, \binits{P.}},
\bauthor{\bsnm{Bartholy}, \binits{J.}},
\bauthor{\bsnm{Pongracz}, \binits{R.}},
\bauthor{\bsnm{Keller}, \binits{D.E.}},
\bauthor{\bsnm{Fischer}, \binits{A.M.}},
\bauthor{\bsnm{Cardoso}, \binits{R.M.}},
\bauthor{\bsnm{Soares}, \binits{P.M.M.}},
\bauthor{\bsnm{Czernecki}, \binits{B.}},
\bauthor{\bsnm{Pagé}, \binits{C.}}:
\batitle{An intercomparison of a large ensemble of statistical downscaling
  methods over {Europe}: {Results} from the {VALUE} perfect predictor
  cross-validation experiment}.
\bjtitle{International Journal of Climatology}
\bvolume{39}(\bissue{9}),
\bfpage{3750}--\blpage{3785}
(\byear{2019})
\doiurl{10.1002/joc.5462} .
Accessed 2025-02-25
\end{barticle}
\endbibitem

\bibitem[\protect\citeauthoryear{Guo et~al.}{2024}]{guo_does_2024}
\begin{barticle}
\bauthor{\bsnm{Guo}, \binits{C.}},
\bauthor{\bsnm{Ning}, \binits{N.}},
\bauthor{\bsnm{Guo}, \binits{H.}},
\bauthor{\bsnm{Tian}, \binits{Y.}},
\bauthor{\bsnm{Bao}, \binits{A.}},
\bauthor{\bsnm{De~Maeyer}, \binits{P.}}:
\batitle{Does {ERA5}-{Land} {Effectively} {Capture} {Extreme} {Precipitation}
  in the {Yellow} {River} {Basin}?}
\bjtitle{Atmosphere}
\bvolume{15}(\bissue{10}),
\bfpage{1254}
(\byear{2024})
\doiurl{10.3390/atmos15101254} .
Accessed 2025-02-25
\end{barticle}
\endbibitem

\bibitem[\protect\citeauthoryear{Holthuijzen
  et~al.}{2022}]{holthuijzen_robust_2022}
\begin{barticle}
\bauthor{\bsnm{Holthuijzen}, \binits{M.}},
\bauthor{\bsnm{Beckage}, \binits{B.}},
\bauthor{\bsnm{Clemins}, \binits{P.J.}},
\bauthor{\bsnm{Higdon}, \binits{D.}},
\bauthor{\bsnm{Winter}, \binits{J.M.}}:
\batitle{Robust bias-correction of precipitation extremes using a novel hybrid
  empirical quantile-mapping method: {Advantages} of a linear correction for
  extremes}.
\bjtitle{Theoretical and Applied Climatology}
\bvolume{149}(\bissue{1-2}),
\bfpage{863}--\blpage{882}
(\byear{2022})
\doiurl{10.1007/s00704-022-04035-2} .
Accessed 2023-01-23
\end{barticle}
\endbibitem

\bibitem[\protect\citeauthoryear{Haruna et~al.}{2023}]{haruna_modeling_2023}
\begin{barticle}
\bauthor{\bsnm{Haruna}, \binits{A.}},
\bauthor{\bsnm{Blanchet}, \binits{J.}},
\bauthor{\bsnm{Favre}, \binits{A.-C.}}:
\batitle{Modeling {Intensity}-{Duration}-{Frequency} {Curves} for the {Whole}
  {Range} of {Non}-{Zero} {Precipitation}: {A} {Comparison} of {Models}}.
\bjtitle{Water Resources Research}
\bvolume{59}(\bissue{6}),
\bfpage{2022}--\blpage{033362}
(\byear{2023})
\doiurl{10.1029/2022WR033362} .
Accessed 2025-01-21
\end{barticle}
\endbibitem

\bibitem[\protect\citeauthoryear{Henckes et~al.}{2018}]{henckes_benefit_2018}
\begin{barticle}
\bauthor{\bsnm{Henckes}, \binits{P.}},
\bauthor{\bsnm{Knaut}, \binits{A.}},
\bauthor{\bsnm{Obermüller}, \binits{F.}},
\bauthor{\bsnm{Frank}, \binits{C.}}:
\batitle{The benefit of long-term high resolution wind data for electricity
  system analysis}.
\bjtitle{Energy}
\bvolume{143},
\bfpage{934}--\blpage{942}
(\byear{2018})
\doiurl{10.1016/j.energy.2017.10.049} .
Accessed 2023-05-03
\end{barticle}
\endbibitem

\bibitem[\protect\citeauthoryear{Husak et~al.}{2007}]{husak_use_2007}
\begin{barticle}
\bauthor{\bsnm{Husak}, \binits{G.J.}},
\bauthor{\bsnm{Michaelsen}, \binits{J.C.}},
\bauthor{\bsnm{Funk}, \binits{C.C.}}:
\batitle{Use of the {Gamma} distribution to represent monthly rainfall in
  {Africa} for drought monitoring applications}.
\bjtitle{International Journal of Climatology}
\bvolume{27}(\bissue{7}),
\bfpage{935}--\blpage{944}
(\byear{2007})
\doiurl{10.1002/joc.1441} .
Accessed 2024-06-12
\end{barticle}
\endbibitem

\bibitem[\protect\citeauthoryear{Joly et~al.}{2010}]{joly_types_2010}
\begin{barticle}
\bauthor{\bsnm{Joly}, \binits{D.}},
\bauthor{\bsnm{Brossard}, \binits{T.}},
\bauthor{\bsnm{Cardot}, \binits{H.}},
\bauthor{\bsnm{Cavailhes}, \binits{J.}},
\bauthor{\bsnm{Hilal}, \binits{M.}},
\bauthor{\bsnm{Wavresky}, \binits{P.}}:
\batitle{Les types de climats en {France}, une construction spatiale}.
\bjtitle{Cybergeo: European Journal of Geography}
(\byear{2010})
\doiurl{10.4000/cybergeo.23155} .
Accessed 2023-08-30
\end{barticle}
\endbibitem

\bibitem[\protect\citeauthoryear{Katiraie-Boroujerdy
  et~al.}{2020}]{katiraie-boroujerdy_bias_2020}
\begin{barticle}
\bauthor{\bsnm{Katiraie-Boroujerdy}, \binits{P.-S.}},
\bauthor{\bsnm{Rahnamay~Naeini}, \binits{M.}},
\bauthor{\bsnm{Akbari~Asanjan}, \binits{A.}},
\bauthor{\bsnm{Chavoshian}, \binits{A.}},
\bauthor{\bsnm{Hsu}, \binits{K.-l.}},
\bauthor{\bsnm{Sorooshian}, \binits{S.}}:
\batitle{Bias {Correction} of {Satellite}-{Based} {Precipitation} {Estimations}
  {Using} {Quantile} {Mapping} {Approach} in {Different} {Climate} {Regions} of
  {Iran}}.
\bjtitle{Remote Sensing}
\bvolume{12}(\bissue{13}),
\bfpage{2102}
(\byear{2020})
\doiurl{10.3390/rs12132102} .
Accessed 2025-02-24
\end{barticle}
\endbibitem

\bibitem[\protect\citeauthoryear{Khan}{2018}]{khan_exponentiated_2018}
\begin{barticle}
\bauthor{\bsnm{Khan}, \binits{S.A.}}:
\batitle{Exponentiated {Weibull} regression for time-to-event data}.
\bjtitle{Lifetime Data Analysis}
\bvolume{24}(\bissue{2}),
\bfpage{328}--\blpage{354}
(\byear{2018})
\doiurl{10.1007/s10985-017-9394-3} .
Accessed 2024-01-29
\end{barticle}
\endbibitem

\bibitem[\protect\citeauthoryear{Khan et~al.}{2007}]{khan_spatio-temporal_2007}
\begin{botherref}
\oauthor{\bsnm{Khan}, \binits{S.}},
\oauthor{\bsnm{Kuhn}, \binits{G.}},
\oauthor{\bsnm{Ganguly}, \binits{A.R.}},
\oauthor{\bsnm{Erickson~III}, \binits{D.J.}},
\oauthor{\bsnm{Ostrouchov}, \binits{G.}}:
Spatio-temporal variability of daily and weekly precipitation extremes in
  {South} {America}.
Water Resources Research
\textbf{43}(11)
(2007)
\doiurl{10.1029/2006WR005384} .
Accessed 2024-10-29
\end{botherref}
\endbibitem

\bibitem[\protect\citeauthoryear{Lafon et~al.}{2013}]{lafon_bias_2013}
\begin{barticle}
\bauthor{\bsnm{Lafon}, \binits{T.}},
\bauthor{\bsnm{Dadson}, \binits{S.}},
\bauthor{\bsnm{Buys}, \binits{G.}},
\bauthor{\bsnm{Prudhomme}, \binits{C.}}:
\batitle{Bias correction of daily precipitation simulated by a regional climate
  model: a comparison of methods}.
\bjtitle{International Journal of Climatology}
\bvolume{33}(\bissue{6}),
\bfpage{1367}--\blpage{1381}
(\byear{2013})
\doiurl{10.1002/joc.3518} .
Accessed 2023-05-03
\end{barticle}
\endbibitem

\bibitem[\protect\citeauthoryear{Li et~al.}{2021}]{li_bias_2021}
\begin{barticle}
\bauthor{\bsnm{Li}, \binits{B.}},
\bauthor{\bsnm{Huang}, \binits{Y.}},
\bauthor{\bsnm{Du}, \binits{L.}},
\bauthor{\bsnm{Wang}, \binits{D.}}:
\batitle{Bias {Correction} for {Precipitation} {Simulated} by {RegCM4} over the
  {Upper} {Reaches} of the {Yangtze} {River} {Based} on the {Mixed}
  {Distribution} {Quantile} {Mapping} {Method}}.
\bjtitle{Atmosphere}
\bvolume{12}(\bissue{12}),
\bfpage{1566}
(\byear{2021})
\doiurl{10.3390/atmos12121566} .
Accessed 2023-03-28
\end{barticle}
\endbibitem

\bibitem[\protect\citeauthoryear{Langousis
  et~al.}{2016}]{langousis_assessing_2016}
\begin{barticle}
\bauthor{\bsnm{Langousis}, \binits{A.}},
\bauthor{\bsnm{Mamalakis}, \binits{A.}},
\bauthor{\bsnm{Deidda}, \binits{R.}},
\bauthor{\bsnm{Marrocu}, \binits{M.}}:
\batitle{Assessing the relative effectiveness of statistical downscaling and
  distribution mapping in reproducing rainfall statistics based on climate
  model results}.
\bjtitle{Water Resources Research}
\bvolume{52}(\bissue{1}),
\bfpage{471}--\blpage{494}
(\byear{2016})
\doiurl{10.1002/2015WR017556} .
Accessed 2024-10-17
\end{barticle}
\endbibitem

\bibitem[\protect\citeauthoryear{Lavaysse
  et~al.}{2012}]{lavaysse_statistical_2012}
\begin{barticle}
\bauthor{\bsnm{Lavaysse}, \binits{C.}},
\bauthor{\bsnm{Vrac}, \binits{M.}},
\bauthor{\bsnm{Drobinski}, \binits{P.}},
\bauthor{\bsnm{Lengaigne}, \binits{M.}},
\bauthor{\bsnm{Vischel}, \binits{T.}}:
\batitle{Statistical downscaling of the {French} {Mediterranean} climate:
  assessment for present and projection in an anthropogenic scenario}.
\bjtitle{Natural Hazards and Earth System Sciences}
\bvolume{12}(\bissue{3}),
\bfpage{651}--\blpage{670}
(\byear{2012})
\doiurl{10.5194/nhess-12-651-2012} .
Accessed 2025-01-21
\end{barticle}
\endbibitem

\bibitem[\protect\citeauthoryear{Maraun}{2013}]{maraun_bias_2013}
\begin{botherref}
\oauthor{\bsnm{Maraun}, \binits{D.}}:
Bias {Correction}, {Quantile} {Mapping}, and {Downscaling}: {Revisiting} the
  {Inflation} {Issue}
(2013)
\doiurl{10.1175/JCLI-D-12-00821.1} .
Accessed 2025-02-24
\end{botherref}
\endbibitem

\bibitem[\protect\citeauthoryear{Mamalakis
  et~al.}{2017}]{mamalakis_parametric_2017}
\begin{barticle}
\bauthor{\bsnm{Mamalakis}, \binits{A.}},
\bauthor{\bsnm{Langousis}, \binits{A.}},
\bauthor{\bsnm{Deidda}, \binits{R.}},
\bauthor{\bsnm{Marrocu}, \binits{M.}}:
\batitle{A parametric approach for simultaneous bias correction and
  high-resolution downscaling of climate model rainfall}.
\bjtitle{Water Resources Research}
\bvolume{53}(\bissue{3}),
\bfpage{2149}--\blpage{2170}
(\byear{2017})
\doiurl{10.1002/2016WR019578} .
Accessed 2024-10-17
\end{barticle}
\endbibitem

\bibitem[\protect\citeauthoryear{Muñoz-Sabater
  et~al.}{2021}]{munoz-sabater_era5-land_2021}
\begin{barticle}
\bauthor{\bsnm{Muñoz-Sabater}, \binits{J.}},
\bauthor{\bsnm{Dutra}, \binits{E.}},
\bauthor{\bsnm{Agustí-Panareda}, \binits{A.}},
\bauthor{\bsnm{Albergel}, \binits{C.}},
\bauthor{\bsnm{Arduini}, \binits{G.}},
\bauthor{\bsnm{Balsamo}, \binits{G.}},
\bauthor{\bsnm{Boussetta}, \binits{S.}},
\bauthor{\bsnm{Choulga}, \binits{M.}},
\bauthor{\bsnm{Harrigan}, \binits{S.}},
\bauthor{\bsnm{Hersbach}, \binits{H.}},
\bauthor{\bsnm{Martens}, \binits{B.}},
\bauthor{\bsnm{Miralles}, \binits{D.G.}},
\bauthor{\bsnm{Piles}, \binits{M.}},
\bauthor{\bsnm{Rodríguez-Fernández}, \binits{N.J.}},
\bauthor{\bsnm{Zsoter}, \binits{E.}},
\bauthor{\bsnm{Buontempo}, \binits{C.}},
\bauthor{\bsnm{Thépaut}, \binits{J.-N.}}:
\batitle{{ERA5}-{Land}: a state-of-the-art global reanalysis dataset for land
  applications}.
\bjtitle{Earth System Science Data}
\bvolume{13}(\bissue{9}),
\bfpage{4349}--\blpage{4383}
(\byear{2021})
\doiurl{10.5194/essd-13-4349-2021} .
Accessed 2023-05-02
\end{barticle}
\endbibitem

\bibitem[\protect\citeauthoryear{Mudholkar
  et~al.}{1996}]{mudholkar_generalization_1996}
\begin{barticle}
\bauthor{\bsnm{Mudholkar}, \binits{G.S.}},
\bauthor{\bsnm{Srivastava}, \binits{D.K.}},
\bauthor{\bsnm{Kollia}, \binits{G.D.}}:
\batitle{A {Generalization} of the {Weibull} {Distribution} with {Application}
  to the {Analysis} of {Survival} {Data}}.
\bjtitle{Journal of the American Statistical Association}
\bvolume{91}(\bissue{436}),
\bfpage{1575}--\blpage{1583}
(\byear{1996})
\doiurl{10.2307/2291583} .
Accessed 2024-01-29
\end{barticle}
\endbibitem

\bibitem[\protect\citeauthoryear{Michelangeli
  et~al.}{2009}]{michelangeli_probabilistic_2009}
\begin{botherref}
\oauthor{\bsnm{Michelangeli}, \binits{P.-A.}},
\oauthor{\bsnm{Vrac}, \binits{M.}},
\oauthor{\bsnm{Loukos}, \binits{H.}}:
Probabilistic downscaling approaches: {Application} to wind cumulative
  distribution functions.
Geophysical Research Letters
\textbf{36}(11)
(2009)
\doiurl{10.1029/2009GL038401} .
Accessed 2022-10-21
\end{botherref}
\endbibitem

\bibitem[\protect\citeauthoryear{Mao et~al.}{2015}]{mao_stochastic_2015}
\begin{barticle}
\bauthor{\bsnm{Mao}, \binits{G.}},
\bauthor{\bsnm{Vogl}, \binits{S.}},
\bauthor{\bsnm{Laux}, \binits{P.}},
\bauthor{\bsnm{Wagner}, \binits{S.}},
\bauthor{\bsnm{Kunstmann}, \binits{H.}}:
\batitle{Stochastic bias correction of dynamically downscaled precipitation
  fields for {Germany} through {Copula}-based integration of gridded
  observation data}.
\bjtitle{Hydrology and Earth System Sciences}
\bvolume{19}(\bissue{4}),
\bfpage{1787}--\blpage{1806}
(\byear{2015})
\doiurl{10.5194/hess-19-1787-2015} .
Accessed 2025-01-21
\end{barticle}
\endbibitem

\bibitem[\protect\citeauthoryear{Martinez-Villalobos and
  Neelin}{2019}]{martinez-villalobos_why_2019}
\begin{barticle}
\bauthor{\bsnm{Martinez-Villalobos}, \binits{C.}},
\bauthor{\bsnm{Neelin}, \binits{J.D.}}:
\batitle{Why {Do} {Precipitation} {Intensities} {Tend} to {Follow} {Gamma}
  {Distributions}?}
\bjtitle{Journal of the Atmospheric Sciences}
\bvolume{76}(\bissue{11}),
\bfpage{3611}--\blpage{3631}
(\byear{2019})
\doiurl{10.1175/JAS-D-18-0343.1} .
Accessed 2023-03-28
\end{barticle}
\endbibitem

\bibitem[\protect\citeauthoryear{Naveau et~al.}{2016}]{naveau_modeling_2016}
\begin{barticle}
\bauthor{\bsnm{Naveau}, \binits{P.}},
\bauthor{\bsnm{Huser}, \binits{R.}},
\bauthor{\bsnm{Ribereau}, \binits{P.}},
\bauthor{\bsnm{Hannart}, \binits{A.}}:
\batitle{Modeling jointly low, moderate, and heavy rainfall intensities without
  a threshold selection}.
\bjtitle{Water Resources Research}
\bvolume{52}(\bissue{4}),
\bfpage{2753}--\blpage{2769}
(\byear{2016})
\doiurl{10.1002/2015WR018552} .
Accessed 2023-09-20
\end{barticle}
\endbibitem

\bibitem[\protect\citeauthoryear{Pelosi}{2023}]{pelosi_performance_2023}
\begin{barticle}
\bauthor{\bsnm{Pelosi}, \binits{A.}}:
\batitle{Performance of the {Copernicus} {European} {Regional} {Reanalysis}
  ({CERRA}) dataset as proxy of ground-based agrometeorological data}.
\bjtitle{Agricultural Water Management}
\bvolume{289},
\bfpage{108556}
(\byear{2023})
\doiurl{10.1016/j.agwat.2023.108556} .
Accessed 2025-01-21
\end{barticle}
\endbibitem

\bibitem[\protect\citeauthoryear{Prein et~al.}{2016}]{prein_precipitation_2016}
\begin{barticle}
\bauthor{\bsnm{Prein}, \binits{A.F.}},
\bauthor{\bsnm{Gobiet}, \binits{A.}},
\bauthor{\bsnm{Truhetz}, \binits{H.}},
\bauthor{\bsnm{Keuler}, \binits{K.}},
\bauthor{\bsnm{Goergen}, \binits{K.}},
\bauthor{\bsnm{Teichmann}, \binits{C.}},
\bauthor{\bsnm{Fox~Maule}, \binits{C.}},
\bauthor{\bsnm{Meijgaard}, \binits{E.}},
\bauthor{\bsnm{Déqué}, \binits{M.}},
\bauthor{\bsnm{Nikulin}, \binits{G.}},
\bauthor{\bsnm{Vautard}, \binits{R.}},
\bauthor{\bsnm{Colette}, \binits{A.}},
\bauthor{\bsnm{Kjellström}, \binits{E.}},
\bauthor{\bsnm{Jacob}, \binits{D.}}:
\batitle{Precipitation in the {EURO}-{CORDEX} 0.11° and 0.44° simulations:
  high resolution, high benefits?}
\bjtitle{Climate Dynamics}
\bvolume{46}(\bissue{1}),
\bfpage{383}--\blpage{412}
(\byear{2016})
\doiurl{10.1007/s00382-015-2589-y} .
Accessed 2023-05-03
\end{barticle}
\endbibitem

\bibitem[\protect\citeauthoryear{Piani et~al.}{2010}]{piani_statistical_2010}
\begin{barticle}
\bauthor{\bsnm{Piani}, \binits{C.}},
\bauthor{\bsnm{Haerter}, \binits{J.O.}},
\bauthor{\bsnm{Coppola}, \binits{E.}}:
\batitle{Statistical bias correction for daily precipitation in regional
  climate models over {Europe}}.
\bjtitle{Theoretical and Applied Climatology}
\bvolume{99}(\bissue{1}),
\bfpage{187}--\blpage{192}
(\byear{2010})
\doiurl{10.1007/s00704-009-0134-9} .
Accessed 2023-01-03
\end{barticle}
\endbibitem

\bibitem[\protect\citeauthoryear{Reiter et~al.}{2018}]{reiter_does_2018}
\begin{barticle}
\bauthor{\bsnm{Reiter}, \binits{P.}},
\bauthor{\bsnm{Gutjahr}, \binits{O.}},
\bauthor{\bsnm{Schefczyk}, \binits{L.}},
\bauthor{\bsnm{Heinemann}, \binits{G.}},
\bauthor{\bsnm{Casper}, \binits{M.}}:
\batitle{Does applying quantile mapping to subsamples improve the bias
  correction of daily precipitation?}
\bjtitle{International Journal of Climatology}
\bvolume{38}(\bissue{4}),
\bfpage{1623}--\blpage{1633}
(\byear{2018})
\doiurl{10.1002/joc.5283} .
Accessed 2025-02-24
\end{barticle}
\endbibitem

\bibitem[\protect\citeauthoryear{Rivoire
  et~al.}{2021}]{rivoire_comparison_2021}
\begin{barticle}
\bauthor{\bsnm{Rivoire}, \binits{P.}},
\bauthor{\bsnm{Martius}, \binits{O.}},
\bauthor{\bsnm{Naveau}, \binits{P.}}:
\batitle{A {Comparison} of {Moderate} and {Extreme} {ERA}-5 {Daily}
  {Precipitation} {With} {Two} {Observational} {Data} {Sets}}.
\bjtitle{Earth and Space Science}
\bvolume{8}(\bissue{4}),
\bfpage{2020}--\blpage{001633}
(\byear{2021})
\doiurl{10.1029/2020EA001633} .
Accessed 2023-10-13
\end{barticle}
\endbibitem

\bibitem[\protect\citeauthoryear{Sangati and
  Borga}{2009}]{sangati_influence_2009}
\begin{barticle}
\bauthor{\bsnm{Sangati}, \binits{M.}},
\bauthor{\bsnm{Borga}, \binits{M.}}:
\batitle{Influence of rainfall spatial resolution on flash flood modelling}.
\bjtitle{Natural Hazards and Earth System Sciences}
\bvolume{9}(\bissue{2}),
\bfpage{575}--\blpage{584}
(\byear{2009})
\doiurl{10.5194/nhess-9-575-2009} .
Accessed 2023-05-17
\end{barticle}
\endbibitem

\bibitem[\protect\citeauthoryear{Semenov
  et~al.}{1998}]{semenov_comparison_1998}
\begin{barticle}
\bauthor{\bsnm{Semenov}, \binits{M.A.}},
\bauthor{\bsnm{Brooks}, \binits{R.J.}},
\bauthor{\bsnm{Barrow}, \binits{E.M.}},
\bauthor{\bsnm{Richardson}, \binits{C.W.}}:
\batitle{Comparison of the {WGEN} and {LARS}-{WG} stochastic weather generators
  for diverse climates}.
\bjtitle{Climate Research}
\bvolume{10}(\bissue{2}),
\bfpage{95}--\blpage{107}
(\byear{1998})
\doiurl{10.3354/cr010095} .
Accessed 2025-01-21
\end{barticle}
\endbibitem

\bibitem[\protect\citeauthoryear{Strohmenger
  et~al.}{2024}]{strohmenger_koppengeiger_2024}
\begin{barticle}
\bauthor{\bsnm{Strohmenger}, \binits{L.}},
\bauthor{\bsnm{Collet}, \binits{L.}},
\bauthor{\bsnm{Andréassian}, \binits{V.}},
\bauthor{\bsnm{Corre}, \binits{L.}},
\bauthor{\bsnm{Rousset}, \binits{F.}},
\bauthor{\bsnm{Thirel}, \binits{G.}}:
\batitle{Köppen–{Geiger} climate classification across {France} based on an
  ensemble of high-resolution climate projections}.
\bjtitle{Comptes Rendus. Géoscience}
\bvolume{356}(\bissue{G1}),
\bfpage{67}--\blpage{82}
(\byear{2024})
\doiurl{10.5802/crgeos.263} .
Accessed 2025-01-21
\end{barticle}
\endbibitem

\bibitem[\protect\citeauthoryear{Schmidli
  et~al.}{2006}]{schmidli_downscaling_2006}
\begin{barticle}
\bauthor{\bsnm{Schmidli}, \binits{J.}},
\bauthor{\bsnm{Frei}, \binits{C.}},
\bauthor{\bsnm{Vidale}, \binits{P.L.}}:
\batitle{Downscaling from {GCM} precipitation: a benchmark for dynamical and
  statistical downscaling methods}.
\bjtitle{International Journal of Climatology}
\bvolume{26}(\bissue{5}),
\bfpage{679}--\blpage{689}
(\byear{2006})
\doiurl{10.1002/joc.1287} .
Accessed 2023-01-03
\end{barticle}
\endbibitem

\bibitem[\protect\citeauthoryear{Sharma
  et~al.}{2022}]{sharma_exponentiated_2022}
\begin{barticle}
\bauthor{\bsnm{Sharma}, \binits{V.K.}},
\bauthor{\bsnm{Singh}, \binits{S.V.}},
\bauthor{\bsnm{Shekhawat}, \binits{K.}}:
\batitle{Exponentiated {Teissier} distribution with increasing, decreasing and
  bathtub hazard functions}.
\bjtitle{Journal of Applied Statistics}
\bvolume{49}(\bissue{2}),
\bfpage{371}--\blpage{393}
(\byear{2022})
\doiurl{10.1080/02664763.2020.1813694} .
Accessed 2024-01-29
\end{barticle}
\endbibitem

\bibitem[\protect\citeauthoryear{Shayeghi
  et~al.}{2024}]{shayeghi_assessing_2024}
\begin{barticle}
\bauthor{\bsnm{Shayeghi}, \binits{A.}},
\bauthor{\bsnm{Ziveh}, \binits{A.R.}},
\bauthor{\bsnm{Bakhtar}, \binits{A.}},
\bauthor{\bsnm{Teymoori}, \binits{J.}},
\bauthor{\bsnm{Hanel}, \binits{M.}},
\bauthor{\bsnm{Vargas~Godoy}, \binits{M.R.}},
\bauthor{\bsnm{Markonis}, \binits{Y.}},
\bauthor{\bsnm{AghaKouchak}, \binits{A.}}:
\batitle{Assessing drought impacts on groundwater and agriculture in {Iran}
  using high-resolution precipitation and evapotranspiration products}.
\bjtitle{Journal of Hydrology}
\bvolume{631},
\bfpage{130828}
(\byear{2024})
\doiurl{10.1016/j.jhydrol.2024.130828} .
Accessed 2024-12-03
\end{barticle}
\endbibitem

\bibitem[\protect\citeauthoryear{Trentini et~al.}{2023}]{trentini_novel_2023}
\begin{barticle}
\bauthor{\bsnm{Trentini}, \binits{L.}},
\bauthor{\bsnm{Dal~Gesso}, \binits{S.}},
\bauthor{\bsnm{Venturini}, \binits{M.}},
\bauthor{\bsnm{Guerrini}, \binits{F.}},
\bauthor{\bsnm{Calmanti}, \binits{S.}},
\bauthor{\bsnm{Petitta}, \binits{M.}}:
\batitle{A {Novel} {Bias} {Correction} {Method} for {Extreme} {Events}}.
\bjtitle{Climate}
\bvolume{11}(\bissue{1}),
\bfpage{3}
(\byear{2023})
\doiurl{10.3390/cli11010003} .
Accessed 2023-05-03
\end{barticle}
\endbibitem

\bibitem[\protect\citeauthoryear{Tencaliec
  et~al.}{2019}]{tencaliec_flexible_2019}
\begin{barticle}
\bauthor{\bsnm{Tencaliec}, \binits{P.}},
\bauthor{\bsnm{Favre}, \binits{A.-C.}},
\bauthor{\bsnm{Naveau}, \binits{P.}},
\bauthor{\bsnm{Prieur}, \binits{C.}},
\bauthor{\bsnm{Nicolet}, \binits{G.}}:
\batitle{Flexible semiparametric {Generalized} {Pareto} modeling of the entire
  range of rainfall amount}.
\bjtitle{Environmetrics}
\bvolume{31}(\bissue{2}),
\bfpage{2582}--\blpage{1}
(\byear{2019})
\doiurl{10.1002/env.2582} .
Accessed 2023-10-12
\end{barticle}
\endbibitem

\bibitem[\protect\citeauthoryear{Themeßl
  et~al.}{2012}]{themesl_empirical-statistical_2012}
\begin{barticle}
\bauthor{\bsnm{Themeßl}, \binits{M.J.}},
\bauthor{\bsnm{Gobiet}, \binits{A.}},
\bauthor{\bsnm{Heinrich}, \binits{G.}}:
\batitle{Empirical-statistical downscaling and error correction of regional
  climate models and its impact on the climate change signal}.
\bjtitle{Climatic Change}
\bvolume{112}(\bissue{2}),
\bfpage{449}--\blpage{468}
(\byear{2012})
\doiurl{10.1007/s10584-011-0224-4} .
Accessed 2023-02-08
\end{barticle}
\endbibitem

\bibitem[\protect\citeauthoryear{Vaittinada~Ayar
  et~al.}{2016}]{vaittinada_ayar_intercomparison_2016}
\begin{barticle}
\bauthor{\bsnm{Vaittinada~Ayar}, \binits{P.}},
\bauthor{\bsnm{Vrac}, \binits{M.}},
\bauthor{\bsnm{Bastin}, \binits{S.}},
\bauthor{\bsnm{Carreau}, \binits{J.}},
\bauthor{\bsnm{Déqué}, \binits{M.}},
\bauthor{\bsnm{Gallardo}, \binits{C.}}:
\batitle{Intercomparison of statistical and dynamical downscaling models under
  the {EURO}- and {MED}-{CORDEX} initiative framework: present climate
  evaluations}.
\bjtitle{Climate Dynamics}
\bvolume{46}(\bissue{3}),
\bfpage{1301}--\blpage{1329}
(\byear{2016})
\doiurl{10.1007/s00382-015-2647-5} .
Accessed 2025-01-21
\end{barticle}
\endbibitem

\bibitem[\protect\citeauthoryear{Verrelle
  et~al.}{2021}]{verrelle_cerra-land_2021}
\begin{botherref}
\oauthor{\bsnm{Verrelle}, \binits{A.}},
\oauthor{\bsnm{Glinton}, \binits{M.}},
\oauthor{\bsnm{Bazile}, \binits{E.}},
\oauthor{\bsnm{Moigne}, \binits{P.L.}}:
{CERRA}-{Land} : {A} new land surface reanalysis at 5.5 km resolution over
  {Europe}.
Technical Report EMS2021-492,
Copernicus Meetings
(June 2021).
\url{https://meetingorganizer.copernicus.org/EMS2021/EMS2021-492.html}
Accessed 2024-03-21
\end{botherref}
\endbibitem

\bibitem[\protect\citeauthoryear{Velasquez et~al.}{2020}]{velasquez_new_2020}
\begin{barticle}
\bauthor{\bsnm{Velasquez}, \binits{P.}},
\bauthor{\bsnm{Messmer}, \binits{M.}},
\bauthor{\bsnm{Raible}, \binits{C.C.}}:
\batitle{A new bias-correction method for precipitation over complex terrain
  suitable for different climate states: a case study using {WRF} (version
  3.8.1)}.
\bjtitle{Geoscientific Model Development}
\bvolume{13}(\bissue{10}),
\bfpage{5007}--\blpage{5027}
(\byear{2020})
\doiurl{10.5194/gmd-13-5007-2020} .
Accessed 2023-02-23
\end{barticle}
\endbibitem

\bibitem[\protect\citeauthoryear{Vrac et~al.}{2016}]{vrac_bias_2016}
\begin{barticle}
\bauthor{\bsnm{Vrac}, \binits{M.}},
\bauthor{\bsnm{Noël}, \binits{T.}},
\bauthor{\bsnm{Vautard}, \binits{R.}}:
\batitle{Bias correction of precipitation through {Singularity} {Stochastic}
  {Removal}: {Because} occurrences matter}.
\bjtitle{Journal of Geophysical Research: Atmospheres}
\bvolume{121}(\bissue{10}),
\bfpage{5237}--\blpage{5258}
(\byear{2016})
\doiurl{10.1002/2015JD024511} .
Accessed 2022-11-04
\end{barticle}
\endbibitem

\bibitem[\protect\citeauthoryear{Vigaud
  et~al.}{2013}]{vigaud_probabilistic_2013}
\begin{barticle}
\bauthor{\bsnm{Vigaud}, \binits{N.}},
\bauthor{\bsnm{Vrac}, \binits{M.}},
\bauthor{\bsnm{Caballero}, \binits{Y.}}:
\batitle{Probabilistic downscaling of {GCM} scenarios over southern {India}}.
\bjtitle{International Journal of Climatology}
\bvolume{33}(\bissue{5}),
\bfpage{1248}--\blpage{1263}
(\byear{2013})
\doiurl{10.1002/joc.3509} .
Accessed 2025-01-21
\end{barticle}
\endbibitem

\bibitem[\protect\citeauthoryear{Vautard et~al.}{2015}]{vautard_extreme_2015}
\begin{barticle}
\bauthor{\bsnm{Vautard}, \binits{R.}},
\bauthor{\bsnm{Yiou}, \binits{P.}},
\bauthor{\bsnm{Oldenborgh}, \binits{G.-J.v.}},
\bauthor{\bsnm{Lenderink}, \binits{G.}},
\bauthor{\bsnm{Thao}, \binits{S.}},
\bauthor{\bsnm{Ribes}, \binits{A.}},
\bauthor{\bsnm{Planton}, \binits{S.}},
\bauthor{\bsnm{Dubuisson}, \binits{B.}},
\bauthor{\bsnm{Soubeyroux}, \binits{J.-M.}}:
\batitle{Extreme {Fall} 2014 {Precipitation} in the {Cévennes} {Mountains}}.
\bjtitle{Bulletin of the American Meteorological Society}
(\byear{2015})
\doiurl{10.1175/BAMS-D-15-00088.1} .
Accessed 2025-01-21
\end{barticle}
\endbibitem

\bibitem[\protect\citeauthoryear{Wang et~al.}{2014}]{wang_global_2014}
\begin{barticle}
\bauthor{\bsnm{Wang}, \binits{C.}},
\bauthor{\bsnm{Zhang}, \binits{L.}},
\bauthor{\bsnm{Lee}, \binits{S.-K.}},
\bauthor{\bsnm{Wu}, \binits{L.}},
\bauthor{\bsnm{Mechoso}, \binits{C.R.}}:
\batitle{A global perspective on {CMIP5} climate model biases}.
\bjtitle{Nature Climate Change}
\bvolume{4}(\bissue{3}),
\bfpage{201}--\blpage{205}
(\byear{2014})
\doiurl{10.1038/nclimate2118} .
Accessed 2024-11-14
\end{barticle}
\endbibitem

\bibitem[\protect\citeauthoryear{Xu et~al.}{2015}]{xu_evaluation_2015}
\begin{barticle}
\bauthor{\bsnm{Xu}, \binits{H.}},
\bauthor{\bsnm{Xu}, \binits{C.-Y.}},
\bauthor{\bsnm{Sælthun}, \binits{N.R.}},
\bauthor{\bsnm{Zhou}, \binits{B.}},
\bauthor{\bsnm{Xu}, \binits{Y.}}:
\batitle{Evaluation of reanalysis and satellite-based precipitation datasets in
  driving hydrological models in a humid region of {Southern} {China}}.
\bjtitle{Stochastic Environmental Research and Risk Assessment}
\bvolume{29}(\bissue{8}),
\bfpage{2003}--\blpage{2020}
(\byear{2015})
\doiurl{10.1007/s00477-014-1007-z} .
Accessed 2024-07-09
\end{barticle}
\endbibitem

\bibitem[\protect\citeauthoryear{Şan et~al.}{2023}]{san_daily_2023}
\begin{barticle}
\bauthor{\bsnm{Şan}, \binits{M.}},
\bauthor{\bsnm{Nacar}, \binits{S.}},
\bauthor{\bsnm{Kankal}, \binits{M.}},
\bauthor{\bsnm{Bayram}, \binits{A.}}:
\batitle{Daily precipitation performances of regression-based statistical
  downscaling models in a basin with mountain and semi-arid climates}.
\bjtitle{Stochastic Environmental Research and Risk Assessment}
\bvolume{37}(\bissue{4}),
\bfpage{1431}--\blpage{1455}
(\byear{2023})
\doiurl{10.1007/s00477-022-02345-5} .
Accessed 2024-07-08
\end{barticle}
\endbibitem

\end{thebibliography}

\end{document}